\documentclass[10pt,floatfix,aps,prx,superscriptaddress,twocolumn,showpacs,noeprint,notitlepage,longbibliography]{revtex4-2}

\usepackage{graphicx}
\usepackage{dcolumn}
\usepackage{bm}
\usepackage{tikz-cd}
\usepackage{mathrsfs}
\usepackage{algpseudocode}
\usepackage[caption=false]{subfig}
\usepackage{graphicx,amsfonts,times,bm,amsmath,amssymb,verbatim,ifthen,braket,xcolor,array,comment,enumerate,multirow,soul,lineno,hyperref}
\usepackage{capt-of}
\usepackage{booktabs}
\usepackage[normalem]{ulem}

\newcommand{\He}{\overline{\hat{H}_{\text{eff}}^{(0)}}}

\begin{document}

\title{Strong-Drive Floquet Engineering of Interacting Qudits: From Finite-Duration Controls to Emergent Symmetry}

\author{Ryan Scott}
\affiliation{Department of Physics, Virginia Tech, Blacksburg, Virginia 24061, USA}

\author{V. W. Scarola}
\email[Email address:]{scarola@vt.edu}
\affiliation{Department of Physics, Virginia Tech, Blacksburg, Virginia 24061, USA}

\begin{abstract} Floquet driving uses periodic controls to tailor the behavior of quantum systems, with applications in quantum analogue simulation, sensing, and the protection of quantum information. Most approaches are designed using idealized, instantaneous pulses, even though experiments necessarily use pulses with finite duration and shape. This mismatch becomes especially challenging for interacting $d$-level systems, or qudits, because the number of possible controls grows rapidly with the number of levels. We develop a strong-drive Floquet theory that incorporates experimentally realizable pulse waveforms directly into the design of the effective interactions. The pulse duration, amplitude, and shape therefore become useful control parameters rather than sources of error. We show that systems with more than two levels offer capabilities unavailable in qubit systems: finite-duration driving can create new interactions that are absent from the original system and can substantially change its symmetries. We demonstrate these capabilities for interacting three-level systems. A single pulse transforms a diagonal interaction into a quantum spin-1 model dominated by nematic interactions, while pulse protocols motivated by trapped ultracold polar molecules produce models with enlarged $SU(2)\times U(1)$ and $SU(3)$ symmetries. Numerical tests of both short-time evolution and many-body dynamics confirm the accuracy of the resulting description. Our results provide a scalable analytical framework for designing finite-duration controls in interacting qudit platforms. 
\end{abstract}

\maketitle
\section{Introduction}
\label{sec_introduction}

Periodic driving provides a powerful route to synthesizing effective quantum Hamiltonians with properties that are inaccessible in static systems.  Floquet engineering addresses Hamiltonian design by applying a periodic drive
$\hat{V}(t)$ to a native Hamiltonian $\hat{H}_0$
\cite{BLANES2009,LESKES2010,Goldman2014,BUKOV2015,ECKARDT2015}:
\begin{equation}
    \hat{H}(t)=\hat{H}_0+\hat{V}(t).
    \label{eq_Hamiltonian}
\end{equation}
For $\hat{V}(t+T)=\hat{V}(t)$, with angular frequency:
\begin{align}
    \omega=\frac{2\pi}{T},
\end{align}
Floquet theory provides an effective description of the driven dynamics.
Properly chosen drives can dynamically suppress unwanted terms in
$\hat{H}_0$ \cite{Vandersypen2005,SUTER2016}, realize symmetric operating
points for quantum sensing \cite{DEGEN2017}, and generate many-body
Hamiltonians for quantum analogue simulation
\cite{Georgescu2014,DALEY2022}.  Extending these capabilities to interacting multilevel systems remains an important open challenge.

Multilevel quantum systems provide an expanded control space and access to
many-body phenomena unavailable in two-level architectures. Arrays of such
$d$-level systems, or qudits, are now realized across neutral atoms
\cite{CHAUDHURY2007,ANDERSON2015,AHMED2025}, optically trapped ultracold
polar molecules \cite{PICARD2024a,Hepworth2025,RUTTLEY2025}, molecular
magnets \cite{THIELE2014,GODFRIN2017,HUSSAIN2018,BIARD2021,CHICCO2021},
trapped ions \cite{LEUPOLD2018,RINGBAUER2022,HRMO2023,EDMUNDS2025,LOW2025,LOW2026},
solid-state defects and donors
\cite{KOEHL2011,MAMIN2014,WIDMANN2015,SOLTAMOV2019,ASAAD2020,ADAMBUKULAM2024,FUENTES2024},
bosonic oscillators \cite{BROCK2025,ROY2025}, photonic systems
\cite{LANYON2009,KUES2017,WANG2018,ERHARD2018,CHI2022}, and superconducting
circuits
\cite{NEELEY2009,BIANCHETTI2010,YURTALAN2020,MORVAN2021,KONONENKO2021,BLOK2021,GOSS2022,LIU2023,LUO2023,GOSS2024,NGUYEN2024,WANG2025,CHAMPION2025,TRIPATHI2025}.
Their enlarged local Hilbert spaces support high-dimensional quantum
information processing and provide direct access to multipolar interactions,
synthetic dimensions, and models with non-Abelian symmetries. A central
challenge is therefore to develop Hamiltonian-engineering methods that retain
these capabilities without incurring control complexity that grows
prohibitively with $d$.

\begin{figure}[t]
\includegraphics[width=0.5\textwidth,angle=0]{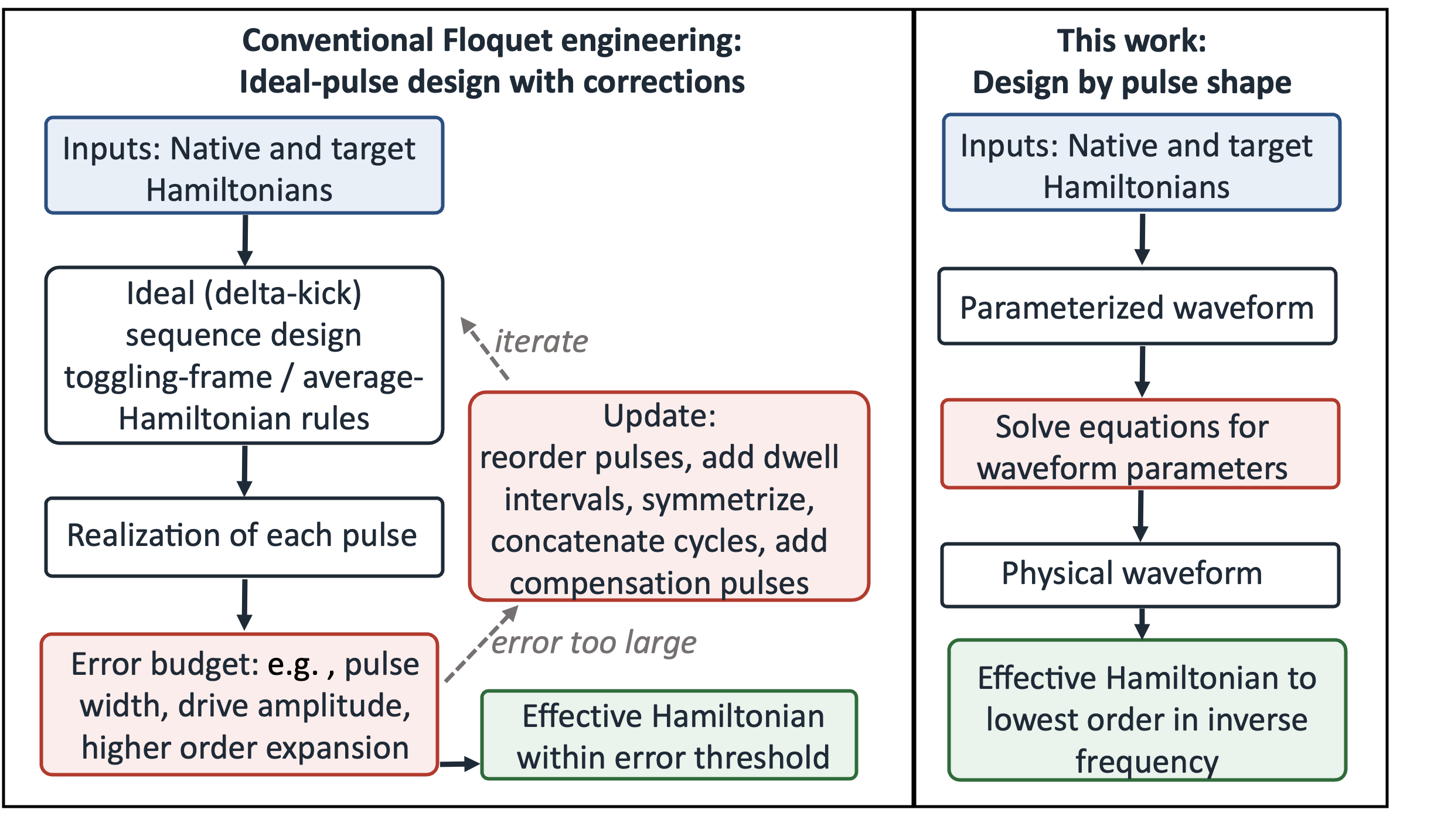}
\caption{
\emph{Floquet-engineering approaches.}
Ideal-pulse approaches first design an effective Hamiltonian using
instantaneous rotations and subsequently account for waveform-dependent
implementation errors. The present approach incorporates a calibrated
finite-duration waveform directly into the leading effective Hamiltonian and
kick operator. The two approaches can be used independently or combined by
embedding waveform-engineered blocks within longer pulse sequences.
}
\label{fig_approaches}
\end{figure}

Most pulse-based constructions first solve the Hamiltonian-design problem
using idealized instantaneous rotations. Experimentally realizable pulses are
then treated as approximations to those rotations, and additional sequence
elements or optimization steps are introduced to compensate for finite pulse
duration, amplitude constraints, and higher-order expansion errors. This
workflow, illustrated in Fig.~\ref{fig_approaches}, has its roots in nuclear
magnetic resonance and underlies many powerful methods for coherent control
and average-Hamiltonian design
\cite{Vandersypen2005,SUTER2016}. It also creates a separation between the
ideal pulse sequence used for theoretical design and the finite-duration
waveform implemented in an experiment.

This separation becomes increasingly restrictive for interacting qudits.
Addressing a larger local operator algebra enlarges the space of possible
control sequences, and correcting the ideal-pulse approximation can require
additional operations as the number of levels increases. Recent work has
developed pulse-sequence, optimization, and dynamical-decoupling methods for
interacting qudit systems
\cite{CHOI2017,OKEEFFE2019,ZHOU2024,TRIPATHI2025,BASSLER2025,READ2025}.
In particular, the growth of pulse overhead and sequence complexity with $d$
has emerged as an important challenge \cite{ZHOU2024}, while constrained
searches for suitable Hamiltonian-engineering sequences become increasingly
demanding as the available control space expands \cite{OKEEFFE2019}.
An analytical description that incorporates 
waveforms directly into the design of
interacting qudit Hamiltonians would therefore provide a useful alternative to
first idealizing the waveform and then correcting its implementation.

\begin{figure}[t]
\includegraphics[width=0.45\textwidth,angle=0]{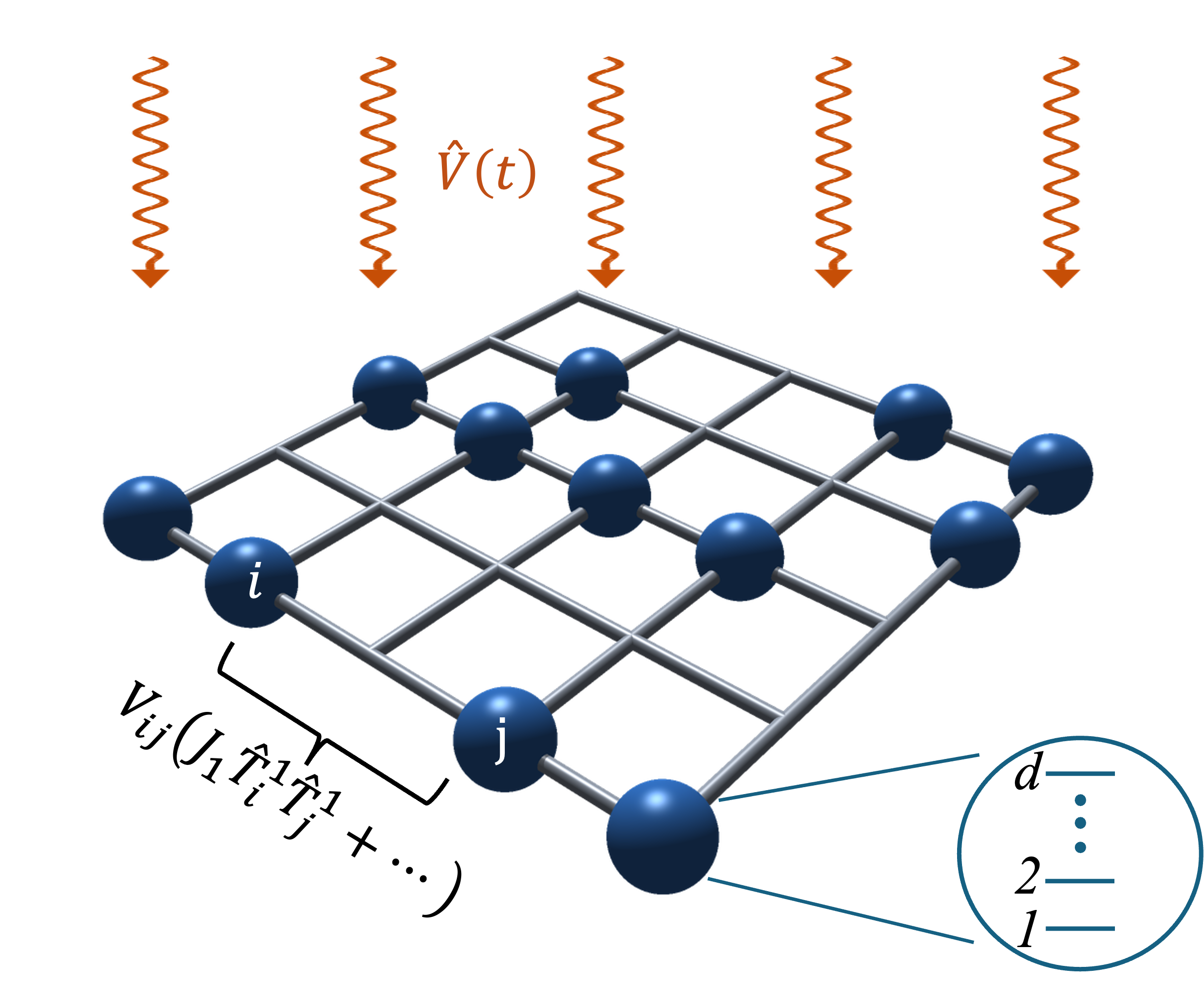}
\caption{
\emph{Global driving of an interacting qudit array.}
Spheres represent $d$-level systems on the vertices of an illustrative square
lattice, with $i$ and $j$ labeling two sites. The dimensionless coupling
$V_{ij}$ specifies the bond dependence of a native interaction
$J_{\beta}\hat{T}_{i}^{\beta}\hat{T}_{j}^{\beta}$, where
$\hat{T}_{i}^{\beta}$ is a local $d$-level generator, e.g., a matrix proportional to a Pauli matrix for $d=2$. Random site occupancy
illustrates a general interaction graph with nonuniform bond strengths. The
wavy lines denote the global time-dependent drive $\hat{V}(t)$.
}
\label{fig_lattice}
\end{figure}

Figure~\ref{fig_lattice} depicts the setting considered here: an interacting
array of qudits subject to a global time-dependent drive. Table~\ref{tab_qudit_platforms}
provides representative platforms that have both populated more than two
on-site levels and observed interactions within the addressed multilevel
manifold. These systems are immediate candidates for waveform-level
Hamiltonian engineering because the native interaction and the required
control modality are already present.

Here we build a strong-drive Floquet framework that incorporates calibrated
finite-duration waveforms directly into the Hamiltonian
design, see Fig.~\ref{fig_approaches}. We
generalize the waveform-engineering approach introduced for $d=2$ in
Ref.~\cite{SCOTT2025} to interacting systems with arbitrary finite local
dimension $d$. At leading order in inverse drive frequency, we resum the
strong drive rather than expanding in its amplitude. The waveform
then enters the effective couplings non-perturbatively through a small set of
bounded pulse-profile averages. Consequently, finite pulse width is not by
itself an error within the leading-order theory. Instead, pulse width,
amplitude, and shape become control parameters that determine the engineered
Hamiltonian.
The arbitrary-$d$ solution is made possible by closure of the nested
commutators in a diagonal, antisymmetric, and symmetric basis of
$\mathfrak{su}(d)$.

An important capability enabled by this construction has no qubit analogue. For
$d=2$, global driving can only redistribute interaction strength among existing
spin channels while preserving a trace sum rule \cite{CHOI2017}. For $d>2$, the enlarged local
operator algebra instead allows finite-duration global controls to generate
same-channel and cross-channel interactions that are absent from the native
Hamiltonian. We derive closed-form coupling equations and finite lookup rules
for designing these new interaction structures at arbitrary finite $d$, without
a numerical search over pulse sequences.

\begin{table}[t]
\renewcommand{\arraystretch}{1.25}
\setlength{\tabcolsep}{3pt}
\caption{Representative experimental platforms hosting localized qudits with
spin-like models and global control. The quantity $d$ is the dimension of the
on-site Hilbert space addressed in the cited work. Each platform has populated
more than two on-site levels and observed an intrinsic inter-qudit interaction
acting within the full multilevel manifold.}
\label{tab_qudit_platforms}
\footnotesize
\begin{tabular*}{\columnwidth}{@{\extracolsep{\fill}} l l c l l @{}}
\hline\hline
Platform &
Encoding &
$d$ &
Global drive &
Refs. \\
\hline
Trapped ions &
Hyperfine states &
3 &
Optical &
\cite{SENKO2015} \\
Diamond vacancies &
Ground triplet &
3 &
Microwave &
\cite{ZHOU2024} \\
Rydberg tweezer array &
Rydberg levels &
4 &
Microwave &
\cite{CHEN2024a,CHEN2025} \\
$^{52}$Cr, optical lattice &
Zeeman manifold &
7 &
Radio &
\cite{DePaz2013a,LEPOUTRE2019} \\
$^{167}$Er, optical lattice &
Hyperfine manifold &
20 &
Radio &
\cite{PATSCHEIDER2020} \\
\hline\hline
\end{tabular*}
\end{table}

We demonstrate these capabilities using interacting qutrits and their mapping
to spin-1 magnetism. First, a single square-pulse block converts a diagonal interaction, of a form relevant to qutrit transmon systems
\cite{GOSS2022,GOSS2024}, into a quantum spin-1 Hamiltonian with dominant
nematic couplings and continuous quadrupolar symmetry. We then consider
dipolar models motivated by trapped ultracold polar molecules
\cite{DeMille2002,Barnett2006,Wall2014,Hermsmeier2024}. For a native anisotropic exchange
model derived from rotational states of RbCs and parameterized using recent
multilevel molecular experiments
\cite{BLACKMORE2020,RUTTLEY2025,Hepworth2025}, a single pulse removes the
native dipole--quadrupole anisotropy and produces an
$SU(2)\times U(1)$-symmetric effective Hamiltonian. Finally, a two-pulse
protocol generates a coupling absent from the native model and tunes an
anisotropic interaction to an $SU(3)$-symmetric
Uimin-Lai-Sutherland-type Hamiltonian
\cite{UIMIN1970,LAI1974,SUTHERLAND1975}. These examples show that compact
finite-duration waveform blocks can generate higher-dimensional symmetry and
new interaction channels rather than merely approximate predetermined ideal
rotations.

We establish the regime of validity in two complementary ways. We use Floquet-prethermal arguments to identify
a long-lived regime governed by the engineered Hamiltonian. We then compare
exact and effective propagators and many-body correlator dynamics. The numerical results verify the predicted inverse-frequency scaling, demonstrate the importance of the leading kick transformation away from stroboscopic times, and show enhanced stroboscopic accuracy for a symmetric two-pulse block.

The framework complements rather than replaces ideal-pulse and
sequence-optimization approaches. Waveform-engineered blocks can be
concatenated, incorporated into longer sequences, or used to quantify the
error incurred when a finite-duration control is replaced by an ideal kick.
The resulting workflow connects a native interacting-qudit Hamiltonian and an
experimentally realizable waveform directly to a leading effective
Hamiltonian and kick operator with a controlled inverse-frequency error.
It therefore provides a direct route to finite-duration Hamiltonian
engineering for quantum simulation, dynamical
decoupling, and sensing across interacting qudit platforms.

The remainder of the paper is organized as follows.
Section~\ref{sec_definitions_assumptions} defines the native Hamiltonians,
strong-drive pulses, and prethermal conditions. In
Sec.~\ref{sec_general_approach} we derive the leading effective description,
and Sec.~\ref{sec_tunable_effective_hamiltonian} presents the closed-form
one- and two-body coupling equations. Section~\ref{sec_lookup_tables}
constructs the arbitrary-$d$ lookup rules, and Sec.~\ref{sec_sum_rule}
establishes the leading-order two-body trace sum rule.
Sections~\ref{sec_d_2_example} and~\ref{sec_top_level_d_3} apply the formalism
to qubits and qutrits, respectively. Section~\ref{sec_numerical} presents
numerical tests of propagators and many-body dynamics,
Sec.~\ref{sec_code} describes the open-source implementation, and
Sec.~\ref{sec_summary} summarizes the results and outlook.

\section{Definitions and Assumptions}
\label{sec_definitions_assumptions}

We begin by defining relevant variables and listing assumptions underlying our formalism. We assume $\hat{H}(t)$ has the form Eq.~\eqref{eq_Hamiltonian} and that the drive frequency exceeds every energy scale of $\hat{H}_0$.  As we will see, this is a non-trivial assumption even within the sequential,
zero-area, midpoint-antisymmetric pulse class considered here, because the
pulse-shape parameters introduce additional energy scales.  We will use high frequency expansions to find lowest order contributions to the effective theory \cite{Goldman2014,SCOTT2025}.

To derive the effective theory we impose a gauge transformation on the system:
\begin{align}
    |\psi(t)\rangle &\rightarrow |\phi(t)\rangle = e^{i\hat{K}(t)}|\psi(t)\rangle, \label{eq:stateVectorGaugeTransform}\\
    \hat{H}(t) &\rightarrow \hat{H}_{\text{eff}} = e^{i\hat{K}(t)}\hat{H}(t)e^{-i\hat{K}(t)} + i\left(\frac{\partial e^{i\hat{K}(t)}}{\partial t}\right)e^{-i\hat{K}(t)}, \label{eq:hamiltonianGaugeTransform}
\end{align}
where $\hat{K}(t)$ is a kick operator that effectively gauge transforms the original solutions of the Schr\"odinger equation,  $|\psi(t)\rangle$, into $|\phi(t)\rangle$.  $\hat{K}(t)$ and $\hat{H}_{\text{eff}}$ combined define the effective theory at all $t$. 
To make progress we expand both in powers of inverse frequency:
\begin{align}
    \hat{H}_{\text{eff}} &= \sum_{n=0}^{\infty}\omega^{-n}\hat{H}^{(n)}_{\text{eff}},\label{eq:floquetHamiltonianExpansion}\\
    \hat{K} &= \sum_{n=0}^{\infty} \omega^{-n}\hat{K}^{(n)}. \label{eq:kickExpansion}
\end{align}
The central problem is then to find both the lowest order kick operator and effective Hamiltonian so that the expansion is consistent to the same order in inverse frequency.  We will derive the lowest order kick operator, $\hat{K}^{(0)}(t)$, and the time average of $\hat{H}^{(0)}_{\text{eff}}$.  These two quantities will be the output of our theory.

\subsection{Operators and choice for $\hat{H}_0$}
\subsubsection{Models are defined as a graph of generators}
We consider an arbitrary real space graph with each vertex (site) labeled with
$i$ and $j$ admitting an elementary $d$-level system.  Graph edges between sites
$i$ and $j$ are thought of as bonds.  Figure~\ref{fig_lattice} depicts an
example of a square lattice.

If $\mathcal{H}_{\Sigma}$ is the $d$-dimensional Hilbert space of a single site
and there are $N$ sites, the composite Hilbert space is
$\mathcal{H}=\mathcal{H}_{\Sigma}^{\otimes N}$.  The traceless part of any local
observable at site $j$ is an element of $\mathfrak{su}(d)$.  We use the
convention in which $\mathfrak{su}(d)$ is spanned by $d^{2}-1$ traceless
Hermitian matrices.  Operators on $\mathcal{H}$ are built from products of these
on-site generators together with the identity.

We define $d$-level generators.  For each site $j$ we introduce Hermitian,
traceless generators $\hat{T}^{\alpha}_{j}\in\mathfrak{su}(d)$, labeled by Greek
channel indices $\alpha,\beta=1,2,\dots$.  These generators commute off-site:
$[\hat T_i^\alpha,\hat T_j^\beta]=0$ for $i\neq j$.

The on-site ($i=j$) commutation relations establish the Lie algebra.  $\hat{T}^{\beta}_j$ can be written as  $d\times d$ matrices for each site $j$.  Examples include the Pauli matrices for $d=2$.  The matrices are either diagonal, antisymmetric, or symmetric (D.A.S. decomposition).  Specifically, $\hat{T}^{\beta}_j$ denotes one of the following matrices at a site $j$:
\begin{align}
    \text{Diagonal: } \hat D_j^{\,r}&:=
    \frac12\left(|r\rangle_j\langle r|-|r+1\rangle_j\langle r+1|\right)\nonumber \\
    \text{Antisymmetric: }  \hat A_j^{mn}&:= \frac{i}{2}\bigg(|m\rangle_j \langle n| - |n\rangle_j \langle m|\bigg) \nonumber \\
     \text{Symmetric: } \hat S_j^{mn} &:= \frac{1}{2}\bigg(|m\rangle_j\langle n| + |n\rangle_j\langle m|\bigg)
    \label{eq_S_A_D_matrices}
\end{align}
where integers $m$ and $n$ label positions in the matrices such that $1\le m<n\le d$ and $r=1,\dots,d-1$. 

Using these definitions we can derive the on-site commutation relations.  One finds:
\begin{align}\left[\hat{T}^{\alpha}_j,\hat{T}^{\beta}_j \right] =i \sum_{\gamma} c_{\alpha\beta\gamma}\hat{T}^{\gamma}_j,
\end{align}
where $c_{\alpha\beta\gamma}$ are structure constants. Sections~\ref{sec_tunable_effective_hamiltonian} and \ref{sec_lookup_tables} will discuss the structure constants for any finite $d$.  

\subsubsection{The bare (undriven) Hamiltonian has one- and two-body terms} 

Prior to driving, we assume a time-independent Hamiltonian with both on-site one-body terms and with channel-diagonal two-body bond terms:
\begin{align}
 \hat{H}_{0} &= \hat{H}_{0}^{h}+\hat{H}_{0}^{J}  \label{eq_H0_assumed} \\
    \hat{H}_{0}^{h} &:= \sum_{j,\beta} \epsilon_j
      h_{\beta}\,\hat{T}_{j}^{\beta} \nonumber \\
      \hat{H}_0^J &:= \frac{1}{2} \sum_{i\neq j} V_{ij}\sum_{\beta}J_{\beta} \hat{T}_i^{\beta}\hat{T}_j^{\beta}.
    \nonumber
\end{align}
 $h_{\beta}$ are local field strengths and $J_{\beta}$ are bond strengths.  $\epsilon_j$ is a unit-less factor that can shift the local energy from site to site.  Similarly, $V_{ij}$ is a unit-less symmetric adjacency coupling matrix such that $V_{ij}=V_{ji}$.  Examples of $V_{ij}$ include nearest neighbor interactions on a chain, e.g., $V_{ij}\rightarrow \delta_{i,j+1}+\delta_{j,i+1}$, and a dipole-dipole interaction. The interaction strengths and the field strengths are fixed by the physical system of interest.  We will include $h_{\beta}$ in our formalism but set it to zero in discussing examples.

\subsection{Strong Driving Pulse Properties}
\label{sec_pulse_assumption_higher_level}

\subsubsection{Pulses are global and strong}

We assume pulses that are global (acting on all sites at once) and have strong driving such that: 
\begin{align}
\hat{V}(t)=\omega \sum_{\alpha} g_{\alpha}(t) \hat{T}^{\alpha}
\label{eq_pulse_general}
\end{align}
where $\hat{T}^{\alpha} := \sum_j \hat{T}^{\alpha}_j$ acts on all sites and $g_{\alpha}(t)$ is a chosen pulse profile. 
We also assume that the pulses are periodic with period $T$: $\hat{V}(t)=\hat{V}(t+T)$.

Strong driving means that the drive amplitude scales with $\omega$ and can
therefore exceed every characteristic energy in $\hat{H}_{0}$. 
Consider the following approximation to local energy scales:
\begin{align} \mathcal{J}:= \max_i |\epsilon_i| \sum_\beta |h_\beta| + \max_i\sum_{j\neq i}|V_{ij}| \sum_\beta |J_\beta|. 
\label{eq:local_scale}
\end{align}
The
expansion below is controlled by:
\begin{align}
\mathcal{J}/\omega\ll1.
\label{eq:strong_drive_hierarchy}
\end{align}

We impose no perturbative smallness condition on the drive amplitude.
This distinguishes the present
expansion from high-frequency expansions in which the drive is treated as a
perturbation of the same order as $\hat{H}_{0}$: here the drive enters the
effective couplings through a resummation rather than order by order.  In particular, the fact that the drive exceeds a given bond strength is not a
source of error.  It is the defining feature of the strong-driving regime, and it
holds for the strongest and the weakest bonds in $V_{ij}$ alike.

\subsubsection{Pulses are ordered and average to zero}
\label{sec_pulse_properties_general}

Pulse shapes will have two properties on which the formalism relies.  
We will rely on the integral of the pulse profile using:
\begin{align}
    G_{\alpha}(t):=\omega \int_0^t ds\, g_{\alpha}(s).
    \label{eq_define_G_general}
\end{align}
First, we assume the channels act in sequential, non-overlapping time
blocks (time ordering) and that each $g_{\alpha}(t)$ integrates to zero
within its own block.  Consequently $G_{\alpha}(t)$ vanishes outside its
block and the $G_{\alpha}(t)$ have disjoint support.  At most one
$G_{\alpha}(t)$ is therefore nonzero at any instant.  Below, this will be used to give $[\hat{K}^{(0)}(t),\partial\hat{K}^{(0)}(t)/\partial t]=0$.  Disjoint support also gives
$\overline{G_{\alpha}(t)G_{\alpha'}(t)}\propto\delta_{\alpha,\alpha'}$, where the
overline denotes the full-cycle average $\overline{X}:=T^{-1}\int_{0}^{T}dt\,X(t)$.
Second, we assume that 
$G_{\alpha}(t)$ is antisymmetric about
the midpoint of its block, so that $\overline{\{G_{\alpha}(t)\}^p}=0$  for $p$ odd.
We refer to waveforms satisfying these requirements as the sequential,
zero-area, midpoint-antisymmetric pulse class. Within this class the waveform
may otherwise have arbitrary finite duration and shape. Square-pulse examples
are discussed in Appendix~\ref{sec_example_pulses} and will be used
extensively.

\subsection{The Toggling Frame Admits a Prethermal Regime}
\label{sec_prethermal_assumption}

Periodic driving generally causes many-body systems to absorb energy.  At high
frequency, however, this heating can be exponentially slow, producing a
prethermal regime described by a quasi-conserved Hamiltonian
\cite{MORI2016,ABANIN2017c,ABANIN2017d,HO2023}.  For a $k$-local Hamiltonian with local
energy scale $\Lambda$, the heating time is bounded by:
\begin{align}
\tau_{\mathrm{heat}}
\gtrsim
\frac{1}{\Lambda}
\exp\left(\frac{c_h\omega}{k\Lambda}\right),
\label{eq_prethermal_lifetime}
\end{align}
where $c_h>0$ is a numerical constant.  The Hamiltonians considered here contain
at most two-body interactions and therefore have $k=2$.

To define $\Lambda$, write the undriven Hamiltonian as \cite{HO2023}:
\begin{align}
\hat H_0
=
\sum_{\substack{X\subseteq\{1,\ldots,N\}\\ |X|\leq 2}}
\hat h_X^{\,0},
\label{eq_H0_local_decomposition}
\end{align}
where $X$ is a set of one or two sites and $\hat h_X^{\,0}$ acts
only on $X$.  The relevant local energy scale is:
\begin{align}
\Lambda
:=
\max_i
\sum_{\substack{X\ni i\\ |X|\leq 2}}
\left\Vert\hat h_X^{\,0}\right\Vert,
\label{eq_Lambda_definition}
\end{align}
where 
$\Vert\hat{M}\Vert$ denotes the spectral norm (largest singular value) of $\hat{M}$. 
Thus, $\Lambda$ measures the total interaction strength associated with any
one site, rather than the norm of the full many-body Hamiltonian.  A
system-size-independent prethermal bound requires $\Lambda$ to remain finite
in the thermodynamic limit.

The strong drive in Eq.~\eqref{eq_pulse_general} has amplitude
$\mathcal{O}(\omega)$, so applying Eq.~\eqref{eq_Lambda_definition} directly in
the laboratory frame would give a local scale that grows with $\omega$.  $\Lambda$ would increase with frequency, and the
usual high-frequency bound would not be useful.  

$\Lambda$ can be bounded in frequency 
by transforming to the toggling frame.
Define the drive propagator:
$e^{-i\hat K^{(0)}(t)},
$
where only one generator is active at any instant, every pulse block has zero net pulse area, and $\hat K^{(0)}(t)$ vanishes at multiples of the period.  
The corresponding toggling-frame Hamiltonian is
$
e^{i\hat K^{(0)}(t)}
\hat H_0
e^{-i\hat K^{(0)}(t)}.
$
Using this construction we will need to show that the toggling-frame
Hamiltonian contains no term proportional to the drive amplitude, so that
$\Lambda$ evaluated in this frame is independent of $\omega$.
We will construct $\hat K^{(0)}(t)$ explicitly and show that this is true in the following sections.

Additional conditions must be imposed on the spatial couplings. 
For the
 Hamiltonian in Eq.~\eqref{eq_H0_assumed}, we assume the one-body terms are bounded: $\sup_i \vert \epsilon_i\vert<\infty$. 
For the two-body terms, a sufficient  condition for
$\Lambda$ to remain finite is:
\begin{align}
\sup_i\sum_{j\neq i}|V_{ij}|<\infty.
\label{eq_Vij_summability}
\end{align}
In particular, dipolar interactions satisfy
Eq.~\eqref{eq_Vij_summability} in two dimensions.  Three-dimensional dipolar
interactions are marginal and instead require a prethermal result formulated
for power-law interactions, such as Ref.~\cite{MACHADO2020}.

\section{General Approach}
\label{sec_general_approach}

In this section we describe the general formalism underlying our derivation of the effective Hamiltonian and the kick operators.   Appendix~\ref{sec_derivation_lowest_order} shows that results for the lowest order effective Hamiltonian and kick operator derived in Ref.~\onlinecite{SCOTT2025} hold for arbitrary finite $d$
under the assumptions of Sec.~\ref{sec_definitions_assumptions}.  Appendix~\ref{sec_derivation_lowest_order} also derives the higher order equations that need to be satisfied for further accuracy or higher order terms to be included in the effective Hamiltonian.  The results in this section apply to arbitrary finite-duration waveforms within
the sequential, zero-area, midpoint-antisymmetric pulse class defined in
Sec.~\ref{sec_pulse_properties_general}.  The waveform dependence is evaluated non-perturbatively in the accumulated pulse area within the leading-order Floquet Hamiltonian. The remaining corrections arise from the high-frequency expansion and begin at order $\mathcal{J}/\omega$.

We aim to find the kick operator and time-averaged effective Hamiltonian  to lowest order in $\mathcal{J}/\omega$: 
\begin{align}
 \hat{K} &=\hat{K}^{(0)}
    +\mathcal{O}(\omega^{-1}) \nonumber \\
    \overline{\hat{H}_{\text{eff}}} &=  \He+\mathcal{O}(\omega^{-1}). 
\end{align}
We insert the lowest order kick operator into Eq.~\eqref{eq:hamiltonianGaugeTransform} and expand the exponential.  Appendix~\ref{sec_derivation_lowest_order} shows that under the assumption of zero-average, time-ordered pulses we obtain:
\begin{align}
    \He &= \hat{H}_0+
    \sum_{p=1}^{\infty}\frac{i^p}{p!}\overline{\left[\left[\hat{K}^{(0)},\hat{H}_0\right]\right]_p}
    \label{eq_H0_general}
\end{align}
where $[[...]]_p$ denotes a $p$th-order nested commutator defined as:
\begin{align}
[[\hat{A},\hat{B}]]_0&= \hat{B}, \nonumber\\ 
[[\hat{A},\hat{B}]]_1&=  [ \hat{A},\hat{B}], \nonumber\\ 
[[\hat{A},\hat{B}]]_2&=  [\hat{A}, [ \hat{A},\hat{B}]], \nonumber\\ 
[[\hat{A},\hat{B}]]_3&=  [\hat{A},[\hat{A}, [ \hat{A},\hat{B}]]], \nonumber\\ 
\vdots \nonumber
\end{align}

Appendix~\ref{sec_derivation_lowest_order} uses the assumptions of Sec.~\ref{sec_definitions_assumptions} to show that a consistent set of kick operator solutions leading to $\He$ can be found:
\begin{align}
\hat{K}^{(0)}(t) &=\sum_{\alpha} G_{\alpha}(t)\hat{T}^{\alpha} 
\label{eq_kick_zero_order}.
\end{align} 
Inserting $\hat{K}^{(0)}(t)$ into Eq.~\eqref{eq_H0_general} yields a general form for the effective Hamiltonian:
\begin{align}
    \He &= \hat{H}_0+
    \sum_{l=1}^{\infty} \sum_{\alpha} \frac{(-1)^l}{(2l)!}
    \overline{\{G_{\alpha}(t)\}^{2l}}\left[\left[\hat{T}^{\alpha},\hat{H}_0\right]\right]_{2l}
    \label{eq_heff_general_for_use}
\end{align}
where the waveform assumptions are that the $G_{\alpha}(t)$ occupy disjoint
blocks, vanish at their block boundaries [because each $g_{\alpha}(t)$ has zero
area], and have vanishing odd-power averages. This expression therefore
constructs $\He$ for arbitrary finite-duration waveforms within the pulse class defined in
Sec.~\ref{sec_pulse_properties_general}. Appendix~\ref{sec_example_pulses} discusses pulse examples.  Section~\ref{sec_tunable_effective_hamiltonian} uses Eqs.~\eqref{eq_kick_zero_order} and~\eqref{eq_heff_general_for_use} to derive closed form effective Hamiltonians. 

\subsection{Prethermal Behavior}

The assumed form for the kick operators satisfies the prethermal assumptions of Sec.~\ref{sec_prethermal_assumption}.  To see this, first note that the pulses are non-overlapping, so only one generator is active at any instant. Moreover, every pulse block has zero net pulse area. The propagator therefore returns to the identity after each complete period, $\exp{[-i\hat{K}^{(0)}(n_tT)]}=\hat{\mathbb I}$ for integer $n_t$. Consequently, the toggling-frame Hamiltonian contains no term proportional to the drive amplitude and is $T$-periodic.

Because the drive is generated by global one-body operators, its propagator factorizes over sites.  We can therefore compute each term in Eq.~\eqref{eq_H0_local_decomposition} separately. Conjugation by the factorized drive preserves both the support and the spectral norm of every term in the local decomposition of Eq.~\eqref{eq_H0_local_decomposition}. Consequently, for this decomposition, the toggling-frame local energy scale is equal to that of the undriven Hamiltonian and, in particular, is independent of $\omega$ and of the pulse parameters.  The optimally truncated toggling-frame Floquet Hamiltonian is quasi-conserved throughout the prethermal regime, and its leading term is $\He$. The difference between the quasi-conserved Hamiltonian and this leading term is locally of relative order $\Lambda/\omega$.

\subsection{Repeated Channels and Multi-Cycle Sequences}
\label{sec_multiblock}

The approach discussed above can be generalized to multiple cycles to enlarge the space of addressable effective Hamiltonians.  
The derivation of Eqs.~\eqref{eq_kick_zero_order} and
\eqref{eq_heff_general_for_use} used three properties of the pulse train:
the $G_{\alpha}(t)$ have disjoint support, each has zero cycle average, and
at most one generator is active at any instant.  None requires the index
$\alpha$ to label distinct generators.  

We may therefore add more cycles to further engineer the effective Hamiltonian.  We can relabel the sequence by
a block index $b=1,\dots,B$ with associated channel $\alpha(b)$, permitting a
given generator to be pulsed more than once per period.
Assigning $\alpha\to b$, the results from Sec.~\ref{sec_general_approach} carry over. 
Equation~\eqref{eq_heff_general_for_use} becomes:
\begin{align}
   \He = \hat{H}_0 +
    \sum_{p=1}^{\infty}\frac{i^p}{p!}
    \sum_{b}
    \overline{\{G_{b}(t)\}^p}
\left[\left[\hat{T}^{\alpha(b)},\hat{H}_0\right]\right]_p .
    \label{eq_Heff_multiblock}
\end{align}

At the retained leading order, we may therefore concatenate and average cycles independently of the order in which the blocks are executed. Block ordering generally enters the subleading inverse-frequency corrections.  Concatenating
$n_c$ cycles, each of duration $T$ and each separately satisfying the
assumptions of Sec.~\ref{sec_definitions_assumptions}, therefore yields a
single sequence of period $n_cT$ with:
\begin{align}
   \He\big\vert_{\text{Total}}=\frac{1}{n_c}\left[\He\big\vert_1+\He\big\vert_2+\cdots+\He\big\vert_{n_c}\right].
   \label{eq_Heff_mean}
\end{align}
 These results imply that, at leading order, we can cycle average multiple effective Hamiltonians within our formalism.  In the following we focus only on a single cycle $n_{c}=1$ with the understanding that further Hamiltonian engineering is possible for $n_{c}>1$.

\section{Tunable Effective Hamiltonians}
\label{sec_tunable_effective_hamiltonian}

We now turn to a derivation of our central result: effective Hamiltonians following from pulse assumptions in Sec.~\ref{sec_pulse_assumption_higher_level}.  Remarkably, we find that the Lie algebra allows closed form solutions of the same form for any $d$.  The solutions thus provide a versatile formalism, with corrections beginning at
$\mathcal{O}(\omega^{-1})$, for arbitrary finite-duration waveforms within the
sequential, zero-area, midpoint-antisymmetric pulse class defined in
Sec.~\ref{sec_pulse_properties_general}.

We will insert the assumed $\hat{H}_0$, Eq.~\eqref{eq_H0_assumed}, into Eq.~\eqref{eq_heff_general_for_use}.  The pulses of
Eq.~\eqref{eq_pulse_general} are global. The drive-frame transformation
therefore factorizes  over sites and cannot change the number of sites an operator acts
on.  Combined with the linearity of Eq.~\eqref{eq_heff_general_for_use} in
$\hat{H}_{0}$, the one-body and two-body sectors remain separate:
\begin{align}
\He =  \overline{\hat{H}_{\text{eff},h}^{(0)}} +  \overline{\hat{H}_{\text{eff},J}^{(0)}},
\label{eq_Heff_central}
\end{align}
where $\overline{\hat{H}_{\text{eff},h}^{(0)}}$ and $\overline{\hat{H}_{\text{eff},J}^{(0)}}$ are one- and two-body effective Hamiltonians, respectively.  We treat them separately.

We start with the two-body bond terms containing $J_{\beta}$ to derive $\overline{\hat{H}_{\text{eff},J}^{(0)}}$.  By substitution of $\hat{H}_0^J$ we obtain:
\begin{widetext}
\begin{align}
    \overline{\hat{H}_{\text{eff},J}^{(0)}} = \hat{H}_0^J +
    \frac{1}{2} 
    \sideset{}{'}\sum_{ \substack{ i\neq j \\ \alpha, \beta}}
     V_{ij}J_{\beta} 
    \sum_{l=1}^{\infty} 
    \frac{(-1)^l}{(2l)!}
    \overline{\{G_{\alpha}(t)\}^{2l}}
\left[\left[\hat{T}^{\alpha},\hat{T}_i^{\beta}\hat{T}_j^{\beta}\right]\right]_{2l},
    \label{eq_Heff_unresolved}
\end{align}
\end{widetext}
where the prime on the sum applies only to channel indices and indicates $\{\alpha,\beta\}$ such that 
$[\hat T_j^\alpha,\hat T_j^\beta]\neq 0$.  To reduce the nested commutators we can use the fact that $[\hat T_i^\alpha,\hat T_j^\beta]=0$ for $i\neq j$.  We also note that the nested commutators factorize over the two sites using the Leibniz property:
\begin{align}
\left[\left[\hat{T}^{\alpha},\hat{T}_i^{\beta}\hat{T}_j^{\beta}\right]\right]_{2l} &= \sum_{u=0}^{2l}\binom{2l}{u} \left[\left[\hat{T}^{\alpha}_{i},\hat{T}_i^{\beta}\right]\right]_{u}
\left[\left[\hat{T}^{\alpha}_{j},\hat{T}_j^{\beta}\right]\right]_{2l-u},
\label{eq_Leibniz}
\end{align}
thus leaving a sum of on-site nested commutators that must be evaluated.

We can define the nested commutators in Eq.~\eqref{eq_Leibniz} for any $d$ using the definitions in Sec.~\ref{sec_definitions_assumptions}.  
The case $\left[\left[ \hat{T}^{\alpha}_{i},\hat{T}^{\beta}_{i} \right]\right]_0=\hat{T}^{\beta}_{i}$ follows 
from the definition of nested commutators.  
For $u\ge 1$ and any non-commuting ordered pair of D.A.S. generators, at any $d$ we find:
\begin{align}
\left[\left[ \hat{T}^{\alpha}_{j},\hat{T}^{\beta}_{j} \right]\right]_u =
\frac{\phi_{\alpha \beta}^{(\wp)}}{2^{\lambda_{\alpha\beta}+(u-1)\nu_{\alpha \beta}}}
\hat{T}^{\kappa_{\wp}(\alpha,\beta)}_{j},
\label{eq_nested_commutator}
\end{align}
where the parity of $u$ is $\wp$, which is $e$ for $u \pmod{2}=0$ and $o$ for $u \pmod{2}=1$.  Here the indices $\kappa_{\wp}(\alpha,\beta)$ are a label-valued map, and $\phi_{\alpha\beta}^{(e)}=\pm 1$ and $\phi_{\alpha\beta}^{(o)}=\pm i$ are the associated phases.  $\lambda_{\alpha\beta}\in\{0,1\}$ records whether the first nonzero commutator has magnitude $1$ or $1/2$, and $\nu_{\alpha\beta}\in\{0,1\}$ records the ratio of successive commutator magnitudes past the first step.  All possible combinations of the specific values of $\phi_{\alpha\beta}^{(\wp)}$, $\lambda_{\alpha\beta}$, and $\nu_{\alpha\beta}$ will be recorded in lookup tables in Sec.~\ref{sec_lookup_tables}.  Here the output label $\kappa_{\wp}(\alpha,\beta)$ ranges over an enlarged
diagonal set, $\hat D_j^{\,mn}:=\sum_{r=m}^{n-1}\hat D_j^{\,r}$, and will also be defined in
Sec.~\ref{sec_lookup_tables}.

Equation~\eqref{eq_nested_commutator} has a useful feature allowing analytic expressions for $\He$.  We note that the structure constants for the nested expression have either no $u$ dependence (for the case $\nu_{\alpha\beta}=0$) or come with a factor $2^{-u}$ (for the case $\nu_{\alpha\beta}=1$).  This $u$ dependence can be absorbed into other terms in the series to yield analytic closed forms for $\He$. 

We define a useful generating function for expressing the effective couplings:
\begin{align}
\mathcal{C}_{\alpha}(c)
&:=2\,\overline{\sin^{2}\!\left[c\,G_{\alpha}(t)/2\right]} \nonumber \\
&=\sum_{l=1}^{\infty}\frac{(-1)^{l+1}c^{2l}}{(2l)!}
\overline{\{G_{\alpha}(t)\}^{2l}},
\label{eq_generating_function}
\end{align}
where $c$ is a $c$-number.  We can combine  Eqs.~\eqref{eq_Heff_unresolved}-~\eqref{eq_generating_function}.  The Lie algebra structure constants merely contribute to $c$. We 
use the following identities: 
$ \sum_{u=1, u\in o }^{2l-1} 
\binom{2l}{u}=2^{2l-1}$ and $\sum_{u=2, u\in e }^{2l-2}\binom{2l}{u}= 2^{2l-1}-2$. 
Using these relations we find the effective two-body Hamiltonian for any $d$:
\begin{align}
    \overline{\hat{H}_{\text{eff},J}^{(0)}} &= \hat{H}_0^J+
    \frac{1}{2} \sideset{}{'}\sum_{ \substack{ i\neq j \\ \alpha , \beta}}V_{ij}
    \bigg\{
    \tilde{J}_{\alpha,\beta} \left( \hat{T}^{\beta}_{i}\hat{T}^{\kappa_{\text{e}}(\alpha,\beta)}_{j}
    +
    \hat{T}^{\kappa_{\text{e}}(\alpha,\beta)}_{i}
    \hat{T}^{\beta}_{j}
    \right) \nonumber \\
    &+\tilde{J}^{(\text{o})}_{\alpha,\beta} \hat{T}^{\kappa_{\text{o}}(\alpha,\beta)}_{i} \hat{T}^{\kappa_{\text{o}}(\alpha,\beta)}_{j} +\tilde{J}^{(\text{e})}_{\alpha,\beta} \hat{T}^{\kappa_{\text{e}}(\alpha,\beta)}_{i} \hat{T}^{\kappa_{\text{e}}(\alpha,\beta)}_{j}
    \bigg\},
    \label{eq_Heff_central_J}
\end{align}
where the effective couplings for each channel combination become:
\begin{align}
\tilde{J}_{\alpha,\beta}
&:=-\frac{J_{\beta}\phi^{(\mathrm e)}_{\alpha\beta}}
{2^{\lambda_{\alpha\beta}-\nu_{\alpha\beta}}}\,
\mathcal{C}_{\alpha}\!\left(2^{-\nu_{\alpha\beta}}\right),
\nonumber \\
\tilde{J}^{(\mathrm o)}_{\alpha,\beta}
&:=\frac{J_{\beta}}{2^{2\lambda_{\alpha\beta}-2\nu_{\alpha\beta}+1}}\,
\mathcal{C}_{\alpha}\!\left(2^{\,1-\nu_{\alpha\beta}}\right),
\nonumber \\
\tilde{J}^{(\mathrm e)}_{\alpha,\beta}
&:=\frac{J_{\beta}}{2^{2\lambda_{\alpha\beta}-2\nu_{\alpha\beta}+1}}
\left[4\,\mathcal{C}_{\alpha}\!\left(2^{-\nu_{\alpha\beta}}\right)
-\mathcal{C}_{\alpha}\!\left(2^{\,1-\nu_{\alpha\beta}}\right)\right].
\label{eq_Jtilde_general_profile}
\end{align}
The last term in $\tilde{J}^{(\mathrm e)}_{\alpha,\beta}$ comes from removing $u=0$ and $u=2l$ terms from the even-$u$ binomial sum.  Those terms are collected in $\tilde{J}_{\alpha,\beta}$.

We point out two important features of Eq.~\eqref{eq_Heff_central_J}.  First,
$\kappa_{\rm e}(\alpha,\beta)$ need not equal $\beta$, so a bond that enters as
$\hat{T}^{\beta}_{i}\hat{T}^{\beta}_{j}$ acquires cross-channel terms
$\hat{T}^{\beta}_{i}\hat{T}^{\kappa_{\rm e}}_{j}
+\hat{T}^{\kappa_{\rm e}}_{i}\hat{T}^{\beta}_{j}$.  Second, the odd branch
generates an entirely new same-channel bond
$\hat{T}^{\kappa_{\rm o}}_{i}\hat{T}^{\kappa_{\rm o}}_{j}$ on a channel absent
from $\hat{H}_{0}$.  The drive therefore does not merely renormalize the bare
couplings; it populates channel pairs that were empty.  The example in Sec.~\ref{sec_d_3_dipole_pulse_SU3} exploits this.

We now focus on the one-body terms containing $h_{\beta}$ to derive $\overline{\hat{H}_{\text{eff},h}^{(0)}}$. Because the off-site generators commute, the nested commutators in
Eq.~\eqref{eq_heff_general_for_use} collapse onto each site:
\begin{align}
    \left[\left[\hat{T}^{\alpha},\hat{H}_{0}^h\right]\right]_{2l}
      = \sum_{j,\beta} \epsilon_{j} h_{\beta}
\left[\left[\hat{T}_{j}^{\alpha},\hat{T}_{j}^{\beta}\right]\right]_{2l}.
    \label{eq_onsite_nested}
\end{align}
We insert
Eq.~\eqref{eq_nested_commutator} and note that every 
commutator retained is of even
order, $u=2l$, so only the even branch $\wp=e$ of
Eq.~\eqref{eq_nested_commutator} contributes. 
We find:
\begin{align}
    \overline{\hat{H}_{\text{eff},h}^{(0)}}
      = \hat{H}_{0}^h
      +  \sideset{}{'}\sum_{j,\alpha,\beta} \epsilon_j
        \tilde{h}_{\alpha,\beta} \hat{T}_{j}^{\kappa_{e}(\alpha,\beta)},
    \label{eq_Heff_loc}
\end{align}
where:
\begin{align}
\tilde{h}_{\alpha,\beta}
=-\frac{h_{\beta}\phi^{(\mathrm e)}_{\alpha\beta}}
{2^{\lambda_{\alpha\beta}-\nu_{\alpha\beta}}}\,
\mathcal{C}_{\alpha}\!\left(2^{-\nu_{\alpha\beta}}\right).
\label{eq_htilde_general_profile}
\end{align}
The prime on the sum follows from the fact that 
$\phi^{(e)}_{\alpha\beta}=0$ whenever
$[\hat{T}^{\alpha}_{j},\hat{T}^{\beta}_{j}]=0$.

We have therefore derived $\He$, a central result of this work.  The
resummation was over all powers of the
accumulated pulse area, so the pulse shapes enter Eqs.~\eqref{eq_Heff_central_J} and~\eqref{eq_Heff_loc} non-perturbatively.  Furthermore, the kick operators follow from the pulse shape averages using Eq.~\eqref{eq_kick_zero_order}.  The remaining approximation is the truncation of the inverse-frequency expansion after its leading terms; corrections begin at $\mathcal{O}(\omega^{-1})$.
Equations~\eqref{eq_Heff_central_J} and~\eqref{eq_Heff_loc}
allow wide tunability through the effective coupling strengths ($\tilde{h}_{\alpha,\beta}$, $\tilde{J}^{(\wp)}_{\alpha,\beta}$, and $\tilde{J}_{\alpha,\beta}$) but they are bounded in certain cases.   Section~\ref{sec_sum_rule} will discuss bounds implied by symmetry.  Otherwise, Eqs.~\eqref{eq_Heff_central_J} and~\eqref{eq_Heff_loc} define a linear set of equations that set the values of $\mathcal{C}_{\alpha}$.  Once these values are established, we obtain a non-linear set of equations that must be solved for pulse parameters internal to $\mathcal{C}_{\alpha}$.

The functional form of $\mathcal{C}_{\alpha}(c)$ can be found analytically for certain pulse choices.  Note that only three arguments of $\mathcal{C}_{\alpha}(c)$ are needed since $\nu_{\alpha\beta}\in\{0,1\}$.  The effective couplings only depend on the following functions:
\begin{align}
u_{\alpha}&:=-\tfrac12\mathcal{C}_{\alpha}(2)
=-\overline{\sin^{2}G_{\alpha}(t)},
\nonumber \\
v_{\alpha}&:=-\tfrac12\mathcal{C}_{\alpha}(1)
=-\overline{\sin^{2}\!\left[G_{\alpha}(t)/2\right]},
\nonumber \\
w_{\alpha}&:=-\tfrac12\mathcal{C}_{\alpha}(\tfrac12)
=-\overline{\sin^{2}\!\left[G_{\alpha}(t)/4\right]}.
\label{eq_uvw_general}
\end{align}
These functions specify the role of pulse shape at any $d$.  The function $w_\alpha$ enters individual contributions with $\nu_{\alpha\beta}=1$.  As we will see for the $d=3$ case, the $w_\alpha$ dependence cancels after all two-body contributions are collected, and the final couplings can be written solely in terms of $u_\alpha$ and $v_\alpha$.  As an example analytic result, consider a cosine form for $g_{\alpha}(t)$: $a_s\cos(\omega t)$, where $a_s$ is a tunable pulse amplitude.  We then find $\mathcal{C}_{\alpha}(c)\rightarrow 1-\mathbb{J}_0(c a_s)$, where $\mathbb{J}_0(x)$ is a Bessel function of the first kind.  But here we focus on square pulse examples.  

\begin{figure}[t]
\includegraphics[width=0.45\textwidth,angle=0]{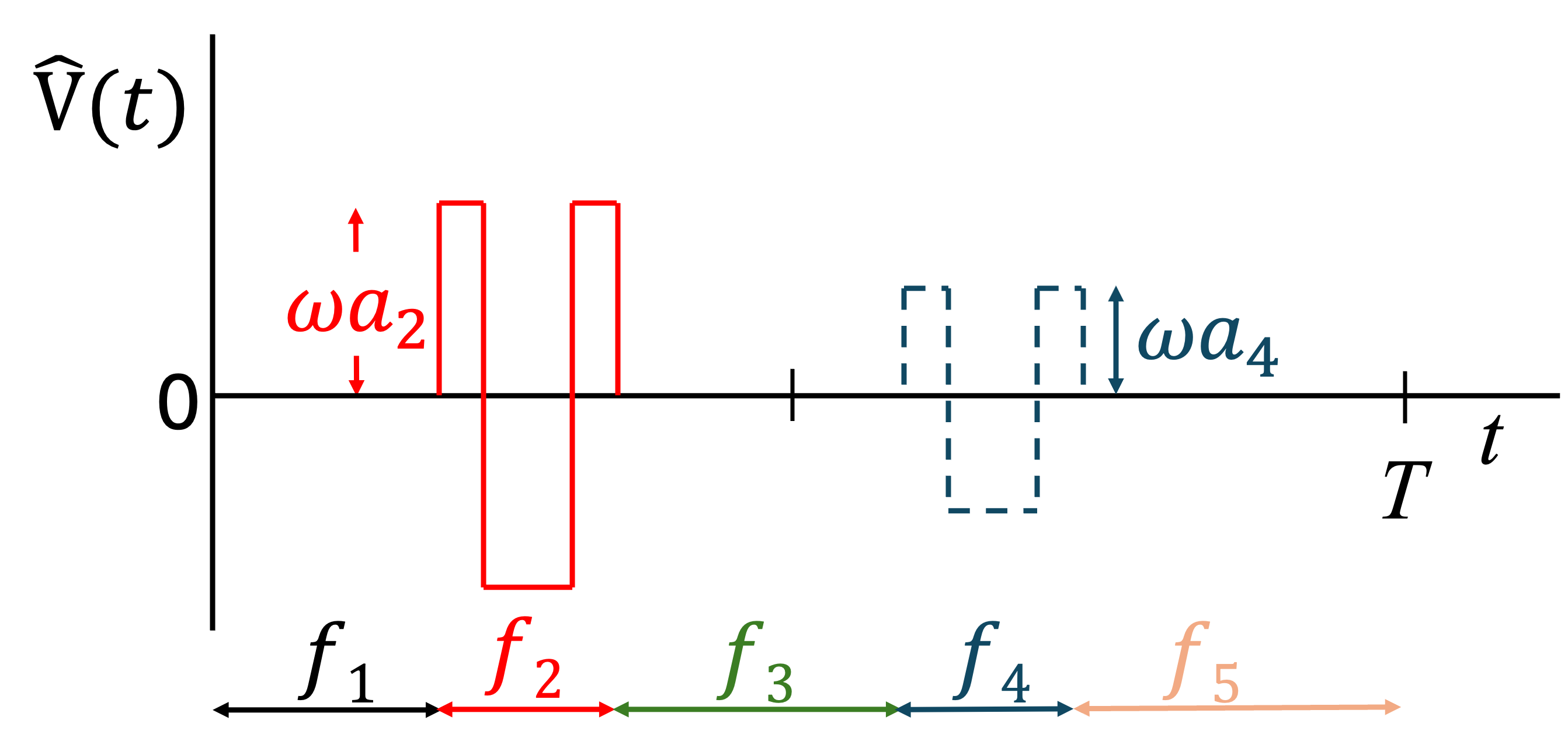}
\caption{\emph{Schematic of an example possible square pulse profile}. $\hat{V}(t)=\omega \sum_{\alpha} g_{\alpha}(t)\hat{T}^{\alpha}$ is shown for only one period, $T$.  Appendix~\ref{sec_example_pulses} defines the pulse details. Here we assume 8 possible generators $\hat{T}^{\alpha}$, i.e., $\alpha=1,2,\cdots,8$, relevant for $d=3$.  In this sequence only pulses on $\hat{T}^{2}$ (solid red line) and $\hat{T}^{4}$ (dashed blue line) are applied.  The remaining space in the period denotes idle time due to the removal of remaining pulse types. Other fractional periods, $f_6=f_7=f_8=0$,  are not shown.  Pulse heights scale with $\omega$ to define strong driving.  The fractional pulse durations $f_{\alpha}$ sum to unity.  $a_{\alpha}$ and $f_{\alpha}$ are incorporated into our formalism as tuning parameters for effective models.   
}
\label{fig_schematic_general_pulse}
\end{figure}

Appendix~\ref{sec_example_pulses} computes $u_{\alpha},v_{\alpha},$ and $w_{\alpha}$ analytically for a general class of square pulses with internal idle time.  The main text uses square pulses with no internal idle time, as shown in Fig.~\ref{fig_schematic_general_pulse}.  In this case we find:
\begin{align}
u_\alpha
&\rightarrow
\frac{f_\alpha}{2}\,\mathrm{sc}_1\!\left(\pi a_\alpha f_\alpha\right), \nonumber \\
v_\alpha
&\rightarrow
\frac{f_\alpha}{2}\,\mathrm{sc}_1\!\left(\frac{\pi a_\alpha f_\alpha}{2}\right), \nonumber \\
w_\alpha
&\rightarrow
\frac{f_\alpha}{2}\,\mathrm{sc}_1\!\left(\frac{\pi a_\alpha f_\alpha}{4}\right),
\label{eq_square_pulse_uvw}
\end{align}  
where $f_{\alpha}$ is the fraction of the period occupied by block $\alpha$, $\omega a_{\alpha}$ is pulse height, and:
\begin{align}
\mathrm{sc}_1(x):=\frac{\sin(x)}{x}-1.
\end{align}
We note that $\mathrm{sc}_1(x)\leq0$.  Other properties of this bounded function are discussed in Appendix~\ref{sec_example_pulses}.  From these expressions we see an example of how we can  tune the effective couplings, Eqs.~\eqref{eq_Jtilde_general_profile} and~\eqref{eq_htilde_general_profile}, with pulse shape parameters instead of just pulse sequencing.

In the following sections we focus on engineering of the two-body terms.  In example applications of our workflow we therefore set $h_{\beta}=0$.  Sections~\ref{sec_d_2_example}-~\ref{sec_numerical} discuss results such that:
\begin{align}
\He \rightarrow \overline{\hat{H}_{\text{eff},J}^{(0)}}
\end{align}
to focus on engineering effective two-body couplings.

\section{Lookup Tables and Workflow}
\label{sec_lookup_tables}

\begin{figure}[t]
\includegraphics[width=0.45\textwidth,angle=0]{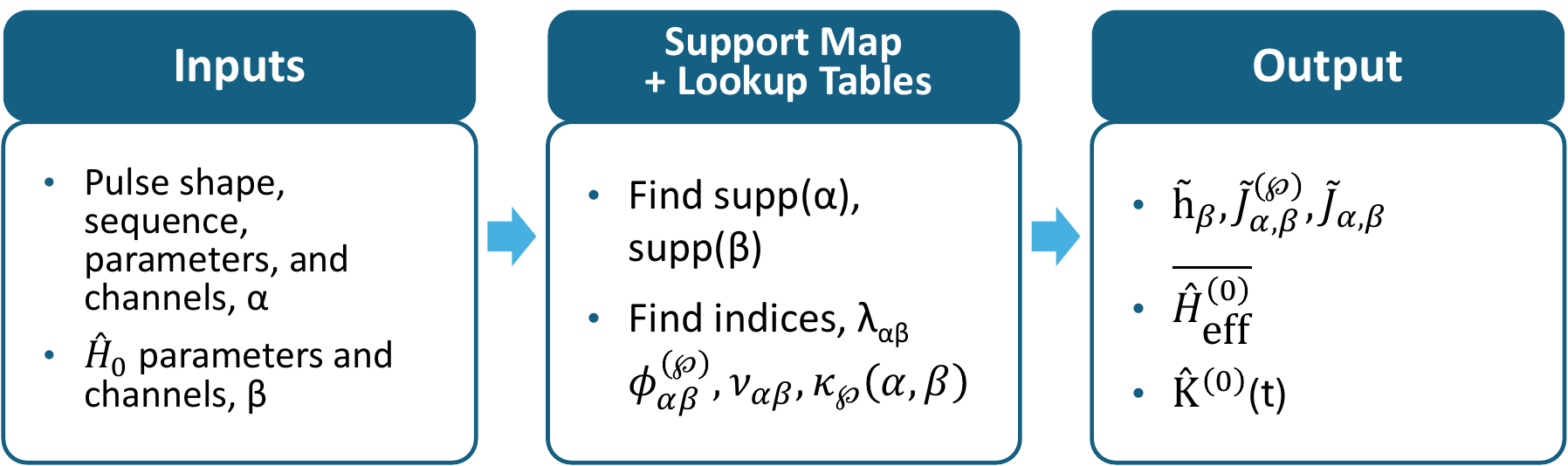}
\caption{
\emph{Technical workflow underlying the use of Eqs.~\eqref{eq_Heff_central_J} and~\eqref{eq_Heff_loc}}.  The first panel defines inputs to $\hat{H}_0+\hat{V}(t)$.  The central panel indicates that the support map for generator matrices must be established.  The map allows use of the lookup table to extract values for indices.  The last panel shows that the output results in a specific effective Hamiltonian and kick operator. 
}
\label{fig_table_workflow}
\end{figure}

In this section we define the workflow needed to construct the effective theory Hamiltonian and kick operators.  Figure~\ref{fig_table_workflow} depicts the workflow.  The kick operators follow directly from the choice of pulse, Eq.~\eqref{eq_kick_zero_order}, but the effective Hamiltonian requires lookup tables to build the indices in Eqs.~\eqref{eq_Heff_central_J} and~\eqref{eq_Heff_loc}. 

We build the lookup tables to define indices $\lambda_{\alpha\beta}$, $\nu_{\alpha\beta}$, $\kappa_{\wp}(\alpha,\beta)$, and $\phi_{\alpha\beta}^{(\wp)}$ given the input generators.  The input generators are found in $\hat{H}_0$ ($\hat{T}^{\beta}_j$ in this section) and from $\hat{V}(t)$ ($\hat{T}^{\alpha}_j$ in this section).  We label local generators by channel indices $\alpha,\beta\in\mathcal A_{\mathrm{in}} $, with:
\begin{align}
\mathcal A_{\mathrm{in}} =\mathcal A_D\cup\mathcal A_A\cup\mathcal A_S.
\end{align}
We start with the input pulse ($\alpha$) generators. The label $\alpha$ encodes both the sector $\tau(\alpha)\in\{D,A,S\}$ and the internal indices. For
$\tau(\alpha)=S$ or $A$, we write $\alpha:=(\tau,m,n)$ with $1\le m<n\le d$. For
$\tau(\alpha)=D$, we write $\alpha:=(D,r)$ with $r=1,\dots,d-1$, where the diagonal sector is fixed to the standard simple-root Cartan basis. The local operators are written:
\begin{align}
\hat T_j^\alpha\in\left\{\hat D_j^{\,r} , \hat A_j^{mn},\hat S_j^{mn}\right\},
\end{align}
using Eq.~\eqref{eq_S_A_D_matrices}.  We also define the support map:
\begin{align}
\mathrm{supp}(\alpha):=
\begin{cases}
\{m,n\}, & \tau(\alpha)\in\{S,A\}, \\
\{r,r+1\}, & \tau(\alpha)=D,
\end{cases}
\label{eq_support_map}
\end{align}
which records the basis labels on which $\hat T_j^\alpha$ acts non-trivially.  Channels labeled by $\beta$ in $\hat{H}_0$ use the same support map. 

We now turn to the output generators needed to construct all terms in Eqs.~\eqref{eq_Heff_central_J} and~\eqref{eq_Heff_loc}.  Because the diagonal sector is represented in the simple-root Cartan basis, it is convenient to enlarge the output label set to:
\begin{align}
\mathcal{A}_{\mathrm{out}}=\mathcal{A}_H \cup \mathcal{A}_A \cup \mathcal {A}_S,
\end{align}
where $\mathcal A_H$ is a bookkeeping sector for diagonal pair-difference operators. If $\gamma=(H,m,n)$ with $1\le m<n\le d$, then
\begin{align}
\hat T_j^{(H,m,n)}
:=
\sum_{r=m}^{n-1}\hat D_j^{\,r}
=
\frac12\left(|m\rangle_j\langle m|-|n\rangle_j\langle n|\right).
\end{align}
Using $\mathcal A_{\mathrm{out}}$, every lookup entry for $\kappa_{\wp}(\alpha,\beta)$ can be used directly as the index of $\hat T_i^{\kappa_{\wp}(\alpha,\beta)}$.

\begin{table}[ht]
\centering
\caption{$d$-level lookup table for same-support channels where the first two columns indicate the input channels for the pulse $\alpha$ and $\hat{H}_0$.  The 2 rightmost columns are derived from Eq.~\eqref{eq_nested_commutator} and list all terms needed for Eqs.~\eqref{eq_Heff_central_J} and~\eqref{eq_Heff_loc} along with $\lambda=\nu=0$.}
\scriptsize
\setlength{\tabcolsep}{3pt}
\begin{tabular}{|l|l|l|l|}
\hline
$\alpha$ & $\beta$ & $(\kappa_{\mathrm o},\phi^{(\mathrm o)})$ & $(\kappa_{\mathrm e},\phi^{(\mathrm e)})$ \\
\hline
$(S,m,n)$ & $(A,m,n)$ & $(H,m,n),\,-i$ & $(A,m,n),\,+1$ \\
\hline
$(A,m,n)$ & $(S,m,n)$ & $(H,m,n),\,+i$ & $(S,m,n),\,+1$ \\
\hline
$(D,r)$ & $(S,r,r+1)$ & $(A,r,r+1),\,-i$ & $(S,r,r+1),\,+1$ \\
\hline
$(D,r)$ & $(A,r,r+1)$ & $(S,r,r+1),\,+i$ & $(A,r,r+1),\,+1$ \\
\hline
$(S,r,r+1)$ & $(D,r)$ & $(A,r,r+1),\,+i$ & $(H,r,r+1),\,+1$ \\
\hline
$(A,r,r+1)$ & $(D,r)$ & $(S,r,r+1),\,-i$ & $(H,r,r+1),\,+1$ \\
\hline
\end{tabular}
\label{table_same_support}
\end{table}

\begin{table}[ht]
\centering
\caption{The same as Table~\ref{table_same_support} but for one-overlap channels with diagonal input $\alpha=(D,r)$ and with $\lambda=\nu=1$. }
\scriptsize
\setlength{\tabcolsep}{3pt}
\begin{tabular}{|l|l|l|l|}
\hline
$\alpha$ & $\beta$ & $(\kappa_{\mathrm o},\phi^{(\mathrm o)})$ & $(\kappa_{\mathrm e},\phi^{(\mathrm e)})$ \\
\hline
$(D,r)$ & $(S,r,n),\ n>r+1$ & $(A,r,n),\,-i$ & $(S,r,n),\,+1$ \\
\hline
$(D,r)$ & $(S,m,r+1),\ m<r$ & $(A,m,r+1),\,-i$ & $(S,m,r+1),\,+1$ \\
\hline
$(D,r)$ & $(S,m,r),\ m<r$ & $(A,m,r),\,+i$ & $(S,m,r),\,+1$ \\
\hline
$(D,r)$ & $(S,r+1,n),\ n>r+1$ & $(A,r+1,n),\,+i$ & $(S,r+1,n),\,+1$ \\
\hline
$(D,r)$ & $(A,r,n),\ n>r+1$ & $(S,r,n),\,+i$ & $(A,r,n),\,+1$ \\
\hline
$(D,r)$ & $(A,m,r+1),\ m<r$ & $(S,m,r+1),\,+i$ & $(A,m,r+1),\,+1$ \\
\hline
$(D,r)$ & $(A,m,r),\ m<r$ & $(S,m,r),\,-i$ & $(A,m,r),\,+1$ \\
\hline
$(D,r)$ & $(A,r+1,n),\ n>r+1$ & $(S,r+1,n),\,-i$ & $(A,r+1,n),\,+1$ \\
\hline
\end{tabular}
\label{table_digonal_input_one_overlap}
\end{table}

\begin{table}[ht]
\centering
\caption{The same as Table~\ref{table_same_support} but for one-overlap channels with diagonal target $\beta=(D,r)$, $\lambda=1$, and $\nu=0$.}
\scriptsize
\setlength{\tabcolsep}{2pt}
\begin{tabular}{|l|l|l|l|}
\hline
$\alpha$ & $\beta$ & $(\kappa_{\mathrm o},\phi^{(\mathrm o)})$ & $(\kappa_{\mathrm e},\phi^{(\mathrm e)})$ \\
\hline
$(S,r,n),\ n>r+1$ & $(D,r)$ & $(A,r,n),\,+i$ & $(H,r,n),\,+1$ \\
\hline
$(S,m,r+1),\ m<r$ & $(D,r)$ & $(A,m,r+1),\,+i$ & $(H,m,r+1),\,+1$ \\
\hline
$(S,m,r),\ m<r$ & $(D,r)$ & $(A,m,r),\,-i$ & $(H,m,r),\,-1$ \\
\hline
$(S,r+1,n),\ n>r+1$ & $(D,r)$ & $(A,r+1,n),\,-i$ & $(H,r+1,n),\,-1$ \\
\hline
$(A,r,n),\ n>r+1$ & $(D,r)$ & $(S,r,n),\,-i$ & $(H,r,n),\,+1$ \\
\hline
$(A,m,r+1),\ m<r$ & $(D,r)$ & $(S,m,r+1),\,-i$ & $(H,m,r+1),\,+1$ \\
\hline
$(A,m,r),\ m<r$ & $(D,r)$ & $(S,m,r),\,+i$ & $(H,m,r),\,-1$ \\
\hline
$(A,r+1,n),\ n>r+1$ & $(D,r)$ & $(S,r+1,n),\,+i$ & $(H,r+1,n),\,-1$ \\
\hline
\end{tabular}
\label{table_digonal_output_one_overlap}
\end{table}

\begin{table}[ht]
\centering
\caption{The same as Table~\ref{table_same_support} but for one-overlap off-diagonal channels. Here $m<n<q$ and $\lambda=\nu=1$.}
\scriptsize
\setlength{\tabcolsep}{3pt}
\begin{tabular}{|l|l|l|l|}
\hline
$\alpha$ & $\beta$ & $(\kappa_{\mathrm o},\phi^{(\mathrm o)})$ & $(\kappa_{\mathrm e},\phi^{(\mathrm e)})$ \\
\hline
$(S,m,n)$ & $(S,n,q)$ & $(A,m,q),\,-i$ & $(S,n,q),\,+1$ \\
\hline
$(S,m,n)$ & $(S,m,q)$ & $(A,n,q),\,-i$ & $(S,m,q),\,+1$ \\
\hline
$(S,m,q)$ & $(S,n,q)$ & $(A,m,n),\,-i$ & $(S,n,q),\,+1$ \\
\hline
$(A,m,n)$ & $(A,n,q)$ & $(A,m,q),\,+i$ & $(A,n,q),\,+1$ \\
\hline
$(A,m,n)$ & $(A,m,q)$ & $(A,n,q),\,-i$ & $(A,m,q),\,+1$ \\
\hline
$(A,m,q)$ & $(A,n,q)$ & $(A,m,n),\,-i$ & $(A,n,q),\,+1$ \\
\hline
$(S,m,n)$ & $(A,n,q)$ & $(S,m,q),\,+i$ & $(A,n,q),\,+1$ \\
\hline
$(S,m,n)$ & $(A,m,q)$ & $(S,n,q),\,+i$ & $(A,m,q),\,+1$ \\
\hline
$(S,m,q)$ & $(A,n,q)$ & $(S,m,n),\,-i$ & $(A,n,q),\,+1$ \\
\hline
$(A,m,n)$ & $(S,n,q)$ & $(S,m,q),\,+i$ & $(S,n,q),\,+1$ \\
\hline
$(A,m,n)$ & $(S,m,q)$ & $(S,n,q),\,-i$ & $(S,m,q),\,+1$ \\
\hline
$(A,m,q)$ & $(S,n,q)$ & $(S,m,n),\,+i$ & $(S,n,q),\,+1$ \\
\hline
\end{tabular}
\label{table_offdigonal_output_one_overlap}
\end{table}

We can now construct lookup tables for any $d$.  We map all generator matrices to D.A.S. matrices, note the support map, and use Eq.~\eqref{eq_nested_commutator}.  To build the lookup tables we note that 
$\kappa_{\wp}(\alpha,\beta)\in\mathcal A_{\mathrm{out}}$
and
$\lambda_{\alpha\beta}\in\{0,1\}$. The value $\lambda_{\alpha\beta}=0$ corresponds to a
first nonzero commutator coefficient of magnitude $1$, while
$\lambda_{\alpha\beta}=1$ corresponds to magnitude $1/2$. Similarly, $\nu_{\alpha\beta}\in\{0,1\}$ records the ratio of successive
commutator magnitudes beyond the first step: $\nu_{\alpha\beta}=0$ corresponds to
successive nested commutators of equal magnitude, while
$\nu_{\alpha\beta}=1$ corresponds to each step being reduced by $1/2$.  All omitted channels
commute, in which case $\phi_{\alpha\beta}^{(\wp)}=0$ and we set
$\lambda_{\alpha\beta}=\nu_{\alpha\beta}=0$ by convention.  

The classification is exhaustive.  Every ordered pair $(\alpha,\beta)$ of
D.A.S.\ generators falls into one of four cases.  If
$\mathrm{supp}(\alpha)\cap\mathrm{supp}(\beta)=\emptyset$ the channels commute.
If the supports coincide, the pair is covered by
Table~\ref{table_same_support}.  If the supports share one index, the
pair is covered by Table~\ref{table_digonal_input_one_overlap} when
$\tau(\alpha)=D$, by Table~\ref{table_digonal_output_one_overlap} when
$\tau(\beta)=D$, and by Table~\ref{table_offdigonal_output_one_overlap} when
$\tau(\alpha),\tau(\beta)\in\{S,A\}$.  The remaining one-overlap case,
$\tau(\alpha)=\tau(\beta)=D$, commutes because both generators are diagonal, and
therefore appears in no table.  We have verified the tables against direct
evaluation of the nested commutator for $d=3,4,5,$ and $6$, covering all
$54$, $168$, $372$, and $690$ non-commuting ordered pairs respectively.

To see how the tables are constructed using the support map, we work out entries of Table~\ref{table_offdigonal_output_one_overlap} explicitly.  Take
$\alpha=(S,m,n)$ and $\beta=(S,n,q)$ with $m<n<q$, so that
$\mathrm{supp}(\alpha)\cap\mathrm{supp}(\beta)=\{n\}$.  Only the shared index
contracts, and
\begin{align}
\left[\hat S_j^{mn},\hat S_j^{nq}\right]
&=\frac14\left(|m\rangle_j\langle q|-|q\rangle_j\langle m|\right)
=-\frac{i}{2}\,\hat A_j^{mq},
\nonumber\\
\left[\hat S_j^{mn},\hat A_j^{mq}\right]
&=\frac{i}{2}\,\hat S_j^{nq},
\label{eq_table4_worked}
\end{align}
so that $[[\hat S_j^{mn},\hat S_j^{nq}]]_{1}=-\tfrac{i}{2}\hat A_j^{mq}$ and
$[[\hat S_j^{mn},\hat S_j^{nq}]]_{2}=\tfrac14\hat S_j^{nq}$.  Comparing with
Eq.~\eqref{eq_nested_commutator}, the magnitude $1/2$ of the first coefficient
gives $\lambda_{\alpha\beta}=1$, the ratio $1/2$ of the second to the first gives
$\nu_{\alpha\beta}=1$, and the operators and phases give
$(\kappa_{\mathrm o},\phi^{(\mathrm o)})=[(A,m,q),-i]$ and
$(\kappa_{\mathrm e},\phi^{(\mathrm e)})=[(S,n,q),+1]$, which is the first row of
the table.  The cycle closes on the two operators $\hat A_j^{mq}$ and
$\hat S_j^{nq}$ and repeats with period two in $u$, which is what allows
Eq.~\eqref{eq_nested_commutator} to hold at every order.

The same computation for the remaining sector and support combinations generates
the rest of Table~\ref{table_offdigonal_output_one_overlap}.  In every case
$\lambda_{\alpha\beta}=\nu_{\alpha\beta}=1$,
$\kappa_{\mathrm e}(\alpha,\beta)=\beta$, and
$\phi^{(\mathrm e)}_{\alpha\beta}=+1$.  The channels obtained by exchanging $\alpha\leftrightarrow\beta$ within this sector
are then fixed by
\begin{align}
\kappa_{\mathrm o}(\beta,\alpha)&=\kappa_{\mathrm o}(\alpha,\beta),
&
\phi^{(\mathrm o)}_{\beta\alpha}&=-\,\phi^{(\mathrm o)}_{\alpha\beta},
\nonumber\\
\kappa_{\mathrm e}(\beta,\alpha)&=\alpha,
&
\phi^{(\mathrm e)}_{\beta\alpha}&=+1 ,
\end{align}
so the twelve entries listed, together with relabeling of the ordered triple
$m<n<q$, generate all one-overlap off-diagonal channels.  Other tables are
generated in a similar fashion.

Tables~\ref{table_digonal_input_one_overlap} and
\ref{table_digonal_output_one_overlap} are written for the simple-root
generators $\hat D^{\,r}$, but they depend on the diagonal generator only
through its support.  Replacing $(D,r)\rightarrow(H,m,n)$ and
$\{r,r+1\}\rightarrow\{m,n\}$ leaves every entry unchanged, so the tables apply with the row conditions read as identifying which index of $\mathrm{supp}(\alpha)$ is shared.  The lookup rules extend directly to any diagonal input generator that is itself a normalized pair-difference operator $\tfrac12(|m\rangle\langle m|-|n\rangle\langle n|)$. A general change of Cartan basis involving arbitrary linear combinations requires corresponding changes in the commutator coefficients.

Given the definition of all possible entries for lookup tables for any $d$, we can now construct all terms in Eqs.~\eqref{eq_Heff_central_J} and~\eqref{eq_Heff_loc}. 
Following Fig.~\ref{fig_table_workflow}, we first select $d$ and build the input generator sets $\hat{T}^\alpha_j$ and $\hat{T}^\beta_j$.  We then map the generators to D.A.S. matrices, Eq.~\eqref{eq_S_A_D_matrices}, and note support structure, Eq.~\eqref{eq_support_map}.  We then use the lookup tables to find values of $\lambda_{\alpha\beta}$, $\nu_{\alpha\beta}$, $\kappa_{\wp}(\alpha,\beta)$, and $\phi_{\alpha\beta}^{(\wp)}$.  We can then use these values and relevant pulse parameters in Eqs.~\eqref{eq_Heff_central_J} and~\eqref{eq_Heff_loc} to find $\He$.  In practice, it is more straightforward to use lookup tables constructed for a specific $d$ of interest.  Section~\ref{sec_code} discusses availability of codes that input $d$ and construct all generators and the specific lookup tables needed.

\section{Sum Rule Constraints on Two-site Hamiltonian Engineering}
\label{sec_sum_rule}

The two-site terms of the effective Hamiltonian obey a trace sum rule. The sum rule constrains the available bond terms for $\He$.  For $d=2$ we will see that this still allows tuning of symmetry but otherwise prevents the appearance of cross terms in $\He$.  Yet for $d>2$, non-trivial cross terms are allowed. 

Reference~\onlinecite{CHOI2017} showed that for ideal pulses, a two-body sum rule 
follows from writing a general bond operator as
$\hat{\mathcal{B}}_{ij} := \sum_{\gamma\delta} J_{\gamma\delta}\,
\hat{T}^{\gamma}_{i}\hat{T}^{\delta}_{j}$.  Then contracting the couplings
against the single-site generator overlaps gives:
\begin{align}
C:=\sum_{\gamma\delta}J_{\gamma\delta}\,
\mathrm{tr}\!\left(\hat{T}^{\gamma}\hat{T}^{\delta}\right),
\label{eq_Q_conserved}
\end{align}
where the trace is over a single $d$-dimensional site space.
$C$ is invariant under these ideal pulses, i.e., $C^{\mathrm{eff}}=C$.

The sum rule holds for the general pulse structure discussed here.  To see this, note that $C$ for general pulses is also invariant at $\mathcal{O}(\omega^{0})$ because the rotation $e^{i\hat{K}^{(0)}}\hat{\mathcal{B}}_{ij}e^{-i\hat{K}^{(0)}}$ preserves $C$, for any $d$,
pulse shape, and cycle count.  For $d=2$ every generator pair shares the same support, so
$\kappa_{\mathrm e}(\alpha,\beta)=\beta$ (Table~\ref{table_same_support}) and
cross terms cannot be generated.  The sum rule then reduces to
$\sum_{\mu}J^{\mathrm{eff}}_{\mu}=\sum_{\mu}J_{\mu}$.
But for $d=3$ cross terms may appear while respecting the sum rule.  We do not find an analogous sum rule for the one-site sector.  We discuss $d=2$ and $d=3$ examples next.

\section{$d=2$ Lookup Table and Example Hamiltonian}
\label{sec_d_2_example}

In this section we exemplify the formalism for the case of $d=2$ (qubits). For $d=2$ there is a single off-diagonal pair $(1,2)$ and a single simple-root
Cartan generator. The three nontrivial local generators are:
\begin{align}
\hat D_j^{\,1}
=
\frac{\hat{\sigma}_j^z}{2},\qquad
\hat A_j^{12}
=
-\frac{\hat{\sigma}_j^y}{2}, \qquad 
\hat S_j^{12}
=
\frac{\hat{\sigma}_j^x}{2}, 
\end{align}
where $\{ \hat{\sigma}^x_j, \hat{\sigma}^y_j, \hat{\sigma}^z_j\}$ are the usual Pauli matrices for a site $j$. 
It is convenient to relabel:
\begin{align}
\hat T_j^z := \hat D_j^{\,1},\qquad 
\hat T_j^y := -\hat A_j^{12}, \qquad 
\hat T_j^x := \hat S_j^{12}.
\label{eq_d_2_relablel}
\end{align}
With this convention we can write the usual $d=2$ commutation relations as 
$
[\hat T_j^\mu,\hat T_j^\nu]=i\,\epsilon_{\mu\nu\rho}\hat T_j^\rho,
$
where
$
\mu,\nu,\rho\in\{x,y,z\}.
$

All nontrivial generators have the same support $\{1,2\}$, so every
non-commuting ordered pair satisfies $\lambda_{\alpha\beta}=0$. Moreover, for
$\alpha\neq\beta$,
\begin{align}
\left[\left[\hat T_j^\alpha,\hat T_j^\beta\right]\right]_{2m+1}
&=
i\,\epsilon_{\alpha\beta\gamma}\,\hat T_j^\gamma,
\nonumber \\
\left[\left[\hat T_j^\alpha,\hat T_j^\beta\right]\right]_{2m}
&=
\hat T_j^\beta,
\end{align}
where $\gamma$ is the unique element of $\{x,y,z\}\setminus\{\alpha,\beta\}$.
The lookup table therefore reduces to Table~\ref{table_d_2_lookup_table}.

\begin{table}[ht]
\centering
\caption{Lookup data for $d=2$ that follows from Table~\ref{table_same_support}. Relabeling from Eq.~\eqref{eq_d_2_relablel} changes some of the signs of $\phi_{\alpha\beta}^{(\mathrm o)}$. Note that here there is no creation of cross channels in Eq.~\eqref{eq_Heff_central_J} since $\kappa_{\mathrm e}(\alpha,\beta)=\beta$. }
\small
\setlength{\tabcolsep}{4pt}
\begin{tabular}{|l|l|c|c|l|l|}
\hline
$\alpha$ & $\beta$ & $\lambda_{\alpha\beta}$ & $\nu_{\alpha\beta}$ & $(\kappa_{\mathrm o}(\alpha,\beta),\phi_{\alpha\beta}^{(\mathrm o)})$ & $(\kappa_{\mathrm e}(\alpha,\beta),\phi_{\alpha\beta}^{(\mathrm e)})$ \\
\hline
$x$ & $y$ & $0$ & $0$ & $(z,+i)$ & $(y,+1)$ \\
\hline
$y$ & $x$ & $0$ & $0$ & $(z,-i)$ & $(x,+1)$ \\
\hline
$y$ & $z$ & $0$ & $0$ & $(x,+i)$ & $(z,+1)$ \\
\hline
$z$ & $y$ & $0$ & $0$ & $(x,-i)$ & $(y,+1)$ \\
\hline
$z$ & $x$ & $0$ & $0$ & $(y,+i)$ & $(x,+1)$ \\
\hline
$x$ & $z$ & $0$ & $0$ & $(y,-i)$ & $(z,+1)$ \\
\hline
\end{tabular}
\label{table_d_2_lookup_table}
\end{table}

To demonstrate the formalism workflow for $d=2$ we must define inputs. 
We consider an anisotropic spin-1/2 model as an input Hamiltonian:
\begin{align}
\hat H_0
=
\frac12\sum_{i\neq j}V_{ij}
\left(
J_x\hat T_i^x\hat T_j^x
+
J_y\hat T_i^y\hat T_j^y
+
J_z\hat T_i^z\hat T_j^z
\right).
\end{align}
Table~\ref{table_d_2_lookup_table} shows that the effective interaction couplings in Eq.~\eqref{eq_Heff_central_J} simplify to: 
\begin{align}
\tilde J_{\alpha,\beta}
&=
2J_{\beta} v_{\alpha},
\nonumber \\
\tilde J^{(\mathrm o)}_{\alpha,\beta}
&=
-J_{\beta}u_{\alpha}, \nonumber\\
\tilde J^{(\mathrm e)}_{\alpha,\beta}
&= J_{\beta}\left[u_{\alpha}-4v_{\alpha} \right]. \nonumber 
\end{align}
The effective Hamiltonian therefore reduces to: 
\begin{align}
\He
=
\frac12\sum_{i\neq j}V_{ij}
\left(
J_x^{\mathrm{eff}}\hat T_i^x\hat T_j^x
+
J_y^{\mathrm{eff}}\hat T_i^y\hat T_j^y
+
J_z^{\mathrm{eff}}\hat T_i^z\hat T_j^z
\right),
\end{align}
with renormalized couplings:
\begin{align}
J_x^{\mathrm{eff}}
&=
J_x+u_y(J_x-J_z)+u_z(J_x-J_y),
\nonumber \\
J_y^{\mathrm{eff}}
&=
J_y+u_x(J_y-J_z)+u_z(J_y-J_x),
\nonumber \\
J_z^{\mathrm{eff}}
&=
J_z+u_x(J_z-J_y)+u_y(J_z-J_x).
\end{align}
In terms of the Pauli matrices this becomes:
\begin{align}
\He
=
\frac18\sum_{i\neq j}V_{ij}
\left(
J_x^{\mathrm{eff}}\sigma_i^x\sigma_j^x
+
J_y^{\mathrm{eff}}\sigma_i^y\sigma_j^y
+
J_z^{\mathrm{eff}}\sigma_i^z\sigma_j^z
\right).
\label{eq_d_2_example_final}
\end{align}
Here we see that the $u_\alpha$ populate a matrix acting on the vector of input interaction strengths $(J_x,J_y,J_z)$ to yield a vector of effective output strengths $(J_x^{\mathrm{eff}},J_y^{\mathrm{eff}},J_z^{\mathrm{eff}})$.   The matrix has unit row and column sums.  $C$ is conserved, leading to
$\sum_\mu J^{\rm eff}_\mu=\sum_\mu J_\mu$.  The drive redistributes weight among
the three channels without changing the total.   

Equation~\eqref{eq_d_2_example_final} still allows tuning of the effective
Hamiltonian to symmetric points.  As an example, consider an anisotropic $XY$
model as a starting point,
$\hat H_0=(1/8)\sum_{i\neq j}V_{ij}(J_x\hat\sigma_i^x\hat\sigma_j^x
+J_y\hat\sigma_i^y\hat\sigma_j^y)$ with $J_x\neq J_y$. 
We can use $x$ and $y$ square 
pulses with heights and durations $\omega a_{\alpha}$ and $f_{\alpha}$, respectively [Fig.~\ref{fig_schematic_general_pulse} and Eqs.~\eqref{eq_square_pulse_uvw}].  These two pulses suffice to reach the $SU(2)$-symmetric Heisenberg model,
$\propto\sum_{i\neq j}V_{ij}(\hat\sigma_i^x\hat\sigma_j^x
+\hat\sigma_i^y\hat\sigma_j^y+\hat\sigma_i^z\hat\sigma_j^z)$, provided
$1/2\le J_y/J_x\lesssim0.66$ or $1.51\lesssim J_y/J_x\le2$.  These bounds arise from the bounds on effective couplings implied by the square pulses of Fig.~\ref{fig_schematic_general_pulse}.  Other pulse forms, e.g., pulses with internal idle time discussed in Appendix~\ref{sec_example_pulses}, can move beyond these bounds.  This $d=2$ example therefore shows how Eq.~\eqref{eq_Heff_central_J} allows the engineering of symmetric effective Hamiltonians.

\section{$d=3$ Lookup Tables and Example Hamiltonians}
\label{sec_top_level_d_3}

We now turn to $d=3$ systems (qutrits) as a second class of examples.  In this section we establish the $d=3$ support map and $d=3$-specific lookup tables.  We then demonstrate use of the workflow to construct a general effective Hamiltonian for $d=3$. Section~\ref{sec_d_3_single_pulse} shows how a single square pulse can drive a classical (diagonal in a product-state basis) interaction into a quantum (off-diagonal) model with dominant quadrupole terms.  Section~\ref{sec_d_3_dipole_pulse_planar} shows that the same pulse can be used to tune an anisotropic model of interacting polar molecules into
one with an enlarged $SU(2)\times U(1)$ symmetry.
Section~\ref{sec_d_3_dipole_pulse_SU3} shows how a two-square pulse sequence can be used to convert an anisotropic dipolar model into an $SU(3)$ symmetric model. 

In this section we identify the generators $\hat{T}^{\alpha}_j$ with $d=3$ generator matrices $\hat{\lambda}_j^{\alpha}$. 
Appendix~\ref{sec_appendix_spin_mapping} defines all 8 of these matrices explicitly.  Appendix~\ref{sec_appendix_spin_mapping} also shows the mapping between $\hat{\lambda}_j^{\alpha}$ matrices and the matrices used in spin-1 magnetism.   

To proceed with the workflow (Fig.~\ref{fig_table_workflow}), we need the support map.  In the $\hat{\lambda}_j^{\alpha}$ basis the support map [Eq.~\eqref{eq_support_map}] becomes:
\begin{align}
\hat{\lambda}^1,\hat{\lambda}^2,\hat{\lambda}^3 &\leftrightarrow \{1,2\},
\nonumber \\
\hat{\lambda}^4,\hat{\lambda}^5,\hat{\lambda}^8 &\leftrightarrow \{1,3\},
\nonumber \\
\hat{\lambda}^6,\hat{\lambda}^7 &\leftrightarrow \{2,3\}.
\end{align}
Accordingly, we define:
\begin{align}
\mathrm{supp}(1)=\mathrm{supp}(2)=\mathrm{supp}(3)=\{1,2\},
\nonumber \\
\mathrm{supp}(4)=\mathrm{supp}(5)=\mathrm{supp}(8)=\{1,3\},
\nonumber \\
\mathrm{supp}(6)=\mathrm{supp}(7)=\{2,3\},
\end{align}
such that the lookup data are given directly in the $\hat{\lambda}^{\alpha}$ basis. 

The general tables fix the diagonal sector to the simple-root Cartan basis, for
which $\mathrm{supp}(D,r)=\{r,r+1\}$, whereas the diagonal pair
$\{\hat{\lambda}^3,\hat{\lambda}^8\}$ used here has $\mathrm{supp}(3)=\{1,2\}$
and $\mathrm{supp}(8)=\{1,3\}$.  This is an admissible input choice because the lookup tables depend on
the diagonal generator only through its support
[Sec.~\ref{sec_lookup_tables}], and it is the choice that renders the diagonal
sector of $\hat H_0$ diagonal on input, so that the input Hamiltonian retains
the form $\tfrac12\sum_{i\neq j}V_{ij}\sum_\beta J_\beta\hat\lambda_i^\beta
\hat\lambda_j^\beta$ assumed in the general derivation.

We choose to enlarge the output label set.  To see why, consider the commutator.  Because the chosen Cartan basis is $\{\hat{\lambda}^3,\hat{\lambda}^8\}$, the commutator
$[\hat{\lambda}^6,\hat{\lambda}^7]=-i\,\hat{\lambda}^h$ does not return one of the input
labels $1,\dots,8$ but instead is proportional to the diagonal pair-difference 
$
\hat{\lambda}_j^h
:=
\hat{\lambda}_j^8-\hat{\lambda}_j^3.
$
We therefore choose the output set to be:
\begin{align}
\mathcal A_{\mathrm{out}}=\{1,2,3,4,5,6,7,8,h\},
\end{align}
with
$
\hat T_j^h\rightarrow \hat{\lambda}_j^h.
$

Using $\mathcal A_{\mathrm{in}}=\{1,\dots,8\}$, $\mathcal A_{\mathrm{out}}$, and the support map we can use the general lookup tables to build specific $d=3$ tables for convenience.  The $d=3$ tables are recorded in Appendix~\ref{sec_appendix_d_3_lookup_tables}. The tables simplify use of $\He$ for $d=3$.

To build $\He$ for $d=3$, we start with: 
\begin{align}
\hat H_0
=
\frac12\sum_{i\neq j}V_{ij}\sum_{\beta\in \mathcal A_{\mathrm{in}}}J_{\beta} \hat \lambda_i^{\beta}\hat \lambda_j^{\beta}.
\end{align}
After inserting the $d=3$ lookup tables into the general formula for
$\He$ and collecting repeated channels, the
result can be written as:
\begin{align}
\He
&=
\frac12\sum_{i\neq j}V_{ij}
\Bigg[
\sum_{\beta=1}^8J_{\beta}^{\mathrm{eff}}\hat \lambda_i^{\beta}\hat \lambda_j^{\beta}
+J^{\mathrm{eff}}_{38}\left(\hat \lambda_i^3\hat \lambda_j^8+\hat \lambda_i^8\hat \lambda_j^3\right)
\Bigg].
\label{eq_d_3_effective_H}
\end{align}
The explicit forms for $J_{\beta}^{\mathrm{eff}}$ and $J^{\mathrm{eff}}_{38}$ are written in App.~\ref{sec_appendix_jeff_definition}.  We verified that these effective couplings obey the sum rule discussed in Sec.~\ref{sec_sum_rule}.  Nonetheless, the sum rule allows a cross-channel term to appear (unlike the $d=2$ case).  Cross terms only arise from the diagonal-target case of Table~\ref{table_digonal_output_one_overlap}, where $\kappa_{\mathrm e}(\alpha,\beta)\neq\beta$ and the output lies in the diagonal sector.  For $d=3$ that sector is spanned
by $\hat\lambda^3$ and $\hat\lambda^8$.

The following three sections use Eq.~\eqref{eq_d_3_effective_H} with specific choices for pulses to engineer symmetry.  
The $d=3$ examples below demonstrate different capabilities.  To distinguish them, let $\hat N_m :=\sum_j|m\rangle_j\langle m|$ count the number of
sites occupying level $m$.  We will consider starting Hamiltonians that conserve every
$\hat N_m$.  The driving in Sec.~\ref{sec_d_3_single_pulse} will 
lead to an effective model that does not conserve $\hat N_m$. 
But the driving in Sec.~\ref{sec_d_3_dipole_pulse_SU3} is different. It will enlarge the
symmetry of the starting Hamiltonian from two conserved quantities to full $SU(3)$ while remaining within
the class of models that conserves $\hat N_m$.

\subsection{$d=3$ with a single pulse driving a diagonal term}
\label{sec_d_3_single_pulse}

In this example we start with a static Hamiltonian $\hat{H}_0$ that is purely diagonal.  We seek to add a driving field to create an effective model with a nematic symmetry in interactions and with off-diagonal (quantum) terms.  We specialize to the $d=3$ starting Hamiltonian in the $\hat{\lambda}^{\beta}_j$ basis using: 
\begin{align}
\hat H_0
=
\frac{J_3}{2}\sum_{i\neq j}V_{ij}\hat{\lambda}_i^3\hat\lambda_j^3,
\label{eq_H0_single_pulse}
\end{align}
where we imagine a 3-level system where the strongest coupling is diagonal and between the nearest levels in this basis.   Such diagonal couplings can be found in qutrit transmon devices \cite{GOSS2022, GOSS2024}.

Equation~\eqref{eq_H0_single_pulse} is anisotropic in spin-1 operators.  To see this we rewrite it
in terms of conventional spin-1 and quadrupolar matrices using the mapping in
Appendix~\ref{sec_appendix_spin_mapping}:
\begin{align}
\hat{\lambda}^3_j =\frac{1}{4}\left(\hat{S}^z_j+ 3\hat{Q}^0_j\right),
\label{eq_lambda3_vs_SQ}
\end{align}
so that
\begin{align}
\hat H_0
=
\frac{J_3}{32}\sum_{i\neq j}V_{ij}
\Big[
\hat{S}^z_i\hat{S}^z_j
+3\bigl(\hat{S}^z_i\hat{Q}^0_j+\hat{Q}^0_i\hat{S}^z_j\bigr)
+9\,\hat{Q}^0_i\hat{Q}^0_j
\Big].
\label{eq_H0_single_pulse_spin}
\end{align}
Equation~\eqref{eq_H0_single_pulse_spin} retains only $z$ components and
mixes the dipolar operator $\hat{S}^z_j$ with the quadrupolar operator
$\hat{Q}^0_j$ through the cross terms
$\hat{S}^z_i\hat{Q}^0_j+\hat{Q}^0_i\hat{S}^z_j$, so it breaks $SU(2)$
spin-rotation symmetry.  It is likewise far from the $SU(3)$ symmetric point which
Sec.~\ref{sec_d_3_dipole_pulse_SU3} reaches from a different starting model.

\begin{figure}[t]
\includegraphics[width=0.45\textwidth,angle=0]{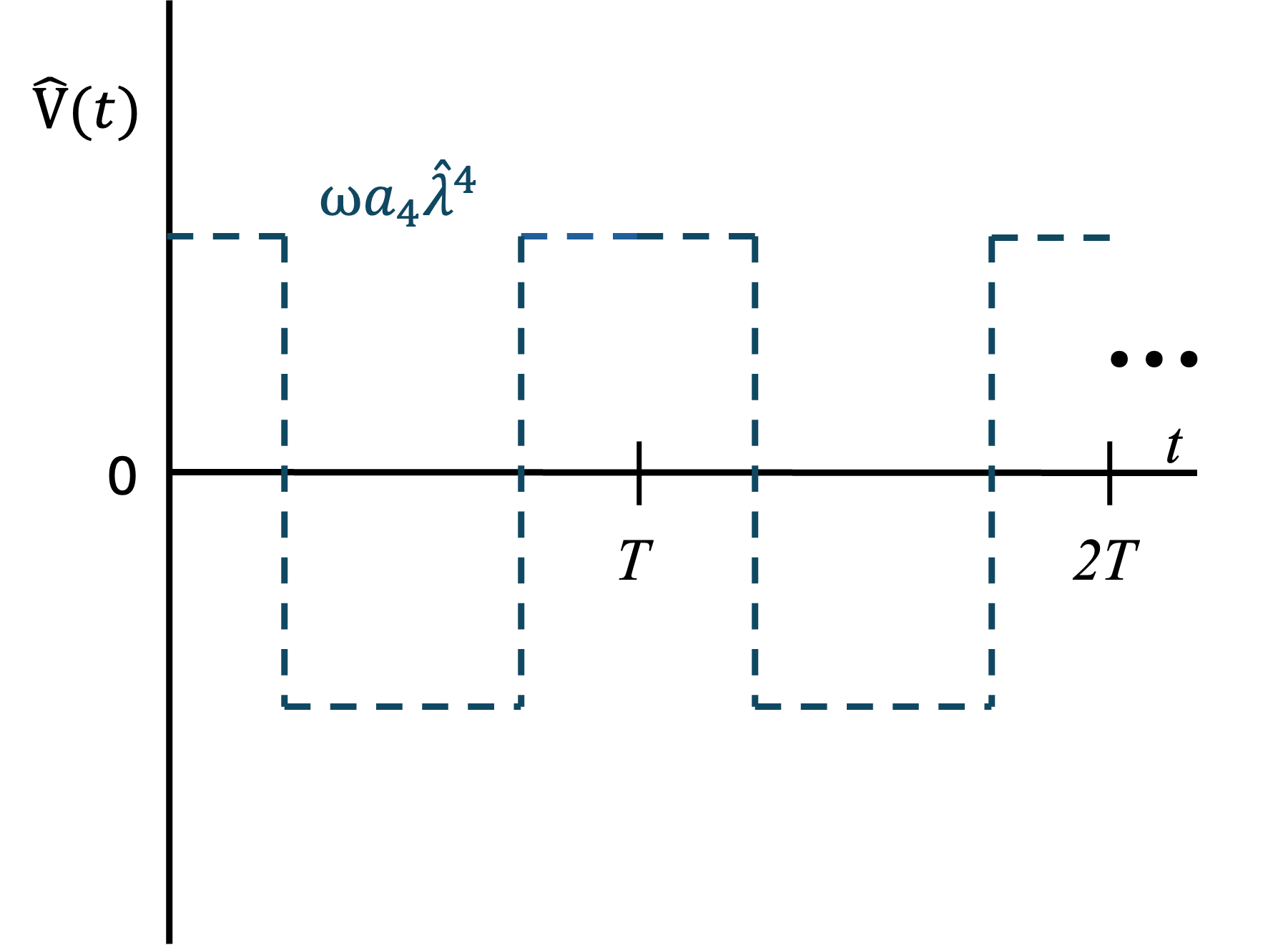}
\caption{
\emph{Single pulse}. Schematic of the pulse used to engineer Eq.~\eqref{eq_H0_single_pulse} into Eq.~\eqref{eq_a4_only_eff}.  The pulse is $\omega a_4 \hat{\lambda}^4$.
}
\label{fig_schematics_single_4_pulse}
\end{figure}

To engineer symmetry of the effective model and generate off-diagonal terms, we consider square pulses. We choose the following pulse parameters:
\begin{align}
a_4\neq 0,
\qquad
 a_\alpha=0 \text{ for } \alpha\neq 4,
\qquad
 f_4=1.
\end{align}
Figure~\ref{fig_schematics_single_4_pulse} shows a schematic of the pulse shape.  The square pulse sequence specifies 
$
u_4
$ and 
$
v_4
$
[see Appendix~\ref{sec_example_pulses} and Eqs.~\eqref{eq_square_pulse_uvw}]
while $
u_\alpha=v_\alpha=0$ for 
$\alpha\neq 4$
since $a_\alpha=0$ for all $\alpha\neq 4$.   

We insert square pulse values for $u_{\alpha}$ and $v_{\alpha}$ into Eq.~\eqref{eq_d_3_effective_H}.  We also re-express the $\hat{\lambda}_j^{\alpha}$ generators in terms of spin-1 operators using Eq.~\eqref{eq_lambda3_vs_SQ} and:
\begin{align}
\hat{\lambda}^5_j&=-\frac12\,\hat{Q}^{xy}_j,
\nonumber\\
\hat{\lambda}^8_j&=\frac12\,\hat{S}^z_j. 
\end{align}
We then find:
\begin{align}
\He
&=
\frac12\sum_{i\neq j}V_{ij}
\Big[
J_{zz}^{\mathrm{eff}}\,\hat{S}_i^z\hat{S}_j^z
+
J_{z0}^{\mathrm{eff}}\bigl(\hat{S}_i^z\hat{Q}_j^0+\hat{Q}_i^0\hat{S}_j^z\bigr)
 \nonumber \\
&+ J_{00}^{\mathrm{eff}}\,\hat{Q}_i^0\hat{Q}_j^0 
+
J_{xy}^{\mathrm{eff}}\,\hat{Q}_i^{xy}\hat{Q}_j^{xy}
\Big],
\label{eq_a4_only_eff}
\end{align}
where
\begin{align}
J_{zz}^{\mathrm{eff}} &= \frac{J_3}{32}\left(1+\frac{\sin\pi a_4}{\pi a_4}\right), \nonumber \\
J_{z0}^{\mathrm{eff}} &= \frac{3J_3}{8}\,\frac{\sin(\pi a_4/2)}{\pi a_4}, \nonumber \\
J_{00}^{\mathrm{eff}} &= \frac{9}{16}\,J_3, \nonumber \\
J_{xy}^{\mathrm{eff}} &= \frac{J_3}{32}\left(1-\frac{\sin\pi a_4}{\pi a_4}\right).
\end{align}
If $a_4$ is chosen to be a multiple of 2, we find:
\begin{align}
\He
&\rightarrow
\frac{J_3}{64}\sum_{i\neq j}V_{ij}
\Big[
\left(\hat{S}_i^z \hat{S}_j^z+\hat{Q}_i^{xy}\hat{Q}_j^{xy}\right)
+18\,\hat{Q}_i^0\hat{Q}_j^0
\Big].
\label{eq_symmetry_a4_eff}
\end{align}

Equation~\eqref{eq_symmetry_a4_eff} reveals a model with a different symmetry
than the starting Hamiltonian, $\hat{H}_0$.  It has dominant nematic terms,
$\hat{Q}_i^0\hat{Q}_j^0$, but with quantum fluctuations induced by
$\hat{Q}_i^{xy}\hat{Q}_j^{xy}$.  The pulse tuned the symmetry beyond conventional spin exchange.  To see this,
note that $\hat{H}_0$ is built from a single diagonal generator and therefore
conserves two independent quantities, $\sum_j\hat{S}^z_j$ and
$\sum_j\hat{Q}^0_j$.  The drive exchanges one of them for another.
Equation~\eqref{eq_symmetry_a4_eff} no longer commutes with
$\sum_j\hat{S}^z_j$, but it does commute with $\sum_j\hat{Q}^{x^2-y^2}_j$,
which $\hat{H}_0$ does not, and it continues to commute with
$\sum_j\hat{Q}^0_j$.  The two surviving quantities commute with each other, so
the effective model again has two independent continuous symmetries, but they
are not the same two.  Both generators of $\hat{H}_0$ are diagonal, whereas
$\hat{Q}^{x^2-y^2}_j=2\hat{\lambda}^4_j$ is off diagonal in the
$\hat{\lambda}^{\alpha}_j$ basis, Appendix~\ref{sec_appendix_spin_mapping}.  The
drive therefore replaces a dipolar symmetry with a quadrupolar one rather than
enlarging the symmetry.  Table~\ref{tab_single_pulse_assumptions} summarizes the
choices of the initial model and the pulse used to arrive at
Eq.~\eqref{eq_symmetry_a4_eff}.

\begin{table}[htbp]
\caption{Example pulse choices that tune Eq.~\eqref{eq_H0_single_pulse} to \eqref{eq_symmetry_a4_eff}.}
\label{tab_single_pulse_assumptions}
\begin{tabular}{|c|c|}
\hline
\multicolumn{2}{|c|}{$\hat H_0$ couplings in Eq.~\eqref{eq_H0_single_pulse}} \\
\hline
$J_3$ & $1$ \\
$J_{\beta\neq 3}$  & $0$ \\
\hline
\hline
\multicolumn{2}{|c|}{$\hat{V}(t)$ square pulse parameters} \\
\hline
$a_4$ & $2$ \\
$a_{\alpha\neq4}$ & $0$ \\
$f_4$ & $1$ \\
$f_{\alpha\neq4}$ & $0$ \\
\hline
\end{tabular}
\end{table}

\subsection{Anisotropic dipole-dipole interactions driven to symmetric points}
\label{sec_d_3_dipole_pulse_top_level}

In this section we choose to start from a $d=3$ version of a dipole-dipole Hamiltonian to demonstrate the formalism and its use in engineering symmetry.  We start with an anisotropic model of rigid dipoles.  Example systems include optically trapped ultracold polar molecules with flip-flop terms dominating two-body interactions \cite{Barnett2006,Wall2014}. For $d=3$, a dipole-dipole Hamiltonian can be written:
\begin{align}
\hat H_0^{\rm dd}
={}&\frac12\sum_{i\neq j}V^{\rm dd}_{ij}
\Big[
J_{12}\bigl(\hat{\lambda}_i^1\hat\lambda_j^1+\hat{\lambda}_i^2\hat\lambda_j^2\bigr)
+J_{45}\bigl(\hat{\lambda}_i^4\hat\lambda_j^4+\hat{\lambda}_i^5\hat\lambda_j^5\bigr) 
\nonumber
\\
&+J_{67}\bigl(\hat{\lambda}_i^6\hat\lambda_j^6+\hat{\lambda}_i^7\hat\lambda_j^7\bigr) 
+J_3\hat{\lambda}_i^3\hat\lambda_j^3
+J_8\hat{\lambda}_i^8\hat\lambda_j^8
\nonumber
\\
&+J_{38}\bigl(\hat{\lambda}_i^3\hat\lambda_j^8+\hat{\lambda}_i^8\hat\lambda_j^3\bigr)
\Big],
\label{eq:H0_start_sparse_su3_fixed}
\end{align}
where \begin{align}
V_{ij}^{\rm dd}=\frac{1-3\cos^{2}\theta_{ij}}{\vert \bm{R}_{ij} \vert^{3}}.
\label{eq_Vdd}
\end{align}  
$\bm{R}_{ij}$ is the dimensionless inter-site displacement and $\theta_{ij}$ is the angle between
$\bm{R}_{ij}$ and the quantization axis.  We take the dipoles to be polarized
along that axis, so that $\theta_{ij}$ is fixed by the lattice geometry and the
field direction rather than by the internal state.   
For a rigid-rotor model of polar molecules, the 3 levels represent rotational angular momentum.   A single field-induced permanent dipole operator contributes to
the diagonal sector.  Off-diagonal terms capture dipole-allowed transitions preserving the component of the rotational quantum number projected onto the lab-frame $z$ axis, $\Delta m_N=0$.  

\subsubsection{Anisotropic RbCs interactions driven to an enlarged $SU(2)\times U(1)$ symmetry}
\label{sec_d_3_dipole_pulse_planar}

In this section we focus on a model motivated by a specific experimental setup that has recently been realized \cite{BLACKMORE2020,RUTTLEY2025,Hepworth2025}: microwave addressing of trapped ultracold $^{87}$Rb$^{133}$Cs molecules.  Appendix~\ref{sec_appendix_RbCs_model} discusses expected inter-molecule interaction parameters between three rotational states of the molecule forming a qutrit.  The native model is anisotropic and can be driven into a symmetry point.

The native Hamiltonian takes the form assumed in
Eq.~\eqref{eq:H0_start_sparse_su3_fixed} with only two nonzero channel pairs:
\begin{align}
\hat H_0^{\rm RbCs}
=
\frac12\sum_{i\neq j}V_{ij}
\Big[&
J_{12}
\left(
\hat\lambda_i^1\hat\lambda_j^1+\hat\lambda_i^2\hat\lambda_j^2
\right)
\nonumber \\
&+
J_{67}
\left(
\hat\lambda_i^6\hat\lambda_j^6+\hat\lambda_i^7\hat\lambda_j^7
\right)
\Big].
\label{eq_RbCs_native}
\end{align}
Explicit computation shows that the rigid-rotor matrix elements lead to ratios that are unequal:
\begin{align}
\frac{J_{67}}{J_{12}}=\frac45. 
\label{eq_RbCs_ratio}
\end{align}

Equation~\eqref{eq_RbCs_ratio} leads to an anisotropic spin-1 model of magnetism.  Using the mappings of  Appendix~\ref{sec_appendix_spin_mapping}, we find:
\begin{align}
\hat H_0^{\rm RbCs}
&=
\frac{J_{12}+J_{67}}{16}
\sum_{i\neq j}V_{ij}
\Big[
\hat S^x_i\hat S^x_j+\hat S^y_i\hat S^y_j
\nonumber \\
&\qquad\qquad\qquad
+\hat Q^{xz}_i\hat Q^{xz}_j+\hat Q^{yz}_i\hat Q^{yz}_j
\Big]
\nonumber \\
&+
\frac{J_{12}-J_{67}}{16}
\sum_{i\neq j}V_{ij}
\Big[
\hat S^x_i\hat Q^{xz}_j+\hat Q^{xz}_i\hat S^x_j
\nonumber \\
&\qquad\qquad\qquad
+\hat S^y_i\hat Q^{yz}_j+\hat Q^{yz}_i\hat S^y_j
\Big].
\label{eq_RbCs_native_SQ}
\end{align}
The first bracket is symmetric between transverse dipole and
transverse quadrupole exchange.  The entire anisotropy resides in the dipole-quadrupole cross coupling in the 
second bracket.   

We now use our theory to remove the cross coupling in Eq.~\eqref{eq_RbCs_native_SQ}
within a single cycle.  We apply one square-pulse block on the
$\hat\lambda^4$ channel and no other, $u_\alpha=v_\alpha=0$ for
$\alpha\neq4$, i.e., the same pulse as Fig.~\ref{fig_schematics_single_4_pulse}.  Inserting the native couplings into Eq.~\eqref{eq_d_3_effective_H}
collapses the effective couplings to
\begin{align}
J_{12}^{\rm eff}
&=
J_{12}+v_{4}
\left(J_{12}-J_{67}\right),
\nonumber \\
J_{67}^{\rm eff}
&=
J_{67}-v_{4}
\left(J_{12}-J_{67}\right),
\label{eq_RbCs_lambda4_Jeff}
\end{align}
with every remaining effective channel vanishing identically.  The
$u_4$ dependence multiplies only native couplings that are zero.  The
pulse redistributes weight between the two exchange channels and
generates nothing else.  The anisotropy therefore tunes continuously,
\begin{align}
\frac{J_{12}^{\rm eff}-J_{67}^{\rm eff}}{2}
=
\left(1+2v_{4}\right)\frac{J_{12}-J_{67}}{2}.
\label{eq_RbCs_dJ_tuning}
\end{align}
The symmetric point $v_{4}=-1/2$ is met by the same
pulse-area condition as in Sec.~\ref{sec_d_3_single_pulse}: choosing
$f_4=1$ and $a_4=2$ in Eqs.~\eqref{eq_square_pulse_uvw} gives
$v_4=-1/2$.  

The effective model
becomes:
\begin{align}
\He
\rightarrow
\frac{J_{12}+J_{67}}{16}
\sum_{i\neq j}V_{ij}
\Big[&
\hat S^x_i\hat S^x_j+\hat S^y_i\hat S^y_j
\nonumber \\
&+\hat Q^{xz}_i\hat Q^{xz}_j+\hat Q^{yz}_i\hat Q^{yz}_j
\Big],
\label{eq_RbCs_channel_symmetric_SQ}
\end{align}
where 
the pulse erased the dipole-quadrupole cross terms of
Eq.~\eqref{eq_RbCs_native_SQ} and leaves the symmetric
bracket intact.  The engineered model retains the conservation of every $\hat N_m$ but is
now also invariant under global rotations generated by $\sum_j\hat\lambda^4_j$
and $\sum_j\hat\lambda^5_j$.  Since $[\hat\lambda^4,\hat\lambda^5]=-i\hat\lambda^8$,
the drive has enlarged the native $U(1)\times U(1)$ symmetry to $SU(2)\times U(1)$:
an $SU(2)$ acting on the $\{|{+}1\rangle,|{-}1\rangle\}$ doublet, generated in
spin-1 language by $\{\hat S^z,\hat Q^{x^2-y^2},\hat Q^{xy}\}$, together with the
$U(1)$ generated by $\hat N_2$.  Equation~\eqref{eq_RbCs_channel_symmetric_SQ} is
the spin-1 analogue of a spin-1/2 XY (flip-flop) model, with the transverse
quadrupole exchange entering on equal footing with the transverse dipole exchange.  Table~\ref{tab_single_pulse_RBCS_assumptions} summarizes the parameters in the native Hamiltonian and the pulse.

\begin{table}[htbp]
\caption{Example pulse choices that tune Eq.~\eqref{eq_RbCs_native} to \eqref{eq_RbCs_channel_symmetric_SQ}.}
\label{tab_single_pulse_RBCS_assumptions}
\begin{tabular}{|c|c|}
\hline
\multicolumn{2}{|c|}{$\hat H_0$ couplings in Eq.~\eqref{eq_RbCs_native}} \\
\hline
$J_{12}$ & $1$ \\
$J_{67}$ & $4/5$ \\
$J_{45}=J_3=J_8=J_{38}$  & $0$ \\
\hline
\hline
\multicolumn{2}{|c|}{$\hat{V}(t)$ square pulse parameters} \\
\hline
$a_4$ & $2$ \\
$a_{\alpha\neq4}$ & $0$ \\
$f_4$ & $1$ \\
$f_{\alpha\neq4}$ & $0$ \\
\hline
\end{tabular}
\end{table}

\subsubsection{Anisotropic interactions driven to the $SU(3)$ symmetric point}
\label{sec_d_3_dipole_pulse_SU3}

We next consider an illustrative anisotropic qutrit interaction compatible with the operator structure obtainable from dressed dipoles~\cite{MICHELI2006}.  We do not derive the assumed coupling ratios from a specific microwave-dressing protocol.  The example is intended to demonstrate that the waveform construction can generate the missing diagonal cross channel and reach the $SU(3)$-symmetric point.

As a demonstration we specialize to assume 
\begin{align}
J_{\rm ex}&:= J_{12}=J_{45}=J_{67}, \nonumber \\
J_{38}&=0,
\label{eq:sparse_KRb_assumptions_su3_fixed}
\end{align}
which is still anisotropic.  To demonstrate symmetry engineering  we show how to Floquet engineer an $SU(3)$ symmetric effective model using the identity:
\begin{align}
&
\sum_{\beta=1}^{8}\hat{\lambda}_i^{\beta}\hat{\lambda}_j^{\beta}
+\frac13\left(\hat{\lambda}_i^3\hat\lambda_j^3+\hat{\lambda}_i^8\hat\lambda_j^8\right)
-\frac23\left(\hat{\lambda}_i^3\hat\lambda_j^8+\hat{\lambda}_i^8\hat\lambda_j^3\right)
\nonumber \\
&=\frac12\left[\mathbf S_i\!\cdot\!\mathbf S_j
+\bigl(\mathbf S_i\!\cdot\!\mathbf S_j\bigr)^2\right]-\frac23, \nonumber
\end{align}
where we used the mappings of Appendix~\ref{sec_appendix_spin_mapping}.  We therefore seek a driving
protocol such that:
\begin{align}
K&:=J^{\text{eff}}_{12}=J^{\text{eff}}_{45}=J^{\text{eff}}_{67}
=\tfrac34 J^{\text{eff}}_{3}=\tfrac34 J^{\text{eff}}_{8}
=-\tfrac32 J^{\text{eff}}_{38}.
\label{eq_su3_conditions}
\end{align}
This requirement is non-trivial because one of the requirements is to realize a $J^{\text{eff}}_{38}$ coupling even though the starting Hamiltonian has $J_{38}=0$. 

Using Eq.~\eqref{eq_d_3_effective_H} we find a $d=3$ effective dipole-dipole Hamiltonian:
\begin{align}
\overline{\hat{H}_{\text{eff}}^{(0)} }
={}&\frac12\sum_{i\neq j}V^{\rm dd}_{ij}
\Big[
J^{\text{eff}}_{12}\bigl(\hat{\lambda}_i^1\hat\lambda_j^1+\hat{\lambda}_i^2\hat\lambda_j^2\bigr)
+J^{\text{eff}}_{45}\bigl(\hat{\lambda}_i^4\hat\lambda_j^4+\hat{\lambda}_i^5\hat\lambda_j^5\bigr)
\nonumber \\
&+J^{\text{eff}}_{67}\bigl(\hat{\lambda}_i^6\hat\lambda_j^6+\hat{\lambda}_i^7\hat\lambda_j^7\bigr)
+J^{\text{eff}}_{3}\hat{\lambda}_i^3\hat\lambda_j^3
+J^{\text{eff}}_{8}\hat{\lambda}_i^8\hat\lambda_j^8
\nonumber \\
&+J^{\text{eff}}_{38}\bigl(\hat{\lambda}_i^3\hat\lambda_j^8+\hat{\lambda}_i^8\hat\lambda_j^3\bigr)
\Big].
\label{eq:Heff_sparse_su3_fixed}
\end{align}
We must design the pulse sequence to tune this to the $SU(3)$ symmetric point: 
\begin{equation}
\overline{\hat{H}_{\text{eff}}^{(0)} }
\propto K\sum_{i\neq j}V^{\rm dd}_{ij}
\Big[
\mathbf S_i\!\cdot\!\mathbf S_j+
\bigl(\mathbf S_i\!\cdot\!\mathbf S_j\bigr)^2
\Big].
\label{eq:sparse_KRb_final_SU3_BLBQH_fixed}
\end{equation}

\begin{figure}[t]
\includegraphics[width=0.45\textwidth,angle=0]{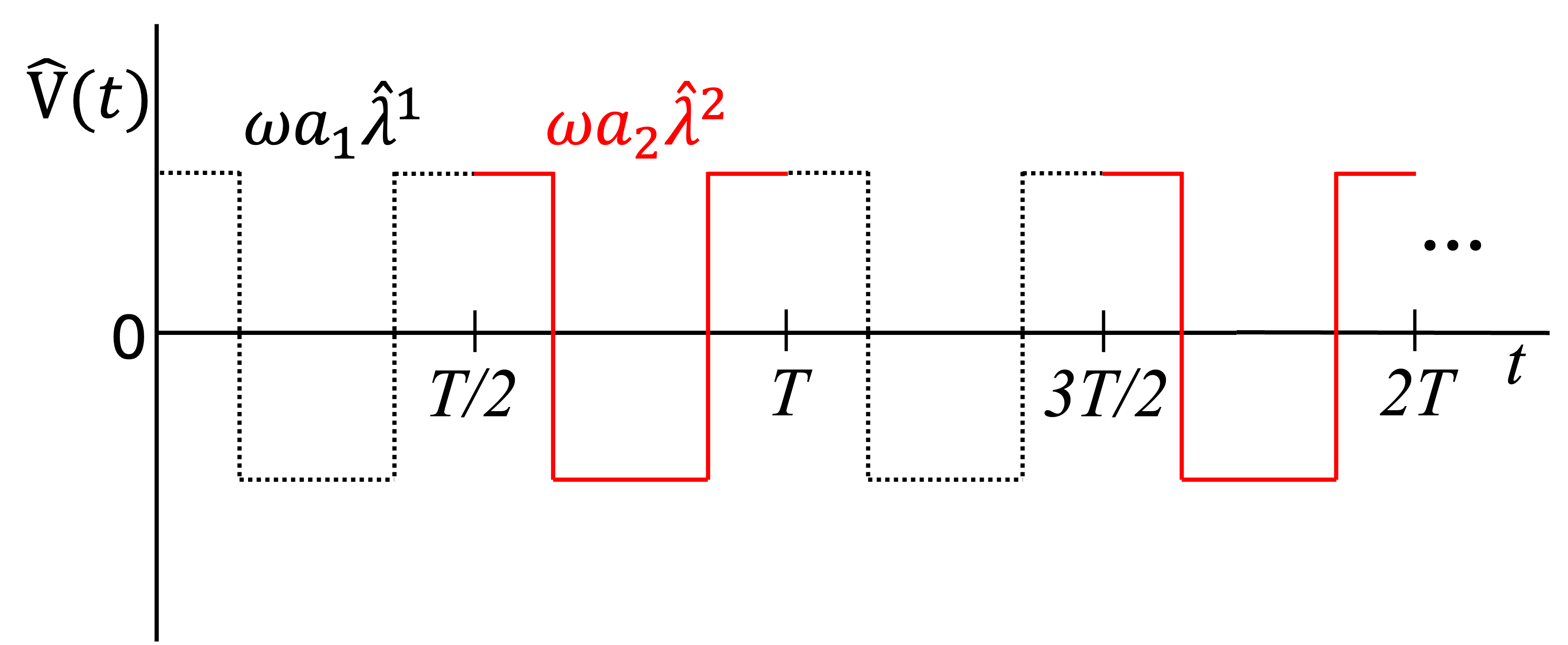}
\caption{
\emph{Double pulse}.  Schematic of an example pulse that can be used to engineer Eq.~\eqref{eq:H0_start_sparse_su3_fixed} into Eq.~\eqref{eq:su3_single_block_BLBQ}.  
The black dotted line indicates a pulse of $\omega a_1 \hat{\lambda}^1$ and the red solid line a pulse of  $\omega a_2 \hat{\lambda}^2$.  The two pulse types are chosen to have equal heights so that $a_A:=a_1=a_2$.  The pulse widths fill one period so that $f_A:=f_1=f_2=1/2$.
}
\label{fig_schematics_two_pulse}
\end{figure}

We find a two-pulse square profile that solves the conditions needed [Eqs.~\eqref{eq_su3_conditions}].  We consider pulses with:
\begin{align}
u_A&:=u_1=u_2,
\nonumber \\
v_A&:=v_1=v_2,
\label{eq:su3_single_block_uv}
\end{align}
while $u_\alpha=v_\alpha=0$ for all $\alpha\neq1,2$.  The general $d=3$ effective
couplings (Appendix~\ref{sec_appendix_jeff_definition}) give:
\begin{align}
J_{12}^{\rm eff}
&= J_{\rm ex}\left(1+u_A\right)-u_A J_3-\tfrac14 u_A J_8,
\nonumber \\
J_{45}^{\rm eff}
&= J_{\rm ex},
\nonumber \\
J_{67}^{\rm eff}
&= J_{\rm ex},
\nonumber \\
J_{3}^{\rm eff}
&= J_3\left(1+2u_A\right)+J_8\left(\tfrac12 u_A-2v_A\right)-2J_{\rm ex}u_A,
\nonumber \\
J_{8}^{\rm eff}
&= J_8,
\nonumber \\
J_{38}^{\rm eff}
&= 2v_A J_8.
\label{eq:su3_single_block_dressed}
\end{align}
Here we note that the $J_{38}^{\rm eff}=2v_AJ_8$ term is generated entirely by the drive from a bare $J_{38}=0$.

To compute parameters we specify square pulses.  We use Eqs.~\eqref{eq_square_pulse_uvw} and choose $a_A:=a_1=a_2$ and all other $a_{\alpha}=0$ with $f_A:=f_1=f_2$. Figure~\ref{fig_schematics_two_pulse} plots an example  pulse solution.  The pulse is obtained by noting that the condition $v_A=-1/4$ is met at
$
a_A=4,
f_A=1/2$.   Here $a_Af_A=2$, so $u_A=v_A=-1/4$.  The constraint $f_1+f_2=1$ is saturated. 
Substituting into Eq.~\eqref{eq:su3_single_block_dressed} gives
$K=J_{12}^{\rm eff}=J_{45}^{\rm eff}=J_{67}^{\rm eff}=1$,
$J_{3}^{\rm eff}=J_{8}^{\rm eff}=4/3$, and $J_{38}^{\rm eff}=-2/3$, in units of
$J_{\rm ex}$.  These satisfy Eq.~\eqref{eq_su3_conditions}.  The
effective Hamiltonian becomes the $SU(3)$ symmetric 
model,
\begin{align}
\overline{\hat{H}_{\text{eff}}^{(0)}}
&=\frac{J_{\rm ex}}{4}\sum_{i\neq j}V^{\rm dd}_{ij}
\Big[
\mathbf S_i\!\cdot\!\mathbf S_j+
\bigl(\mathbf S_i\!\cdot\!\mathbf S_j\bigr)^2
-\frac{4}{3}
\Big].
\label{eq:su3_single_block_BLBQ}
\end{align}
This model is a type of Uimin-Lai-Sutherland model \cite{UIMIN1970,LAI1974,SUTHERLAND1975} conventionally used to study spin-1 magnetism in the limit of nearest neighbor interaction on a chain lattice.  We have therefore shown that an anisotropic qutrit interaction with the operator structure available to dressed dipoles can be driven to the $SU(3)$-symmetric point.  Table~\ref{tab_two_pulse_assumptions} summarizes the set of assumptions leading to the symmetric point.

\begin{table}[htbp]
\caption{Example pulse choices that drive Eq.~\eqref{eq:H0_start_sparse_su3_fixed} to Eq.~\eqref{eq:su3_single_block_BLBQ}. Couplings are
in units of $J_{\rm ex}$.}
\label{tab_two_pulse_assumptions}
\begin{tabular}{|c|c|}
\hline
\multicolumn{2}{|c|}{$\hat H_0^{\rm dd}$ couplings in Eq.~\eqref{eq:H0_start_sparse_su3_fixed}} \\
\hline
$J_{\rm ex}:=J_{12}=J_{45}=J_{67}$ & $1$ \\
$J_{38}$ & $0$ \\
$J_3$ & $2/3$ \\
$J_8$ & $4/3$ \\
\hline
\hline
\multicolumn{2}{|c|}{$\hat{V}(t)$ square pulse parameters} \\
\hline
$a_A:=a_1=a_2$ & $4$ \\
$a_3=a_4=a_5=a_6=a_7=a_8$ & $0$ \\
$f_A:=f_1=f_2$ & $1/2$ \\
$f_3=f_4=f_5=f_6=f_7=f_8$ & $0$ \\
\hline
\end{tabular}
\end{table}

\section{Numerical Error Tests}
\label{sec_numerical}

The central assertion of the formalism constructed here is that our Floquet driving protocol leads to certain effective Hamiltonians and kick operators such that the truncation error scales as $\mathcal{O}(\omega^{-1})$.  In this section we numerically test the $\mathcal{O}(\omega^{-1})$ accuracy assertion.  

We start with a test of relevant bond terms from Sec.~\ref{sec_top_level_d_3}.  We use the spectral norm to define an error estimator and test the error on a single bond for the $d=3$ square pulse examples discussed in the prior sections.  We find that our numerical checks are consistent with the $\mathcal{O}(\omega^{-1})$ scaling for sampling non-stroboscopic times.  We also point out the importance of accurate kick operators. 

We then test the full dynamics of the spin-1 models discussed in Sec.~\ref{sec_top_level_d_3}.  We compare the exact dynamics of local correlators against dynamics derived from the effective theory.  We find excellent agreement for large frequencies and show how the accuracy of the effective dynamics breaks down as frequency is lowered. 

\subsection{Testing Bond Terms}
\label{sec_numerical_test_bond_terms}

To set up the numerical tests we define the exact and effective time propagators for square pulses.  The exact propagator is the time-ordered exponential defined using Eq.~\eqref{eq_Hamiltonian}:
\begin{align}
\hat{U}^{\text{exact}}_{t_0\rightarrow t}
&=\mathcal{T}\exp\left[-i\int_{t_0}^{t}\!ds\,\hat{H}(s)\right],
\label{eq:Udyson}
\end{align}
where $\mathcal{T}$ orders later times to the left.  The square pulses of
Appendix~\ref{sec_example_pulses} are piecewise constant, so the integral can
be performed exactly.  For a partition $t_0<t_1<\cdots<t_m=t$ chosen to contain
every switching time of $\hat V(t)$, with $\delta t_k:=t_k-t_{k-1}$ and factors
ordered so that larger $k$ is to the left, Eq.~\eqref{eq:Udyson} becomes:
\begin{align}
\hat{U}^{\text{exact}}_{t_0\rightarrow t}
=\prod_{k=m}^{1}e^{-i\left[\hat{H}_0+\hat{V}_k\right]\delta t_k},
\label{eq:Uexact}
\end{align}
where $\hat{V}_k$ is $\hat V(t)$ in the time interval $k$. 
This form is exact rather than a Trotter approximation.  No discretization error
enters our error estimates.  A partition that does not resolve the switching
times introduces an error.

To test the effective model we construct the effective time propagator for comparison:
\begin{align}
\hat{U}^{\text{eff}}_{t_0\rightarrow t} = e^{-i\hat{K}^{(0)}(t)}
e^{-i(t-t_0)\overline{\hat{H}_{\text{eff}}^{(0)}}} e^{i \hat{K}^{(0)}(t_0) }.
\end{align}
In our tests we choose $t_0=0$ so that $\hat{K}^{(0)}(t_0)=0$.  But we sample both stroboscopic and non-stroboscopic $t$ to test the accuracy of $\hat{K}^{(0)}(t)$ in $\hat{U}^{\text{eff}}_{t_0\rightarrow t}$. 

\begin{figure}[t]
\includegraphics[width=0.45\textwidth,angle=0]{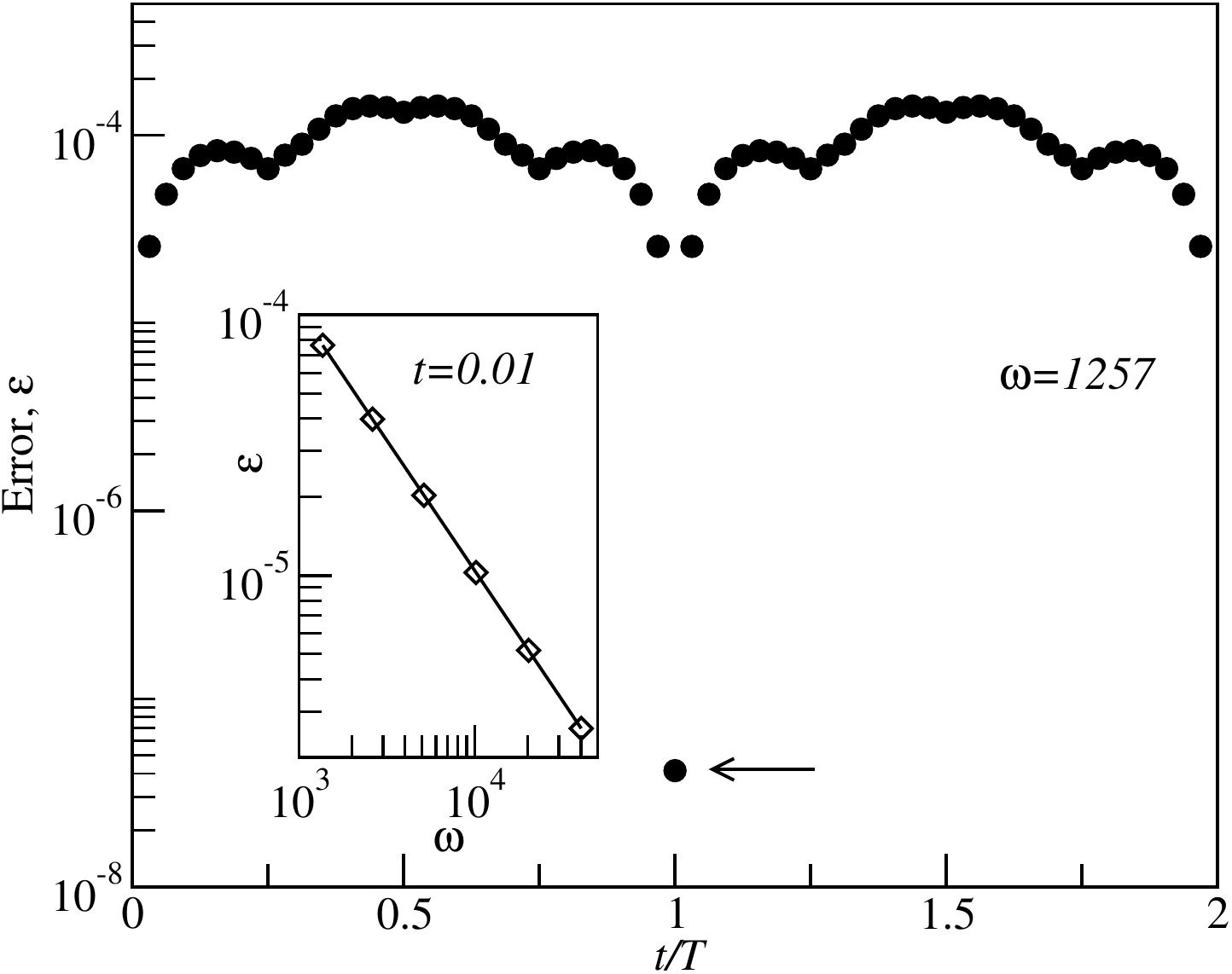}
\caption{\emph{Bond error}. Main: The error in the effective propagator compared to the exact propagator plotted versus time.  The effective model is Eq.~\eqref{eq_symmetry_a4_eff} and exact model, $\hat{H}(t)$, is defined with parameters from Table~\ref{tab_single_pulse_assumptions}.  The drive frequency is fixed to $\omega=1257$.  The arrow shows a stroboscopic point where the kick operator vanishes. These points only occur at integer multiples of the period.  The non-stroboscopic points show higher error but have smoother time-dependence to the error.  Inset: Log-log plot of the error versus the drive frequency for time fixed to $t=0.01$ in units where $J_3=1$.  The slope of the line is $-1$, consistent with  $\mathcal{O}(\omega^{-1})$ scaling of the error.  
}
\label{fig_error_one_pulse}
\end{figure}

We use a spectral norm to define error.  
We can compare the exact and effective propagators using:
\begin{align}
\varepsilon :=\left\|
\hat{U}^{\text{eff}}_{t_0\rightarrow t}
-\hat{U}^{\text{exact}}_{t_0\rightarrow t}
\right\|.
\end{align}
$\varepsilon=0$ implies that the propagators are identical.  Note that $\varepsilon$ is sensitive to phase differences and that $\varepsilon$ is at most 2. 

We first test the numerical error in implementing the $d=3$ protocol from Sec.~\ref{sec_d_3_single_pulse}.  Using the single pulse from Table~\ref{tab_single_pulse_assumptions} we build $\hat{U}^{\text{exact}}_{t_0\rightarrow t}$.  The effective propagator uses Eq.~\eqref{eq_symmetry_a4_eff}.  We work in units of $J_3$ and choose $T=0.005$ (corresponding to $\omega=1257$) and sample both stroboscopic and non-stroboscopic times.  We test the error on a single $i$-$j$ bond.

The main panel of Fig.~\ref{fig_error_one_pulse} plots the error versus time.  Here we see that at non-stroboscopic times the error is $\sim\mathcal{O}(10^{-4})$ for this frequency.  At non-stroboscopic times, the dominant discrepancy is associated with the neglected subleading correction to the kick operator.
We also see that at stroboscopic times (arrow at $t=T$) the error drops quickly by about three orders of magnitude.  

The arrow in  Fig.~\ref{fig_error_one_pulse} shows that at first glance it might be optimal to sample at stroboscopic times only.  However, the steep rise means that technical noise in the sampling time translates directly into error.  To estimate the size of the error we expand about a stroboscopic point.  For
$t=T+\delta t$ with $0<\delta t/T\ll1$, the drive contribution to
$\hat U^{\text{exact}}$ is canceled by $\hat K^{(0)}(\delta t)$ to first order in
$\delta t$, leaving
$\varepsilon\simeq\Vert\overline{\hat{H}^{(0)}_{\text{eff}}}-\hat{H}_0\Vert\,\delta t$.
We have verified this numerically. Effective models further from the starting model therefore show a
steeper rise away from stroboscopic times.

We now consider the error scaling with $\omega$. The inset of Fig.~\ref{fig_error_one_pulse} shows the error as a function of drive frequency at fixed physical time, $t=0.01$ for $J_3=1$. For each frequency, the period is chosen so that $t/T=n_c+0.1672$, with integer $n_c$. Thus, the physical evolution time and the intra-cycle sampling phase are both held fixed as $\omega$ is varied.
  Here we find a linear scaling of $\varepsilon$ with increasing pulse frequency (on a log-log plot).  The slope of the line is $-1$.  The results are therefore consistent with the assertion of error in the effective model that scales as $\mathcal{O}(\omega^{-1})$.  We have tested other non-stroboscopic times as well to check for the same scaling.  

\begin{figure}[t]
\includegraphics[width=0.45\textwidth,angle=0]{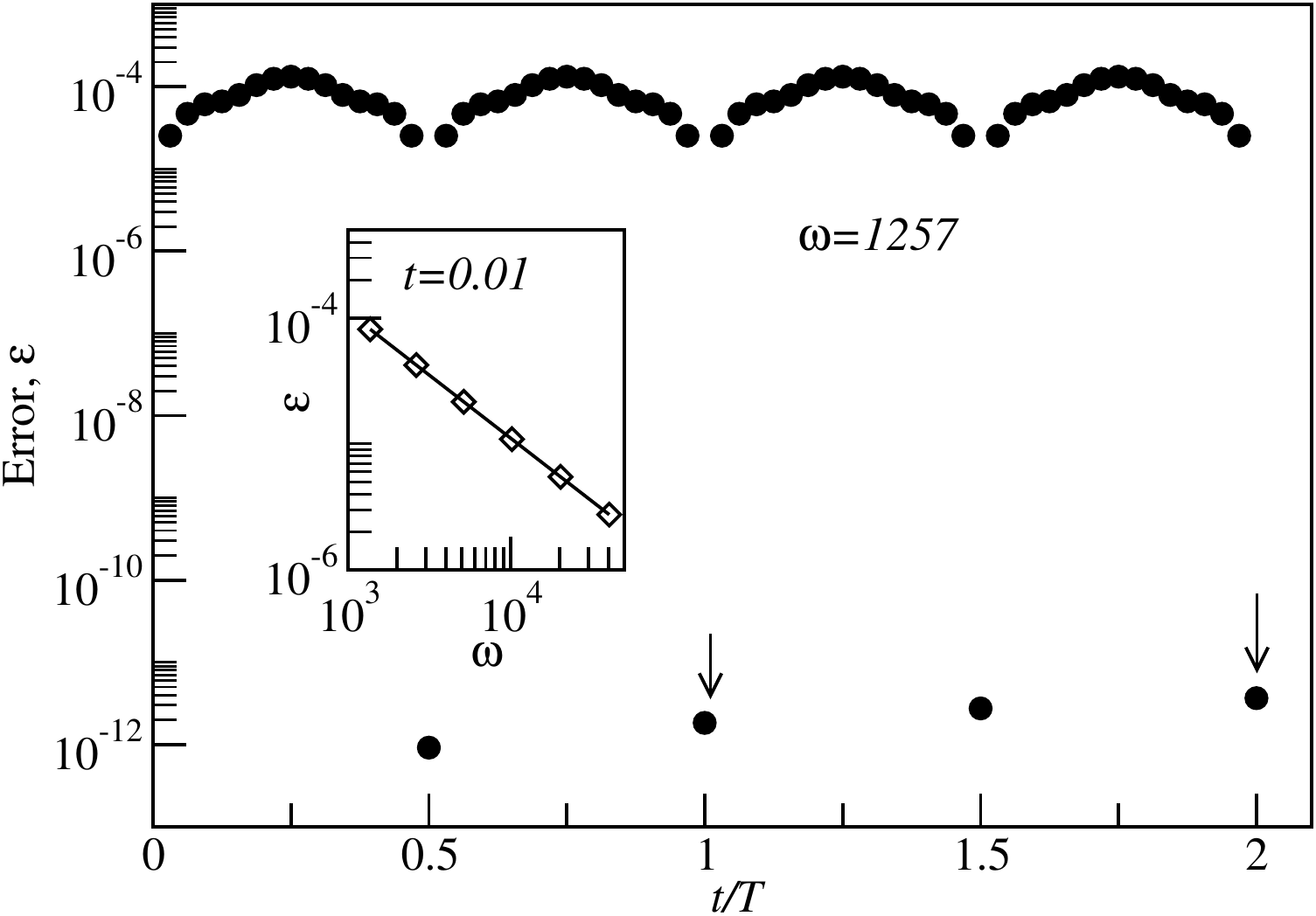}
\caption{The same as Fig.~\ref{fig_error_one_pulse} but for the effective model defined by Eq.~\eqref{eq:su3_single_block_BLBQ} and for the exact model defined using parameters in Table~\ref{tab_two_pulse_assumptions}.  The two low-error points with no arrows, at $t/T=0.5$ and $1.5$, appear because of the structure of the two-pulse profile.
Units are set by $J_{\rm{ex}}=1$.}
\label{fig_error_two_pulse}
\end{figure}

We also test the error of the model derived in Sec.~\ref{sec_d_3_dipole_pulse_SU3}.  Figure~\ref{fig_error_two_pulse} shows the same as Fig.~\ref{fig_error_one_pulse} but compares the exact propagator using the double pulse parameters from Table~\ref{tab_two_pulse_assumptions} against the effective propagator arising from Eq.~\eqref{eq:su3_single_block_BLBQ}.  Here we see a similar precipitous drop in error at stroboscopic times (arrows) but here the drop is about 7 orders of magnitude.  The other two low-error points (at $t=T/2$ and $t=3T/2$) arise because the kick operator vanishes at the internal block boundary and, by the symmetry between the two pulse blocks, the half-sequence average already equals $\He$.  Otherwise the $\sim\mathcal{O}(10^{-4})$ error at non-stroboscopic times and linear error scaling with $\omega$ are similar to the results reported in Fig.~\ref{fig_error_one_pulse}.

The drop is consistent with cancellation of the leading stroboscopic correction.  At
stroboscopic times the kick operator drops out and the residual error is set by
the neglected $\omega^{-1}\hat H^{(1)}_{\text{eff}}$ accumulated over one period.  For the specific two-pulse protocol and native Hamiltonian considered here, the numerical results show cancellation of the leading stroboscopic correction, with $\varepsilon(T)\propto\omega^{-3}$ over the tested frequency range. The enhanced scaling correlates with the channel-exchange structure of the two pulse blocks and is analogous to the improved scaling produced by symmetric sequences in average-Hamiltonian theory. 

The derived form for the kick operator, Eq.~\eqref{eq_kick_zero_order}, plays an important role.  To test the magnitude of its importance, we remove the kick operator in our numerical tests.  We found $\varepsilon\sim 1$, i.e., near maximal error, with the kick operator removed.  We therefore conclude that a good approximate form for the kick operator is vital in accurately interpreting the dynamics of the effective model at all times except stroboscopic times.

\subsection{Testing Correlator Dynamics on Chains}
\label{sec_numerical_dynamics_ed}

In this section we test exact dynamics of local correlation functions against the predictions from our effective theory.  We use exact matrix representations on small chains and construct the exact driven propagator as an ordered product of matrix exponentials over the piecewise-constant pulse segments
discussed in Secs.~\ref{sec_d_3_single_pulse} and~\ref{sec_d_3_dipole_pulse_SU3}.  We find that the effective models show excellent agreement with the exact dynamics at high frequencies, as expected.  We explore the breakdown of agreement as frequency is lowered.  

We start by considering Eq.~\eqref{eq_H0_single_pulse} with the driving parameters listed in Table~\ref{tab_single_pulse_assumptions}.  We put the model on a periodic chain with up to 6 sites and with nearest neighbor interactions.  We construct the exact propagator of $\hat{H}_0+\hat{V}(t)$ under driving.  We compute $\langle \hat{S}^z_i \hat{S}^z_j \rangle$ and  $\langle \hat{Q}^{xy}_i \hat{Q}^{xy}_j \rangle$ as a function of time.

\begin{figure}[t]
\includegraphics[width=0.45\textwidth,angle=0]{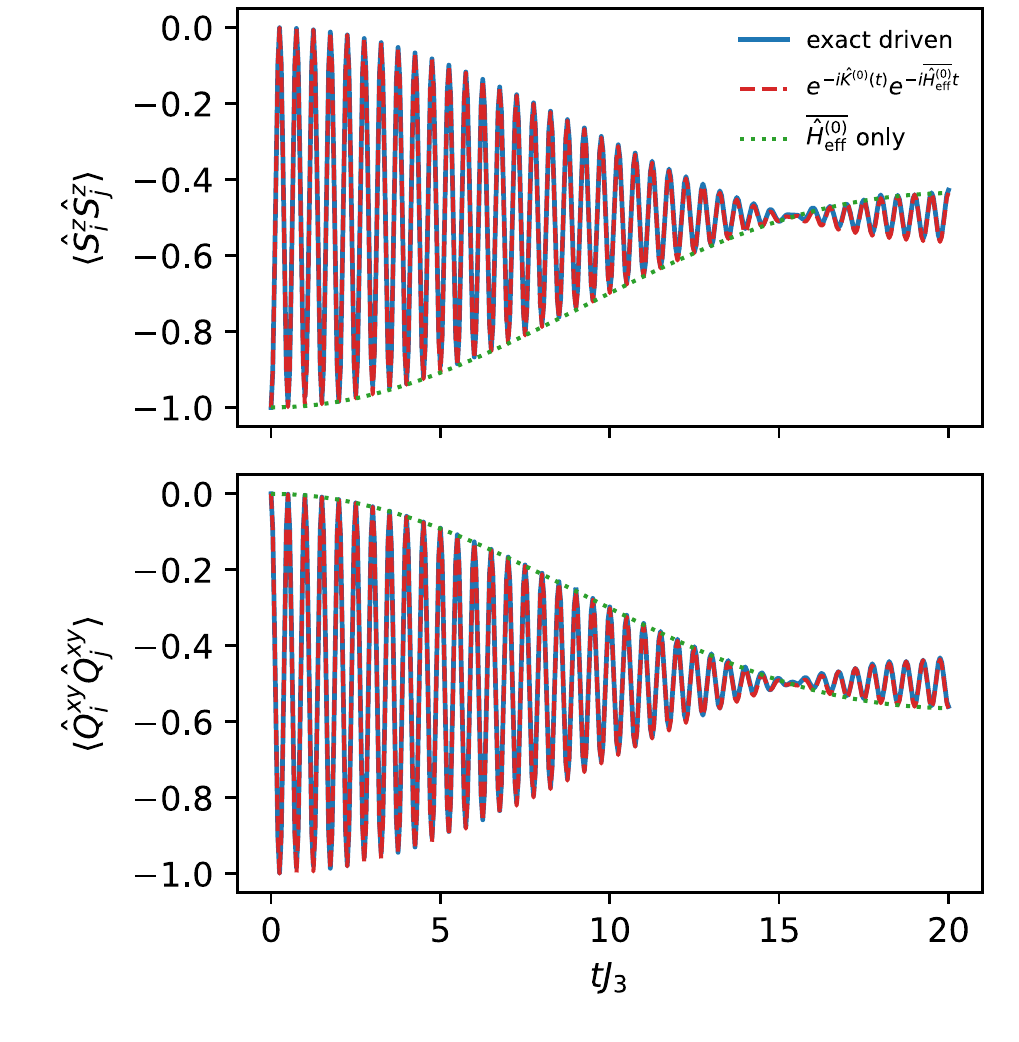}
\caption{\emph{Correlations in a chain}. Comparison of dynamics of nearest neighbor correlators using both exact and effective propagators.  The solid blue line plots the time evolution of exact correlators for $\hat{H}(t)$ defined with the pulse parameters of Table~\ref{tab_single_pulse_assumptions} and $\hat{H}_0$ using Eq.~\eqref{eq_H0_single_pulse}.  We choose $N=6$ sites, nearest neighbor interactions for $V_{ij}$, periodic boundary conditions, $a_4=2$, and a high frequency, $\omega=3.1 J_3$.  The initial state is chosen to be a staggered $S_z$ state.  The dashed red line plots the effective propagator using Eq.~\eqref{eq_symmetry_a4_eff} and the kick operator.  The dotted green line plots the effective result with the kick operator omitted.  Here we see that the dashed and solid lines are indistinguishable for all times.  But the dotted line agrees with the exact result only at stroboscopic times.  
}
\label{fig_ed_dynamics_single_pulse_omega_3p1}
\end{figure}

The solid lines in Fig.~\ref{fig_ed_dynamics_single_pulse_omega_3p1} plot the exact dynamics for nearest-neighbor correlations ($j=i+1$) at $\omega=3.1J_3$. For the periodic nearest-neighbor chain, Eq.~\eqref{eq:local_scale} gives $\mathcal{J}=2J_3$ and hence $\mathcal{J}/\omega\simeq0.65$. Although this value is not deep in the asymptotic regime $\mathcal{J}/\omega\ll1$, the exact and leading-order effective dynamics remain in excellent agreement.
The numerics therefore confirm that the effective dynamics accurately capture the exact dynamics of these correlators.  The dotted line plots the dynamics without the kick operators.  Here we see that the effective Hamiltonian captures the large period envelope oscillations with accurate comparisons at stroboscopic times.  But the small oscillations at non-stroboscopic times are missed without the kick operator included.   

\begin{figure}[t]
\includegraphics[width=0.45\textwidth,angle=0]{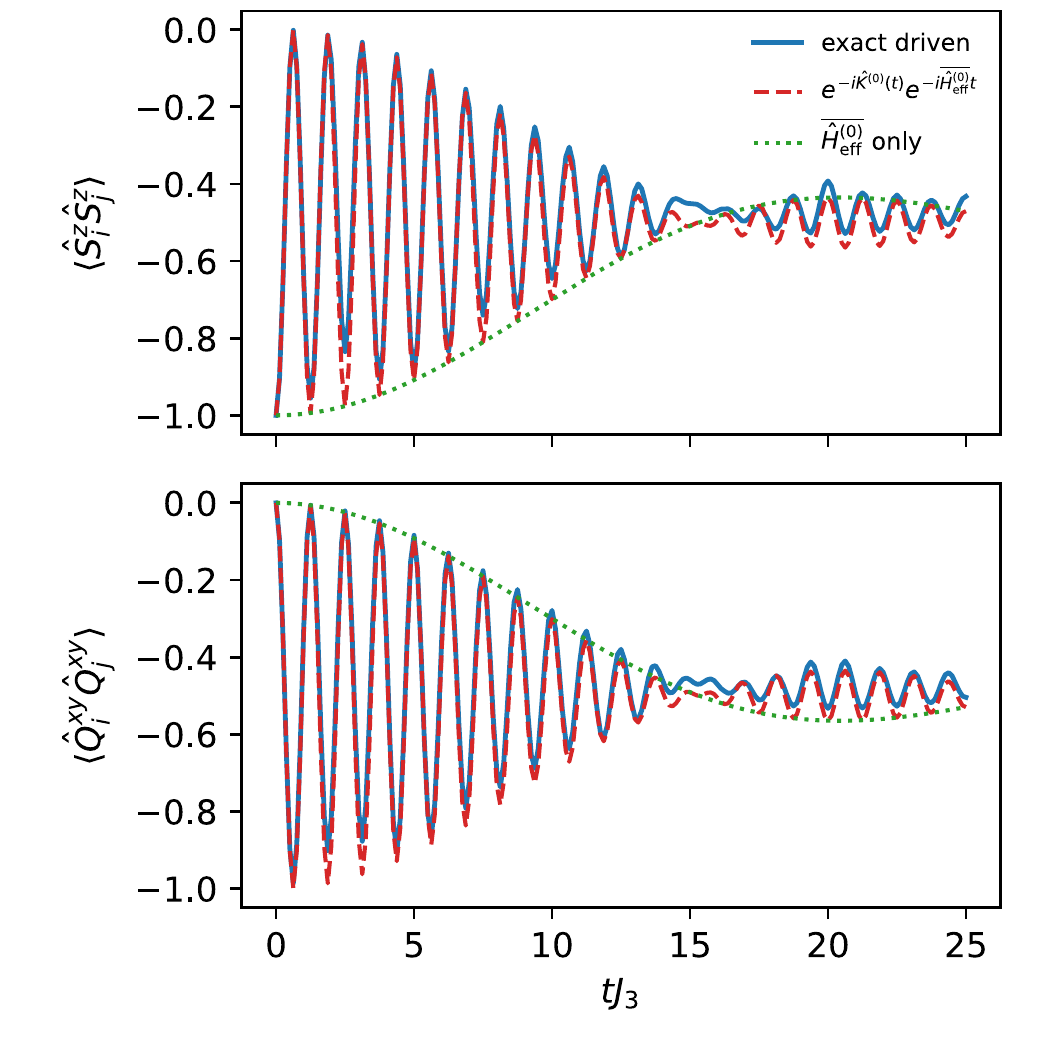}
\caption{The same as Fig.~\ref{fig_ed_dynamics_single_pulse_omega_3p1} but for a lower frequency, $\omega =1.3 J_3$.  Here we see that agreement between the exact and effective results starts to break down, as expected from our high frequency expansion. 
}
\label{fig_ed_dynamics_single_pulse_omega_1p3}
\end{figure}

Figure~\ref{fig_ed_dynamics_single_pulse_omega_1p3} plots the same as Fig.~\ref{fig_ed_dynamics_single_pulse_omega_3p1} but for a lower driving frequency, $\omega=1.3 J_3$.  Here the comparison between the exact and effective dynamics breaks down. For the periodic nearest-neighbor chain, $\mathcal{J}/\omega\simeq1.5$, well outside the sufficient high-frequency condition $\mathcal{J}/\omega\ll1$, so substantial higher-order corrections are expected.

We have tested the high frequency dynamics of other correlators and for other initial states. We have also tested effective dynamics for the model discussed in Sec.~\ref{sec_d_3_dipole_pulse_SU3}.  Here we also find that the effective and exact dynamics agree at high frequencies.  We have therefore used many-body numerics to confirm that the exact dynamics of local correlators are accurately captured by our effective models at high frequencies.  Section~\ref{sec_code} discusses the codes used to produce these results. 

\section{Data Availability}
\label{sec_code}
The codes used to generate the results here are openly available in the Zenodo repository cited in Ref.~\cite{SCOTT2026code}.
We include the following code in both Mathematica and Python formats for convenience: i) codes that input $d$ and produce all generator matrices and user-friendly lookup tables for use in the workflow; ii) the numerical error code for our $d=3$ examples used in Sec.~\ref{sec_numerical_test_bond_terms}; and iii) the exact propagator codes used in Sec.~\ref{sec_numerical_dynamics_ed} to plot correlator dynamics.

\section{Summary and Outlook}
\label{sec_summary}

We constructed Floquet engineering theory for $d$-level systems that can be modeled as two-body and one-body interactions on a graph.  The formalism is defined by a set of renormalized effective couplings for the single- and two-body terms (an effective Hamiltonian) and a kick operator defining the new time-dependent phase acquired by the many-body wavefunction.  The effective model was derived using a high frequency expansion in the strong drive limit.  Our strong driving formalism allows pulse shapes, in addition to pulse sequencing, to engineer the Hamiltonians.  Our theory workflow uses lookup tables for Hamiltonian indices that become more complex as $d$ increases.  We provide codes to define these tables and reproduce the numerical checks performed. 

Our central finding is a formalism that uses pulse shape as a resource rather than an error.  Conventional Floquet engineering treats a pulse of finite width and height as a departure from an ideal that must be corrected with additional pulses \cite{Vandersypen2005,SUTER2016}.  We instead keep the width and height from the start and find that they enter the effective couplings through bounded functions, e.g., $\mathrm{sc}_1$ for square pulses.  Pulse width and height then become tuning knobs on the same footing as the ordering of pulses.  Two consequences follow.  First, the waveform dependence of the leading-order effective model is treated
non-perturbatively for arbitrary finite-duration waveforms within the
sequential, zero-area, midpoint-antisymmetric pulse class defined in
Sec.~\ref{sec_pulse_properties_general}; a pulse that is not short does not by
itself spoil the result. Corrections arise from the high-frequency expansion and begin at order $\mathcal{J}/\omega$.  Second, the theory is low cost.  In the examples considered here, the target leading-order Hamiltonians are obtained using one or two waveform blocks. This provides a compact alternative to ideal-pulse constructions, although a resource-matched comparison with optimized pulse sequences is platform dependent.

We also established conditions under which the engineered model describes the dynamics for a long time.  Strong driving at first appears to conflict with prethermalization because the drive amplitude grows with frequency.  We showed that our pulse assumptions remove this conflict.  
In the toggling frame the drive cancels, while conjugation by the factorized drive preserves the support and norm of each term in the inherited local decomposition. The resulting local energy scale is therefore independent of $\omega$ and of the pulse parameters.  The heating bounds of Refs.~\cite{MORI2016,ABANIN2017c,HO2023} then apply directly.  We caution that the prethermal bound is a sufficient condition and that the $N$ dependence of the propagator error is model dependent.  Models with long-range or disordered $V_{ij}$ that violate Eq.~\eqref{eq_Vij_summability} may show an $\mathcal{O}(\omega^{-1})$ error that grows with $N$.  Such cases require a bound formulated for power-law interactions \cite{MACHADO2020} and should be checked numerically on a case-by-case basis.

The formalism can be extended in several directions.  i) We used global pulses because they are the simplest to implement across platforms. The toggling-frame framework can be extended to site-dependent pulses, although the corresponding coupling formulas and lookup structure must be generalized.  ii) We worked to lowest order.  Appendix~\ref{sec_derivation_lowest_order} records the conditions needed at $\mathcal{O}(\omega^{-1})$ and beyond, and the symmetry of the two-pulse block in Sec.~\ref{sec_numerical} already shows that pulse ordering can push the stroboscopic error to higher order.  iii) We used square pulses with and without internal idle time, but the
closed-form construction applies to arbitrary finite-duration waveforms within
the sequential, zero-area, midpoint-antisymmetric pulse class defined in
Sec.~\ref{sec_pulse_properties_general}. Smooth envelopes within this class may
be preferred on platforms where sharp edges drive leakage out of the $d$-level
manifold.  iv) Pulse shape engineering and pulse sequence engineering \cite{CHOI2017,ZHOU2024} are not in competition.  Our effective Hamiltonians can serve as the building blocks of a conventional sequence, Sec.~\ref{sec_multiblock}, so both tools can be used together.

We expect the most immediate use of our results to be in the three settings that motivated the work: dynamical decoupling \cite{Vandersypen2005,SUTER2016}, tuning to symmetric points for sensing \cite{DEGEN2017,LIN2026}, and building effective models for quantum analogue simulation \cite{Georgescu2014,DALEY2022}.  The qudit platforms in Table~\ref{tab_qudit_platforms} each have a native interaction and a global drive.  Our theory provides a direct route from a measured native model and a realizable pulse to a predicted effective model, with a controlled leading-order error scaling, without first idealizing the pulse.  It would be interesting to use this route to reach spin-1 models with valence-bond solid ground states \cite{AFFLECK1989} on platforms where the native interaction is far from the required symmetric point.

\begin{acknowledgments}
This material is based upon work supported by the
U.S. Department of Energy, Office of Science, Office of
Basic Energy Sciences Energy Frontier Research Centers program under Award Number DE-SC0026289.
\end{acknowledgments}

\appendix

\section{Example of Square Pulsing}
\label{sec_example_pulses}

Choosing a pulse profile for $g_{\alpha}(t)$ in Eq.~\eqref{eq_pulse_general} will define the kick operator and allow engineering of $\He$.  In this section we define a square pulse profile that satisfies the assumptions of Sec.~\ref{sec_pulse_assumption_higher_level}.  We show that the integral, $\overline{\sin^{2}\!\left[c\,G_{\alpha}(t)/2\right]}$, needed to define $\mathcal{C}_{\alpha}(c)$ in Eq.~\eqref{eq_generating_function} can be performed analytically. We record these results as analytic functions for $u_{\alpha}$, $v_{\alpha}$, and $w_{\alpha}$ for use in the Hamiltonian effective couplings. 

We choose square pulses with particular conditions on ordering and averages.  We consider pulse-dependent time fractions $f_\alpha$ satisfying:
\begin{align}
    \sum_\alpha f_\alpha = 1,
    \qquad
    f_\alpha \geq 0.
    \label{eq_fraction_constraint}
\end{align}
Channels with $f_{\alpha}=0$ are absent from the sequence.  Idle time is
described equivalently by a block with $f_{\alpha}>0$ and $a_{\alpha}=0$: every
pulse-dependent coefficient derived below vanishes in either limit.
Define the cumulative fractions:
$    F_\alpha = \sum_{\alpha'=1}^{\alpha} f_{\alpha'}
$
and $F_0=0$. 
Then the pulse channel $\alpha$ occupies the interval
$
    F_{\alpha-1}T < t \le F_\alpha T.
$
Here $T$ is the full cycle time.  We use these definitions to define the square pulses as:
\begin{widetext}
\begin{equation}
    g_\alpha(t)=
    \begin{cases}
        a_\alpha, & F_{\alpha-1}T < t\le \left(F_{\alpha-1}+\dfrac{r_\alpha f_\alpha}{4}\right)T,\\[4pt]
        0, & \left(F_{\alpha-1}+\dfrac{r_\alpha f_\alpha}{4}\right)T < t\le \left(F_{\alpha-1}+\dfrac{\left(2-r_\alpha\right)f_\alpha}{4}\right)T,\\[4pt]
        -a_\alpha, & \left(F_{\alpha-1}+\dfrac{\left(2-r_\alpha\right)f_\alpha}{4}\right)T < t \le \left(F_{\alpha-1}+\dfrac{\left(2+r_\alpha\right)f_\alpha}{4}\right)T, \\[4pt]
        0, & \left(F_{\alpha-1}+\dfrac{\left(2+r_\alpha\right)f_\alpha}{4}\right)T < t\le \left(F_{\alpha-1}+\dfrac{\left(4-r_\alpha\right)f_\alpha}{4}\right)T,\\[4pt]
        a_\alpha, & \left(F_{\alpha-1}+\dfrac{\left(4-r_\alpha\right)f_\alpha}{4}\right)T < t \le F_\alpha T,\\[4pt]
        0, & \text{otherwise.}
    \end{cases}
    \label{eq_assumed_square_pulse_g}
\end{equation}
\end{widetext}
This allocates a total time $r_{\alpha}f_\alpha T$ to the pulse on $\hat{T}^{\alpha}$ with strength $a_\alpha$.  
$r_{\alpha}$ controls idle time internal to the pulse and satisfies $0\leq r_{\alpha} \leq1$.  

\begin{figure}[t]
\includegraphics[width=0.45\textwidth,angle=0]{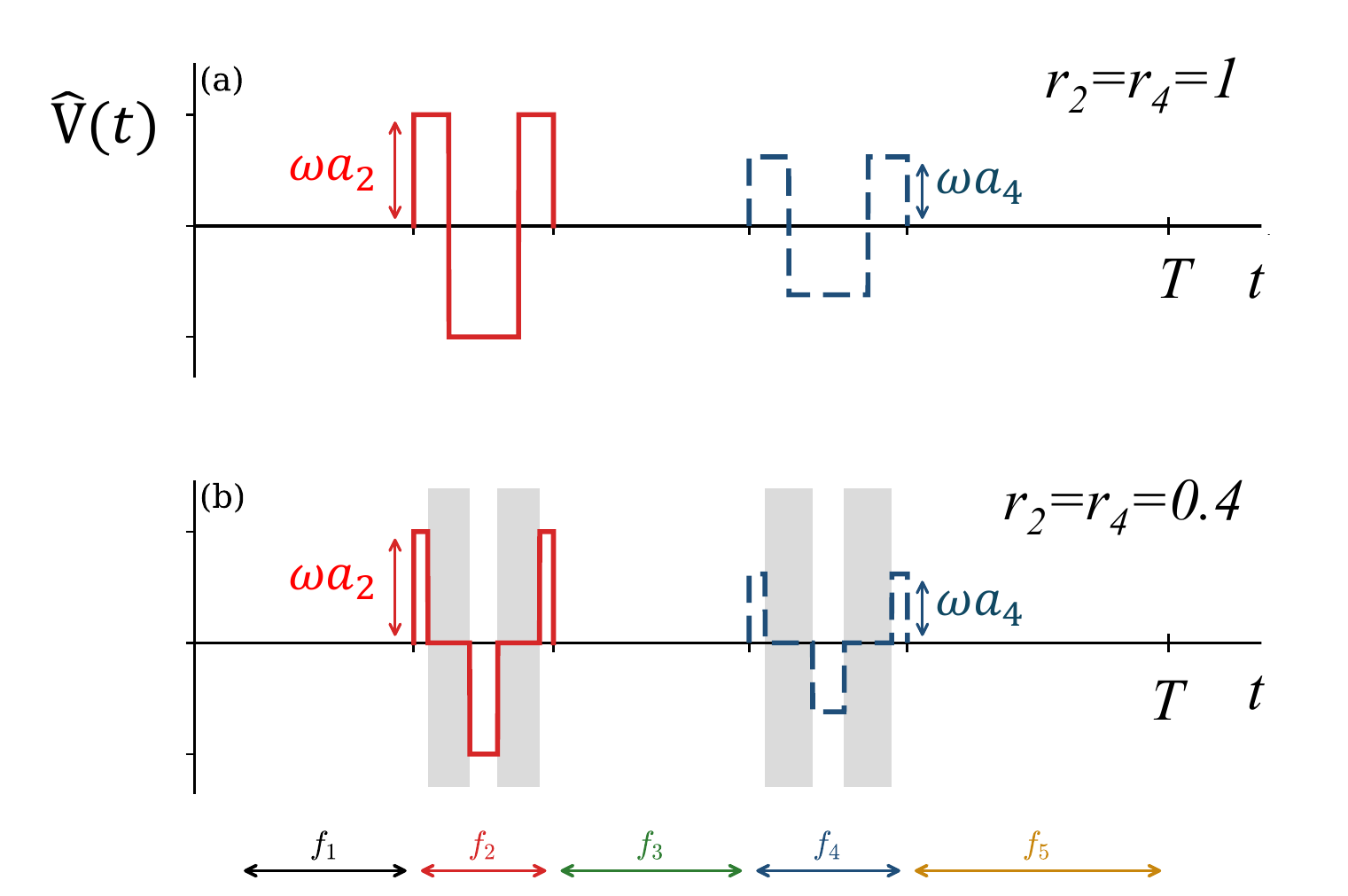}
\caption{(a) The same schematic as Fig.~\ref{fig_schematic_general_pulse} where we plot Eq.~\eqref{eq_assumed_square_pulse_g}.  In this case there is no internal idle time ($r_2=r_4=1$) and $u_{\alpha}$, $v_{\alpha}$, and $w_{\alpha}$ are given by Eqs.~\eqref{eq_square_pulse_uvw}. (b) The same as panel (a) but for a pulse with internal idle time (shaded grey regions), $r_2=r_4=0.4$.  Here $u_{\alpha}$, $v_{\alpha}$, and $w_{\alpha}$ are given by Eqs.~\eqref{eq_uvw_trapezoid}.
}
\label{fig_schematic_general_pulse_with_r}
\end{figure}

Figure~\ref{fig_schematic_general_pulse_with_r}a plots 
the pulse for the case $r_2=r_4=1$ used in the main text, whereas 
Fig.~\ref{fig_schematic_general_pulse_with_r}b plots the pulse originating from Eq.~\eqref{eq_assumed_square_pulse_g}.  Here we choose two pulses with internal idle times such that $r_2=r_4=0.4$.  

We use Eq.~\eqref{eq_define_G_general} to define the pulse averages:
\begin{widetext}
\begin{equation}
    G_\alpha(t)=
    \begin{cases}
        0, & t \le F_{\alpha-1}T,\\[4pt]
        \omega a_\alpha\!\left(t-F_{\alpha-1}T\right),
        & F_{\alpha-1} T < t \le \left(F_{\alpha-1}+\dfrac{r_\alpha f_\alpha}{4}\right)T,\\[4pt]
        \omega a_\alpha\dfrac{r_\alpha f_\alpha T}{4},
        & \left(F_{\alpha-1}+\dfrac{r_\alpha f_\alpha}{4}\right)T < t \le \left(F_{\alpha-1}+\dfrac{\left(2-r_\alpha\right)f_\alpha}{4}\right)T,\\[4pt]
        \omega a_\alpha\!\left(\dfrac{f_\alpha T}{2}-t+F_{\alpha-1}T\right),
        & \left(F_{\alpha-1}+\dfrac{\left(2-r_\alpha\right)f_\alpha}{4}\right)T < t \le \left(F_{\alpha-1}+\dfrac{\left(2+r_\alpha\right)f_\alpha}{4}\right)T,\\[4pt]
        -\omega a_\alpha\dfrac{r_\alpha f_\alpha T}{4},
        & \left(F_{\alpha-1}+\dfrac{\left(2+r_\alpha\right)f_\alpha}{4}\right)T < t \le \left(F_{\alpha-1}+\dfrac{\left(4-r_\alpha\right)f_\alpha}{4}\right)T,\\[4pt]
        \omega a_\alpha\!\left(t-F_\alpha T\right),
        & \left(F_{\alpha-1}+\dfrac{\left(4-r_\alpha\right)f_\alpha}{4}\right)T < t\le F_\alpha T,\\[4pt]
        0, & t > F_\alpha T .
    \end{cases}
    \label{eq_trapezoid_G}
\end{equation}
\end{widetext}

Equation~\eqref{eq_assumed_square_pulse_g} retains both properties on which the
formalism relies: $g_{\alpha}$ integrates to zero within its own block, so the
$G_{\alpha}$ have disjoint support, and $G_{\alpha}$ remains antisymmetric about
the midpoint of its block, so that $\overline{\{G_\alpha(t)\}^p}=0$ for $p$ odd and:
\begin{align}
    \overline{\{G_\alpha(t)\}^{2l}}=
     \dfrac{\left(\pi a_\alpha r_\alpha\right)^{2l} f_\alpha^{2l+1}}{2^{2l}}
        \left(1-r_\alpha+\dfrac{r_\alpha}{2l+1}\right).
    \label{eq_average_power_of_G_trapezoid}
\end{align}
We use $G_\alpha(t)$ to compute the three needed $\mathcal{C}_{\alpha}(c)$ integrals.  We find:
\begin{align}
    u_{\alpha}
    &=
    f_{\alpha}\left[
    \frac{r_{\alpha}}{2}\,
    \mathrm{sc}_1\!\left(\pi a_{\alpha}f_{\alpha}r_{\alpha}\right)
    -\left(1-r_{\alpha}\right)
    \sin^{2}\!\left(\frac{\pi a_{\alpha}f_{\alpha}r_{\alpha}}{2}\right)
    \right],
    \nonumber \\
    v_{\alpha}
    &=
    f_{\alpha}\left[
    \frac{r_{\alpha}}{2}\,
    \mathrm{sc}_1\!\left(\frac{\pi a_{\alpha}f_{\alpha}r_{\alpha}}{2}\right)
    -\left(1-r_{\alpha}\right)
    \sin^{2}\!\left(\frac{\pi a_{\alpha}f_{\alpha}r_{\alpha}}{4}\right)
    \right],
    \nonumber \\
    w_{\alpha}
    &=
    f_{\alpha}\left[
    \frac{r_{\alpha}}{2}\,
    \mathrm{sc}_1\!\left(\frac{\pi a_{\alpha}f_{\alpha}r_{\alpha}}{4}\right)
    -\left(1-r_{\alpha}\right)
    \sin^{2}\!\left(\frac{\pi a_{\alpha}f_{\alpha}r_{\alpha}}{8}\right)
    \right].
    \label{eq_uvw_trapezoid}
\end{align}

\begin{figure}[t]
\includegraphics[width=0.45\textwidth,angle=0]{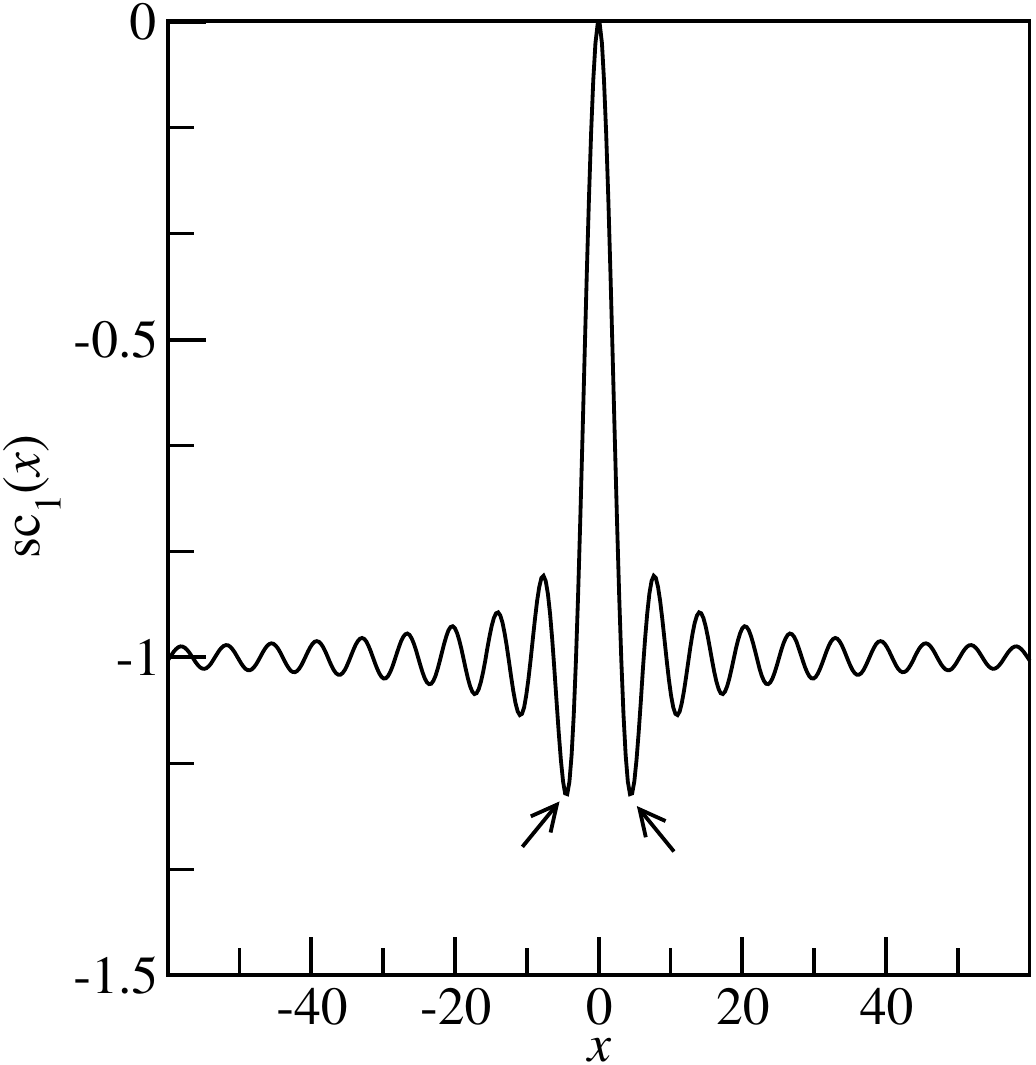}
\caption{\emph{The $\mathrm{sc}_1$ function}. Plot of $\mathrm{sc}_1(x)$ versus its argument.  Note $\mathrm{sc}_1(x)\leq 0$. The arrows indicate the global minima of $\approx -1.2172$ at $x\approx \pm4.4934$.  
The effective strengths $\tilde{h}_{\alpha,\beta}$, $\tilde{J}_{\alpha,\beta}$ and $\tilde{J}^{(\wp)}_{\alpha,\beta}$ 
in Eqs.~\eqref{eq_htilde_general_profile} and~\eqref{eq_Jtilde_general_profile}
all depend on $\mathrm{sc}_1$ functions for the case of square pulses via Eqs.~\eqref{eq_square_pulse_uvw}.  The oscillations show that a wide range of effective couplings are possible.  The global minimum bounds their values. 
}
\label{fig_sc1}
\end{figure}

For the square pulses with no internal idle time ($r_{\alpha}=1$) we recover Eqs.~\eqref{eq_square_pulse_uvw} discussed in the main text.  Figure~\ref{fig_sc1} 
plots $\mathrm{sc}_1(x)$ to show how the couplings will vary with pulse strength $a_{\alpha}$ for $r_{\alpha}=1$.  We see that even for very large $a_{\alpha}$,  $\tilde{h}_{\alpha,\beta}$, $\tilde{J}^{(\wp)}_{\alpha,\beta}$, and $\tilde{J}_{\alpha,\beta}$ remain bounded because $\mathrm{sc}_1(x)$ is bounded.

\section{Derivation of lowest order terms in high frequency expansion of the kick operator}
\label{sec_derivation_lowest_order}

In this appendix we show that Eq.~\eqref{eq_kick_zero_order}, together with setting $\hat K^{(1)}(t)=0$ within the leading-order ansatz, yields a consistent construction of the leading kick operator and cycle-averaged effective Hamiltonian. This choice does not assert that the exact micromotion lacks corrections of order $\omega^{-1}$.  We follow an approach established for $d=2$ in Refs.~\cite{Goldman2014,SCOTT2025} but with key differences.  We: i) Apply simplifications from the pulse assumptions of Sec.~\ref{sec_pulse_assumption_higher_level}; ii) We establish the framework needed for future work to go to higher order in inverse frequency; iii) We work out the effective Hamiltonian and keep both weak and strong driving explicit for use in future work (the main text discusses only strong driving); and iv) We show that the nested commutation relations derived here for arbitrary finite $d$, Eq.~\eqref{eq_nested_commutator}, still allow a relatively simple solution to the high frequency expansion problem.  

We start by analyzing the two terms on the right side of Eq.~\eqref{eq:hamiltonianGaugeTransform}.  By expanding the exponentials in the first and second terms we find: 
\begin{align}
    e^{i\hat{K}(t)} \hat{H} e^{-i\hat{K}(t)} &= \sum_{p=0}^{\infty} \frac{i^p}{p!} \left[\left[\hat{K},\hat{H}\right]\right]_p \label{eq:KderivsOfH}
    \end{align}
    and
    \begin{align}
    \left(\frac{\partial e^{i\hat{K}(t)}}{\partial t}\right)e^{-i \hat{K}(t)}& = \sum_{p=0}^{\infty} \frac{i^{p+1}}{(p+1)!} \left[\left[\hat{K}(t),\frac{\partial \hat{K}(t)}{\partial t}\right]\right]_p, \label{eq:KderivsOfdKdt}
\end{align}
respectively.  In the following we drop the time dependence of operators for brevity.  If we combine Eqs.~\eqref{eq:floquetHamiltonianExpansion},~\eqref{eq:kickExpansion},~\eqref{eq:KderivsOfH}, and~\eqref{eq:KderivsOfdKdt} in Eq.~\eqref{eq:hamiltonianGaugeTransform}, we obtain a general expansion for the effective Hamiltonian:
\begin{widetext}
    \begin{align}
        \sum_{n=0}^{\infty} \omega^{-n} \hat{H}_{\text{eff}}^{(n)} &= \sum_{n=0}^{\infty} \frac{i^n}{n!} \sum_{q_1,...,q_n = 0}^{\infty} \omega^{-(q_1 + ... + q_n)}\left[\hat{K}^{(q_1)},...\left[\hat{K}^{(q_n)},\hat{H}\right]...\right] \nonumber \\
        &+ i\sum_{q=0}^{\infty}\sum_{n=1}^{\infty} \omega^{-q}\frac{i^n}{n!}\left\{ \sum_{q_2,...,q_n=0}^{\infty} \omega^{-(q_2 + ... + q_n)}\left[\hat{K}^{(q_2)},...\left[\hat{K}^{(q_n)},\frac{\partial \hat{K}^{(q)}}{\partial t}\right]...\right]\right\}. 
        \label{eq:FullEffectiveHamiltonian}
    \end{align}
\end{widetext}
We separate the driving potential into a part that is small compared to $\omega$ and a part that is of order $\omega$,
$
    \hat{V}  = \hat{V}_1  + \omega \hat{V}_2 . 
$
We can therefore write
$
    \hat{H}(t)  = \hat{H}_0 + \hat{V}_1 + \omega \hat{V}_2.
$
We insert these into Eq.~\eqref{eq:FullEffectiveHamiltonian}.  To make progress, we use the notation:
\begin{widetext}
    \begin{align}
        D_{\{0,1,...,n\}}^n[-] &:= \bigg(\left[\hat{K}^{(n)},\left[\hat{K}^{(0)},...,\left[\hat{K}^{(0)},-\right]...\right]\right] + \left[\hat{K}^{(0)},\left[\hat{K}^{(n)},...,\left[\hat{K}^{(0)},-\right]...\right]\right] + ... + \left[\hat{K}^{(0)},\left[\hat{K}^{(0)},...,\left[\hat{K}^{(n)},-\right]...\right]\right]\bigg) \nonumber\\
         &+ \bigg(\left[\hat{K}^{(n-1)},\left[\hat{K}^{(1)},\left[\hat{K}^{(0)},...,\left[\hat{K}^{(0)},-\right]...\right]\right]\right] + \left[\hat{K}^{(n-1)},\left[\hat{K}^{(0)},\left[\hat{K}^{(1)},...,\left[\hat{K}^{(0)},-\right]...\right]\right]\right] + ... \nonumber\\
         &+ \left[\hat{K}^{(0)},\left[\hat{K}^{(n-1)},\left[\hat{K}^{(1)},...,\left[\hat{K}^{(0)},-\right]...\right]\right]\right] + ...\bigg) + ... + \left[\hat{K}^{(1)},\left[\hat{K}^{(1)},...,\left[\hat{K}^{(1)},-\right]...\right]\right]
    \end{align}
\end{widetext}
so that the orders of $\hat{K}$ in the ($n^{th}$ order) Lie derivatives need to add to $n$ in each term appearing in the sum, and the sum is over all permutations. $D^m_{\{0,1,...,n\}}$ with $m < n$ just amounts to truncating the terms in $D^n_{\{0,1,...,n\}}$ so that the sums of the orders still add to $n$ but the derivative is only of order $m$.

In particular, a single commutator forces the orders to be carried by one factor,
\begin{equation}
    D^{1}_{\{0,1,...,n\}}[-] = \left[\hat{K}^{(n)},-\right] .
    \label{eq:D_first_order}
\end{equation}
With this notation we can express the expansion in orders of $\omega$. Furthermore, if $m = 0$ and $n > 0$, then $D^{0}_{\{0,...n\}} = 0$, since we must always satisfy the rule that the sum of the corrections must be $p+1$ when $p > -1$.  If we grade these by orders of $\omega$ (putting $\hat{H}' = \hat{H}_0 + \hat{V}_1$), Eq.~\eqref{eq:FullEffectiveHamiltonian} yields:

{\allowdisplaybreaks
\begin{align}
    \mathcal{O}(\omega):\ &
        \omega \hat{V}_2
        + i \omega D^1_{\{0\}}[\hat{V}_2]
        + i^2\, \partial_t \hat{K}^{(0)} \nonumber\\
    &+ \frac{i^2 \omega}{2!}D^2_{\{0\}}[\hat{V}_2]
        + \frac{i^3}{2!}D^1_{\{0\}}\big[\partial_t \hat{K}^{(0)}\big]
        + \cdots \nonumber\\
    &+ \omega\frac{i^n}{n!}D^n_{\{0\}}[\hat{V}_2]
        + \frac{i^{n+1}}{n!}D^{n-1}_{\{0\}}\big[\partial_t \hat{K}^{(0)}\big]
        + \cdots \\[4pt]
    \mathcal{O}(1):\ &
        \hat{H}'
        + i D^1_{\{0\}}[\hat{H}']
        + iD^1_{\{0,1\}}[\hat{V}_2]
        + \frac{i^2}{\omega}\, \partial_t \hat{K}^{(1)} \nonumber\\
    &+ \frac{i^2}{2!}\Big(D^2_{\{0\}}[\hat{H}']
        + D^2_{\{0,1\}}[\hat{V}_2]\Big) \nonumber\\
    &+ \frac{i^3}{2!\,\omega}\Big(
        D^1_{\{0,1\}}\big[\partial_t \hat{K}^{(0)}\big]
        + D^1_{\{0\}}\big[\partial_t \hat{K}^{(1)}\big]\Big)
        + \cdots \\[4pt]
    \mathcal{O}(\omega^{-1}):\ &
        \frac{i}{\omega}\Big(D^1_{\{0,1\}}[\hat{H}']
        + D^1_{\{0,1,2\}}[\hat{V}_2]\Big)
        + \frac{i^2}{\omega^2}\, \partial_t \hat{K}^{(2)} \nonumber\\
    &+ \frac{i^2}{2!\,\omega}\Big(D^2_{\{0,1\}}[\hat{H}']
        + D^2_{\{0,1,2\}}[\hat{V}_2]\Big) \nonumber\\
    &+ \frac{i^3}{2!\,\omega^2}\Big(
        D^1_{\{0\}}\big[\partial_t \hat{K}^{(2)}\big]
        + D^1_{\{0,1\}}\big[\partial_t \hat{K}^{(1)}\big] \nonumber\\
    &\qquad\qquad
        + D^1_{\{0,1,2\}}\big[\partial_t \hat{K}^{(0)}\big]\Big)
        + \cdots \\
    &\ \ \vdots \nonumber
\end{align}}

These can be expressed in closed form as
{\allowdisplaybreaks
\begin{align}
    \mathcal{O}(\omega):\ &
        \omega \sum_{n=0}^{\infty} \frac{i^n}{n!}D^{n}_{\{0\}}[\hat{V}_2] \nonumber\\
    &+ i \sum_{n=1}^{\infty} \frac{i^n}{n!}
        D^{n-1}_{\{0\}}\big[\partial_t \hat{K}^{(0)}\big]
        \label{eq:OrderOmega}\\[4pt]
    \mathcal{O}(1):\ &
        \sum_{n=0}^{\infty}\frac{i^n}{n!} D^n_{\{0\}}[\hat{H}']
        + \sum_{n=1}^{\infty} \frac{i^n}{n!}D^n_{\{0,1\}}[\hat{V}_2] \nonumber\\
    &+ \frac{i}{\omega}\sum_{n=1}^{\infty} \frac{i^{n+1}}{(n+1)!}
        D^{n}_{\{0,1\}}\big[\partial_t \hat{K}^{(0)}\big] \nonumber\\
    &+ \frac{i}{\omega}\sum_{n=0}^{\infty} \frac{i^{n+1}}{(n+1)!}
        D^{n}_{\{0\}}\big[\partial_t \hat{K}^{(1)}\big]
        \label{eq:OrderOne}\\[4pt]
    \mathcal{O}(\omega^{-1}):\ &
        \frac{1}{\omega}\sum_{n=1}^{\infty}\frac{i^n}{n!} D^n_{\{0,1\}}[\hat{H}']
        + \frac{1}{\omega}\sum_{n=1}^{\infty} \frac{i^n}{n!}D^n_{\{0,1,2\}}[\hat{V}_2] \nonumber\\
    &+ \frac{i}{\omega^2}\sum_{n=1}^{\infty} \frac{i^n}{n!}\Big(
        D^{n}_{\{0,1,2\}}\big[\partial_t \hat{K}^{(0)}\big]
        + D^{n}_{\{0,1\}}\big[\partial_t \hat{K}^{(1)}\big] \nonumber\\
    &\qquad\qquad\qquad\quad
        + D^{n-1}_{\{0\}}\big[\partial_t \hat{K}^{(2)}\big]\Big)
        \label{eq:OrderOmegaInverse}\\
    &\ \ \vdots \nonumber\\[4pt]
    \mathcal{O}(\omega^{-p}):\ &
        \frac{1}{\omega^{p}}\sum_{n=1}^{\infty}\frac{i^n}{n!} D^n_{\{0,\dots,p\}}[\hat{H}']
        + \frac{1}{\omega^p}\sum_{n=1}^{\infty} \frac{i^n}{n!}D^n_{\{0,\dots,p+1\}}[\hat{V}_2] \nonumber\\
    &+ \frac{i}{\omega^{p+1}}\sum_{n=1}^{\infty} \frac{i^n}{n!}\Big(
        D^{n}_{\{0,\dots,p+1\}}\big[\partial_t \hat{K}^{(0)}\big]
        + \cdots \nonumber\\
    &\qquad\qquad\qquad\quad
        + D^{n-1}_{\{0\}}\big[\partial_t \hat{K}^{(p+1)}\big]\Big)
        \label{eq:OrderOmegaMinusP}\\
    &\ \ \vdots \nonumber
\end{align}}
where the term $\mathcal{O}(\omega^{-1})$ will be thought of as a correction term going forward.  Equations~\eqref{eq:OrderOmegaMinusP} define higher order conditions that must be satisfied to go beyond the lowest order approach discussed in the main text.

To obtain the lowest order results discussed in the main text we impose the
single condition at $\mathcal{O}(\omega)$, Eq.~\eqref{eq:OrderOmega}, which
determines $\hat{K}^{(0)}$, and then read off $\hat{H}^{(0)}_{\text{eff}}$ from
the $\mathcal{O}(1)$ block, Eq.~\eqref{eq:OrderOne}.
We set $\hat K^{(1)}=0$ as a simplifying ansatz for determining a consistent leading-order solution $\hat K^{(0)}$. With this choice, the terms containing $\hat K^{(1)}$ or $\partial_t\hat K^{(1)}$ drop out; for example, Eq.~\eqref{eq:D_first_order} gives $D^{1}_{\{0,1\}}[\hat V_2] =[\hat K^{(1)},\hat V_2]=0$. This choice does not imply that the complete subleading micromotion vanishes. Rather, the residual time dependence of the toggling-frame Hamiltonian generally produces corrections to the kick operator of order $\omega^{-1}$, which are not retained in the present leading-order treatment. The $\mathcal{O}(\omega^{-1})$ equation, Eq.~\eqref{eq:OrderOmegaInverse}, contains the higher-order conditions required to determine these corrections consistently.

Since $\hat{H}_{\text{eff}}$ must remain finite as $\omega\to\infty$, we require that terms
that grow linearly in $\omega$ must cancel. This is a condition on
$\hat{K}^{(0)}$, and it is the only such condition at this order.  Equation~\eqref{eq:OrderOmega} therefore yields:
\begin{equation}
    \omega \sum_{n=0}^{\infty} \frac{i^n}{n!}D^{n}_{\{0\}}[\hat{V}_2]
    + i \sum_{n = 1}^{\infty} \frac{i^n}{n!}
    D^{n-1}_{\{0\}}\left[\frac{\partial \hat{K}^{(0)}}{\partial t}\right] = 0 ,
    \label{eq:OmegaCondition}
\end{equation}
where $D^{1}_{\{0\}}=[\hat{K}^{(0)},\,\cdot\,]$ and $D^{n}_{\{0\}}$ denotes $n$
nested commutators with $\hat{K}^{(0)}$.

The Lie derivatives are highly non-trivial and depend on $d$.  Rather than resumming these series, we use the pulse assumptions of
Sec.~\ref{sec_definitions_assumptions} directly. Time ordering means that at any instant at most 
one channel is active, so that $\hat{K}^{(0)}$ and $\hat{V}_2$ are both
proportional to the same generator $\hat{T}^{\alpha}$, and therefore
\begin{align}
\left[\hat{K}^{(0)},\hat{V}_2\right]=0,
\qquad
\left[\hat{K}^{(0)},\frac{\partial \hat{K}^{(0)}}{\partial t}\right]=0 .
\label{eq:K0_commutes}
\end{align}
Equation~\eqref{eq:K0_commutes} truncates both series in
Eq.~\eqref{eq:OmegaCondition} term by term: every
$n\ge 1$ term of the first sum contains at least one factor of
$[\hat{K}^{(0)},\hat{V}_2]$, and every $n\ge 2$ term of the second contains at
least one factor of $[\hat{K}^{(0)},\partial_t\hat{K}^{(0)}]$. Only the $n=0$ term
of the first sum and the $n=1$ term of the second survive, leaving
\begin{equation}
    \omega \hat{V}_2 - \frac{\partial \hat{K}^{(0)}}{\partial t} = 0 ,
\end{equation}
that is,
\begin{equation}
    \frac{\partial \hat{K}^{(0)}}{\partial t} = \omega \hat{V}_2 .
    \label{eq:K0DiffEqn}
\end{equation}
Integrating Eq.~\eqref{eq:K0DiffEqn} and using
Eq.~\eqref{eq_pulse_general} gives Eq.~\eqref{eq_kick_zero_order}.  We have therefore shown that Eq.~\eqref{eq_kick_zero_order} satisfies the $\mathcal{O}(\omega)$ block, Eq.~\eqref{eq:OrderOmega}, under the time-ordering and zero average assumptions of Sec.~\ref{sec_definitions_assumptions}.  The approach and assumptions avoided the need to use the $d$-level nested commutator, Eq.~\eqref{eq_nested_commutator}, at this order. 

We now move on to Eq.~\eqref{eq:OrderOne}.  In contrast to the $\mathcal{O}(\omega)$ block, the $\mathcal{O}(1)$ block is not a condition to be imposed: it
defines $\hat{H}^{(0)}_{\text{eff}}$. Of the four sums in
Eq.~\eqref{eq:OrderOne}, the last three each carry a factor of $\hat{K}^{(1)}$ or
$\partial_t\hat{K}^{(1)}$ and therefore vanish, leaving
\begin{equation}
\hat{H}^{(0)}_{\text{eff}}(t)=\sum_{n=0}^{\infty}\frac{i^n}{n!}
D^{n}_{\{0\}}\!\left[\hat{H}'\right]
=e^{i\hat{K}^{(0)}}\hat{H}'e^{-i\hat{K}^{(0)}} .
\label{eq:H0eff_conjugated}
\end{equation}
This expression is still time dependent through $\hat{K}^{(0)}(t)$, and becomes
the time-independent effective Hamiltonian only after the cycle average.  After time averaging, Eq.~\eqref{eq:H0eff_conjugated} yields the general form for the lowest order effective Hamiltonian including both weak and strong driving.

To make contact with the main text, we set the weak driving to zero. 
With
$\hat{H}'=\hat{H}_0$, Eq.~\eqref{eq:H0eff_conjugated} gives:
\begin{equation}
\He=\hat{H}_0+\sum_{p=1}^{\infty}\frac{i^p}{p!}
\overline{\left[\left[\hat{K}^{(0)}(t),\hat{H}_0\right]\right]_p} ,
\end{equation}
in agreement with Eq.~\eqref{eq_H0_general}.  This establishes $\hat K^{(0)}(t)$ and the cycle-averaged effective Hamiltonian consistently at leading order. Subleading corrections to the micromotion, including contributions of order $\omega^{-1}$, are not determined in the present treatment.

We summarize the results found in this appendix.  We use the approach of Refs.~\onlinecite{Goldman2014,SCOTT2025} to generalize to qudits.  After implementing the time ordering and zero average assumptions of  Sec.~\ref{sec_definitions_assumptions}, we derived Eqs.~\eqref{eq_H0_general} and \eqref{eq_kick_zero_order}.  We also derived higher order terms, Eq.~\eqref{eq:OrderOmegaMinusP}, needed to be satisfied for future work.  And finally, we derived the effective Hamiltonian needed to incorporate weak driving.

\section{$d=3$ generator matrices $\hat{\lambda}^{\alpha}_j$ and mapping to spin-1 magnetism}
\label{sec_appendix_spin_mapping}

In this section we define the $d=3$ generator matrices used in Sec.~\ref{sec_top_level_d_3} and map them to operators for spin-1 magnetism.  In Sec.~\ref{sec_top_level_d_3} we identify $\hat T_j^\alpha$ with the following generators for $d=3$:   
\begin{align}
\hat{\lambda}_j^1
&=
\frac{1}{2}
\begin{pmatrix}
0&1&0\\
1&0&0\\
0&0&0
\end{pmatrix},
\nonumber \\
\hat{\lambda}_j^2
&=
\frac{1}{2}
\begin{pmatrix}
0&i&0\\
-i&0&0\\
0&0&0
\end{pmatrix},
\nonumber \\
\hat{\lambda}_j^3
&=
\frac{1}{2}
\begin{pmatrix}
1&0&0\\
0&-1&0\\
0&0&0
\end{pmatrix},
\nonumber \\
\hat{\lambda}_j^4
&=
\frac{1}{2}
\begin{pmatrix}
0&0&1\\
0&0&0\\
1&0&0
\end{pmatrix},
\nonumber \\
\hat{\lambda}_j^5
&=
\frac{1}{2}
\begin{pmatrix}
0&0&i\\
0&0&0\\
-i&0&0
\end{pmatrix},
\nonumber \\
\hat{\lambda}_j^6
&=
\frac{1}{2}
\begin{pmatrix}
0&0&0\\
0&0&1\\
0&1&0
\end{pmatrix},
\nonumber \\
\hat{\lambda}_j^7
&=
\frac{1}{2}
\begin{pmatrix}
0&0&0\\
0&0&i\\
0&-i&0
\end{pmatrix},
\nonumber \\
\hat{\lambda}_j^8
&=
\frac{1}{2}
\begin{pmatrix}
1&0&0\\
0&0&0\\
0&0&-1
\end{pmatrix}.
\end{align}
This defines a non-standard Cartan basis in which $\hat{\lambda}_j^3$ and $\hat{\lambda}_j^8$ are not orthogonal.  The matrices are proportional to the usual Gell-Mann matrices with the exception of the diagonal matrices.  

These matrices can be mapped to spin-1 operators. We work in a spin-1 encoding with local basis on the $j^{\text{th}}$ site:
$\{|{+}1\rangle_j,|0\rangle_j,|{-}1\rangle_j\}$, identified with the qutrit basis
$\{|1\rangle_j,|2\rangle_j,|3\rangle_j\}$ used for the $\hat{\lambda}_j^\alpha$ matrices.
Define the spin-1 operators: 
\begin{align}
\hat{S}^x_j&=\frac{1}{\sqrt2}
\begin{pmatrix}
0&1&0\\
1&0&1\\
0&1&0
\end{pmatrix},
\nonumber\\
\hat{S}^y_j&=\frac{1}{\sqrt2}
\begin{pmatrix}
0&-i&0\\
i&0&-i\\
0&i&0
\end{pmatrix},
\nonumber\\
\hat{S}^z_j&=
\begin{pmatrix}
1&0&0\\0&0&0\\0&0&-1
\end{pmatrix},
\end{align}
which satisfy $[\hat{S}^\mu_j,\hat{S}^\nu_j]=i\epsilon_{\mu\nu\rho}\hat{S}^\rho_j$ and
$\operatorname{tr}(\hat{S}^\mu_j \hat{S}^\nu_j)=2\delta_{\mu\nu}$. We also define the quadrupole operators:
\begin{align}
\hat{Q}^{xy}_j &= \hat{S}^x_j \hat{S}^y_j + \hat{S}^y_j \hat{S}^x_j, \nonumber \\
\hat{Q}^{xz}_j &= \hat{S}^x_j \hat{S}^z_j + \hat{S}^z_j \hat{S}^x_j, \nonumber \\
\hat{Q}^{yz}_j &= \hat{S}^y_j \hat{S}^z_j + \hat{S}^z_j \hat{S}^y_j, \nonumber \\
\hat{Q}^{x^2-y^2}_j &= (\hat{S}^x_j)^2 - (\hat{S}^y_j)^2, \nonumber \\
\hat{Q}^{0}_j &= (\hat{S}^z_j)^2 - \tfrac{2}{3}\,\mathbb{I}_j.
\end{align}
The eight generators map onto spin-1 dipole and quadrupole operators:
\begin{align}
\hat{\lambda}^1_j &= \frac{\sqrt2}{4}\left(\hat{S}^x_j + \hat{Q}^{xz}_j\right), \nonumber \\
\hat{\lambda}^2_j &= -\frac{\sqrt2}{4}\left(\hat{S}^y_j + \hat{Q}^{yz}_j\right), \nonumber \\
\hat{\lambda}^3_j &= \frac{1}{4}\,\hat{S}^z_j + \frac{3}{4}\,\hat{Q}^{0}_j, \nonumber \\
\hat{\lambda}^4_j &= \frac{1}{2}\,\hat{Q}^{x^2-y^2}_j, \nonumber \\
\hat{\lambda}^5_j &= -\frac{1}{2}\,\hat{Q}^{xy}_j, \nonumber \\
\hat{\lambda}^6_j &= \frac{\sqrt2}{4}\left(\hat{S}^x_j - \hat{Q}^{xz}_j\right), \nonumber \\
\hat{\lambda}^7_j &= -\frac{\sqrt2}{4}\left(\hat{S}^y_j - \hat{Q}^{yz}_j\right), \nonumber \\
\hat{\lambda}^8_j &= \frac{1}{2}\,\hat{S}^z_j.
\end{align}

\section{$d=3$ Lookup Tables}
\label{sec_appendix_d_3_lookup_tables}

In this section, the lookup tables, Tables~\ref{tab_d_3_same_support}-~\ref{tab_d_3_one_overlap}, record the values of $\phi_{\alpha\beta}^{(\wp)}$, $\lambda_{\alpha\beta}$, $\nu_{\alpha\beta}$, and $\kappa_{\wp}(\alpha,\beta)$ for use with Eqs.~\eqref{eq_Heff_central_J} and~\eqref{eq_Heff_loc}, specific to $d=3$.

\begin{table}[t]
\centering
\caption{Same-support channels in the $d=3$ $\hat{\lambda}^{\alpha}$ basis.}
\small
\setlength{\tabcolsep}{4pt}
\begin{tabular}{|l|l|c|c|l|l|}
\hline
$\alpha$ & $\beta$ & $\lambda_{\alpha\beta}$ & $\nu_{\alpha\beta}$ & $(\kappa_{\mathrm o}(\alpha,\beta),\phi_{\alpha\beta}^{(\mathrm o)})$ & $(\kappa_{\mathrm e}(\alpha,\beta),\phi_{\alpha\beta}^{(\mathrm e)})$ \\
\hline
$1$ & $2$ & $0$ & $0$ & $(3,-i)$ & $(2,+1)$ \\
\hline
$2$ & $1$ & $0$ & $0$ & $(3,+i)$ & $(1,+1)$ \\
\hline
$3$ & $1$ & $0$ & $0$ & $(2,-i)$ & $(1,+1)$ \\
\hline
$1$ & $3$ & $0$ & $0$ & $(2,+i)$ & $(3,+1)$ \\
\hline
$3$ & $2$ & $0$ & $0$ & $(1,+i)$ & $(2,+1)$ \\
\hline
$2$ & $3$ & $0$ & $0$ & $(1,-i)$ & $(3,+1)$ \\
\hline
$4$ & $5$ & $0$ & $0$ & $(8,-i)$ & $(5,+1)$ \\
\hline
$5$ & $4$ & $0$ & $0$ & $(8,+i)$ & $(4,+1)$ \\
\hline
$8$ & $4$ & $0$ & $0$ & $(5,-i)$ & $(4,+1)$ \\
\hline
$4$ & $8$ & $0$ & $0$ & $(5,+i)$ & $(8,+1)$ \\
\hline
$8$ & $5$ & $0$ & $0$ & $(4,+i)$ & $(5,+1)$ \\
\hline
$5$ & $8$ & $0$ & $0$ & $(4,-i)$ & $(8,+1)$ \\
\hline
$6$ & $7$ & $0$ & $0$ & $(h,-i)$ & $(7,+1)$ \\
\hline
$7$ & $6$ & $0$ & $0$ & $(h,+i)$ & $(6,+1)$ \\
\hline
\end{tabular}
\label{tab_d_3_same_support}
\end{table}

\begin{table}[t]
\centering
\caption{One-overlap channels between supports $\{1,2\}$ and $\{1,3\}$.}
\small
\setlength{\tabcolsep}{4pt}
\begin{tabular}{|l|l|c|c|l|l|}
\hline
$\alpha$ & $\beta$ & $\lambda_{\alpha\beta}$ & $\nu_{\alpha\beta}$ & $(\kappa_{\mathrm o}(\alpha,\beta),\phi_{\alpha\beta}^{(\mathrm o)})$ & $(\kappa_{\mathrm e}(\alpha,\beta),\phi_{\alpha\beta}^{(\mathrm e)})$ \\
\hline
$1$ & $4$ & $1$ & $1$ & $(7,-i)$ & $(4,+1)$ \\
\hline
$1$ & $5$ & $1$ & $1$ & $(6,+i)$ & $(5,+1)$ \\
\hline
$1$ & $8$ & $1$ & $0$ & $(2,+i)$ & $(3,+1)$ \\
\hline
$2$ & $4$ & $1$ & $1$ & $(6,-i)$ & $(4,+1)$ \\
\hline
$2$ & $5$ & $1$ & $1$ & $(7,-i)$ & $(5,+1)$ \\
\hline
$2$ & $8$ & $1$ & $0$ & $(1,-i)$ & $(3,+1)$ \\
\hline
$3$ & $4$ & $1$ & $1$ & $(5,-i)$ & $(4,+1)$ \\
\hline
$3$ & $5$ & $1$ & $1$ & $(4,+i)$ & $(5,+1)$ \\
\hline
$4$ & $1$ & $1$ & $1$ & $(7,+i)$ & $(1,+1)$ \\
\hline
$4$ & $2$ & $1$ & $1$ & $(6,+i)$ & $(2,+1)$ \\
\hline
$4$ & $3$ & $1$ & $0$ & $(5,+i)$ & $(8,+1)$ \\
\hline
$5$ & $1$ & $1$ & $1$ & $(6,-i)$ & $(1,+1)$ \\
\hline
$5$ & $2$ & $1$ & $1$ & $(7,+i)$ & $(2,+1)$ \\
\hline
$5$ & $3$ & $1$ & $0$ & $(4,-i)$ & $(8,+1)$ \\
\hline
$8$ & $1$ & $1$ & $1$ & $(2,-i)$ & $(1,+1)$ \\
\hline
$8$ & $2$ & $1$ & $1$ & $(1,+i)$ & $(2,+1)$ \\
\hline
\end{tabular}
\end{table}

\begin{table}[t]
\centering
\caption{One-overlap channels between supports $\{1,2\}$ and $\{2,3\}$.}
\small
\setlength{\tabcolsep}{4pt}
\begin{tabular}{|l|l|c|c|l|l|}
\hline
$\alpha$ & $\beta$ & $\lambda_{\alpha\beta}$ & $\nu_{\alpha\beta}$ & $(\kappa_{\mathrm o}(\alpha,\beta),\phi_{\alpha\beta}^{(\mathrm o)})$ & $(\kappa_{\mathrm e}(\alpha,\beta),\phi_{\alpha\beta}^{(\mathrm e)})$ \\
\hline
$1$ & $6$ & $1$ & $1$ & $(5,-i)$ & $(6,+1)$ \\
\hline
$1$ & $7$ & $1$ & $1$ & $(4,+i)$ & $(7,+1)$ \\
\hline
$2$ & $6$ & $1$ & $1$ & $(4,+i)$ & $(6,+1)$ \\
\hline
$2$ & $7$ & $1$ & $1$ & $(5,+i)$ & $(7,+1)$ \\
\hline
$3$ & $6$ & $1$ & $1$ & $(7,+i)$ & $(6,+1)$ \\
\hline
$3$ & $7$ & $1$ & $1$ & $(6,-i)$ & $(7,+1)$ \\
\hline
$6$ & $1$ & $1$ & $1$ & $(5,+i)$ & $(1,+1)$ \\
\hline
$6$ & $2$ & $1$ & $1$ & $(4,-i)$ & $(2,+1)$ \\
\hline
$6$ & $3$ & $1$ & $0$ & $(7,-i)$ & $(h,-1)$ \\
\hline
$7$ & $1$ & $1$ & $1$ & $(4,-i)$ & $(1,+1)$ \\
\hline
$7$ & $2$ & $1$ & $1$ & $(5,-i)$ & $(2,+1)$ \\
\hline
$7$ & $3$ & $1$ & $0$ & $(6,+i)$ & $(h,-1)$ \\
\hline
\end{tabular}
\end{table}

\begin{table}[t]
\centering
\caption{One-overlap channels between supports $\{1,3\}$ and $\{2,3\}$.}
\small
\setlength{\tabcolsep}{4pt}
\begin{tabular}{|l|l|c|c|l|l|}
\hline
$\alpha$ & $\beta$ & $\lambda_{\alpha\beta}$ & $\nu_{\alpha\beta}$ & $(\kappa_{\mathrm o}(\alpha,\beta),\phi_{\alpha\beta}^{(\mathrm o)})$ & $(\kappa_{\mathrm e}(\alpha,\beta),\phi_{\alpha\beta}^{(\mathrm e)})$ \\
\hline
$4$ & $6$ & $1$ & $1$ & $(2,-i)$ & $(6,+1)$ \\
\hline
$4$ & $7$ & $1$ & $1$ & $(1,-i)$ & $(7,+1)$ \\
\hline
$5$ & $6$ & $1$ & $1$ & $(1,+i)$ & $(6,+1)$ \\
\hline
$5$ & $7$ & $1$ & $1$ & $(2,-i)$ & $(7,+1)$ \\
\hline
$8$ & $6$ & $1$ & $1$ & $(7,-i)$ & $(6,+1)$ \\
\hline
$8$ & $7$ & $1$ & $1$ & $(6,+i)$ & $(7,+1)$ \\
\hline
$6$ & $4$ & $1$ & $1$ & $(2,+i)$ & $(4,+1)$ \\
\hline
$6$ & $5$ & $1$ & $1$ & $(1,-i)$ & $(5,+1)$ \\
\hline
$6$ & $8$ & $1$ & $0$ & $(7,+i)$ & $(h,+1)$ \\
\hline
$7$ & $4$ & $1$ & $1$ & $(1,+i)$ & $(4,+1)$ \\
\hline
$7$ & $5$ & $1$ & $1$ & $(2,+i)$ & $(5,+1)$ \\
\hline
$7$ & $8$ & $1$ & $0$ & $(6,-i)$ & $(h,+1)$ \\
\hline
\end{tabular}
\label{tab_d_3_one_overlap}
\end{table}

\section{Effective interaction strengths for $d=3$}
\label{sec_appendix_jeff_definition}

This section presents the results for the effective interaction terms between $3$-level systems used in Eq.~\eqref{eq_d_3_effective_H}.  We find:
\begin{align}
J_1^{\mathrm{eff}}
&= J_1\!\left(1+u_2+u_3+v_4+v_5+v_6+v_7+v_8\right)
\nonumber \\
&- (u_3+v_8)J_2 - u_2 J_3
- v_7 J_4 - v_6 J_5
\nonumber \\
&- v_5 J_6 - v_4 J_7 - \tfrac14 u_2\, J_8,
\end{align}

\begin{align}
J_2^{\mathrm{eff}}
&= J_2\!\left(1+u_1+u_3+v_4+v_5+v_6+v_7+v_8\right)
\nonumber \\
&- (u_3+v_8)J_1 - u_1 J_3
- v_6 J_4 - v_7 J_5
\nonumber \\
&- v_4 J_6 - v_5 J_7 - \tfrac14 u_1\, J_8,
\end{align}

\begin{align}
J_3^{\mathrm{eff}}
&= J_3\!\left(1+u_1+u_2+\tfrac14 u_6+\tfrac14 u_7+v_6+v_7\right)
\nonumber \\
&- u_2 J_1 - u_1 J_2 - u_7 J_6 - u_6 J_7
\nonumber \\
&+ \left(\tfrac14 u_1+\tfrac14 u_2+\tfrac14 u_6+\tfrac14 u_7-v_1-v_2-v_6-v_7\right)J_8,
\end{align}

\begin{align}
J_4^{\mathrm{eff}}
&= J_4\!\left(1+u_5+u_8+v_1+v_2+v_3+v_6+v_7\right)
\nonumber \\
&- v_7 J_1 - v_6 J_2 - \tfrac14 u_5\, J_3
- (u_8+v_3)J_5
\nonumber \\
&- v_2 J_6 - v_1 J_7 - u_5 J_8,
\end{align}

\begin{align}
J_5^{\mathrm{eff}}
&= J_5\!\left(1+u_4+u_8+v_1+v_2+v_3+v_6+v_7\right)
\nonumber \\
&- v_6 J_1 - v_7 J_2 - \tfrac14 u_4\, J_3
- (u_8+v_3)J_4
\nonumber \\
&- v_1 J_6 - v_2 J_7 - u_4 J_8,
\end{align}

\begin{align}
J_6^{\mathrm{eff}}
&= J_6\!\left(1+u_7+v_1+v_2+v_3+v_4+v_5+v_8\right)
\nonumber \\
&- v_5 J_1 - v_4 J_2 - \tfrac14 u_7\, J_3
- v_2 J_4 - v_1 J_5
\nonumber \\
&- (v_3+v_8)J_7 - \tfrac14 u_7\, J_8,
\end{align}

\begin{align}
J_7^{\mathrm{eff}}
&= J_7\!\left(1+u_6+v_1+v_2+v_3+v_4+v_5+v_8\right)
\nonumber \\
&- v_4 J_1 - v_5 J_2 - \tfrac14 u_6\, J_3
- v_1 J_4 - v_2 J_5
\nonumber \\
&- (v_3+v_8)J_6 - \tfrac14 u_6\, J_8,
\end{align}

\begin{align}
J_8^{\mathrm{eff}}
&= J_8\!\left(1+u_4+u_5+\tfrac14 u_6+\tfrac14 u_7+v_6+v_7\right)
\nonumber \\
&+ \left(\tfrac14 u_4+\tfrac14 u_5+\tfrac14 u_6+\tfrac14 u_7-v_4-v_5-v_6-v_7\right)J_3
\nonumber \\
&- u_5 J_4 - u_4 J_5 - u_7 J_6 - u_6 J_7,
\end{align}

\begin{align}
J_{38}^{\mathrm{eff}}
&= \left(-\tfrac14 u_6-\tfrac14 u_7+v_4+v_5\right)J_3
\nonumber \\
&+ u_7 J_6 + u_6 J_7
+ \left(-\tfrac14 u_6-\tfrac14 u_7+v_1+v_2\right)J_8.
\end{align}

These equations follow from inserting the
$d=3$ lookup tables of Appendix~\ref{sec_appendix_d_3_lookup_tables} into
Eq.~\eqref{eq_Heff_central_J} and collecting terms channel by channel.  Their
structure tracks the support map and obeys the sum rule from Sec.~\ref{sec_sum_rule}. 

\section{A model of trapped RbCs molecules }
\label{sec_appendix_RbCs_model}

In this section we derive the native Hamiltonian of undressed
$^{87}$Rb$^{133}$Cs molecules from their rotational level structure.  In the main text we express it both in terms of $\hat\lambda^{\alpha}$ operators and the operators relevant for spin-1 magnetism.  We show that there is an intrinsic anisotropy to the interactions.  

We encode a qutrit in the $\Delta m_N=0$ rotational states of an undressed
molecule,
\begin{align}
|1\rangle \equiv |N{=}0,0\rangle,
\quad
|2\rangle \equiv |N{=}1,0\rangle,
\quad
|3\rangle \equiv |N{=}2,0\rangle,
\label{eq_RbCs_encoding}
\end{align}
where $N$ and $\Delta m_N$ label the rotational angular momentum and its
projection.  Here we ignore the hyperfine structure.  Couplings $J$ below carry the unit $\mu_{\text{mol}}^2/(4\pi\epsilon_0 R_0^3)$, where $R_0$ is
the length unit of $\bm R_{ij}$ in Eq.~\eqref{eq_Vdd} and $\mu_{\text{mol}}$ is the dipole moment.  Global microwave control of these states has been
demonstrated \cite{BLACKMORE2020,RUTTLEY2025,Hepworth2025}.

For RbCs
the rotational constant $B/h=490.174$~MHz and centrifugal distortion
$D/h=207.3$~Hz \cite{BLACKMORE2020} place the two encoded
transitions at:
\begin{align}
f_{12}=2B/h-4D/h&\simeq980.347~\text{MHz},
\nonumber \\
f_{23}=4B/h-32D/h&\simeq1960.689~\text{MHz},
\label{eq_RbCs_frequencies}
\end{align}
where Planck's constant, $h$, was restored. 
The rotational ladder is anharmonic:
$f_{23}-f_{12}\simeq980.342$~MHz.  

The electric-dipole matrix elements of a rigid rotor are
\cite{MULLERa}:
\begin{align}
\mu_{N,N+1}
=
\mu_{\text{mol}}\,\frac{N+1}{\sqrt{(2N+1)(2N+3)}},
\label{eq_RbCs_dipole_elements}
\end{align}
where $\mu_{\text{mol}}=1.225$~D is the RbCs dipole moment 
\cite{MOLONY2014}.  The encoding gives
\begin{align}
\mu_{12}=\frac{\mu_{\text{mol}}}{\sqrt3}\simeq0.707~\text{D},
\qquad
\mu_{23}=\frac{2\mu_{\text{mol}}}{\sqrt{15}}\simeq0.633~\text{D},
\label{eq_RbCs_mu_values}
\end{align}
while $\mu_{13}=0$ is enforced by the $\Delta N=\pm1$ selection rule.
The bare rotational eigenstates carry no permanent laboratory-frame
dipole moments.

We project the dipole-dipole interaction between molecules onto the
encoded subspace and retain only resonant exchange processes. 
The native Hamiltonian can be written as in Eq.~\eqref{eq_RbCs_native}
where:
\begin{align}
J_{12}=2|\mu_{12}|^2=\frac{2\mu_{\text{mol}}^2}{3},
\qquad
J_{67}=2|\mu_{23}|^2=\frac{8\mu_{\text{mol}}^2}{15}.
\label{eq_RbCs_J_values}
\end{align}
The vanishing dipole moments eliminate all diagonal couplings,
$J_3=J_8=J_{38}=0$, and the selection rule eliminates the
$1\!\leftrightarrow\!3$ exchange channel, $J_{45}=0$.
Equation~\eqref{eq_RbCs_native} conserves every level occupation
$\hat N_m$.  The interaction is anisotropic with the parameter-free
ratio given by Eq.~\eqref{eq_RbCs_ratio} due 
to the rigid-rotor matrix elements.  Section~\ref{sec_d_3_dipole_pulse_planar} of the main text seeks to remove this anisotropy with Floquet engineering. 
Table~\ref{tab_RbCs_bond_strengths} lists bond strengths at
representative separations for $\theta_{ij}=\pi/2$.

\begin{table}[t]
\centering
\caption{Native RbCs bond strengths
 for in-plane geometry
($\theta_{ij}=\pi/2$), using $\mu_{\text{mol}}=1.225$~D.}
\label{tab_RbCs_bond_strengths}
\begin{tabular}{|c|c|c|}
\hline
$R_{ij}$ & $J_{12}V_{ij}/h$ & $J_{67}V_{ij}/h$  \\
\hline
$0.5~\mu$m & $1.21$~kHz & $0.97$~kHz \\
\hline
$1.0~\mu$m & $151$~Hz & $121$~Hz  \\
\hline
$2.0~\mu$m & $18.9$~Hz & $15.1$~Hz \\
\hline
\end{tabular}
\end{table}

Section~\ref{sec_d_3_dipole_pulse_planar} discusses the application of a square-pulse block on the
$\hat\lambda^4$ channel.  Because the $1\!\leftrightarrow\!3$ transition is
dipole forbidden in an RbCs implementation, the effective $\hat\lambda^4$ control could be generated by a two-photon microwave process. The Hamiltonian used here should then be interpreted as the effective rotating-frame control after eliminating the carrier-scale dynamics.

\bibliography{refs}

\end{document}